\documentclass[12pt,reqno]{article}%{smfart}
\usepackage{times}%\usepackage{mathptmx}
\usepackage{amsmath}
\usepackage{amsthm}
\usepackage{amsfonts}
\usepackage{amssymb}
\usepackage{epic}
\usepackage{eepic,epsfig}
\usepackage{times, macros}
\usepackage{graphics,color}

\usepackage{datetime,mathdots}
\usepackage{float}

\def\be{\begin{equation}}
\def\ee{\end{equation}}
\def\bea{\begin{eqnarray}}
\def\eea{\end{eqnarray}}
\def\bal{\begin{align}}
\def\eal{\end{align}}
\def\nn{\nonumber}

\def\qqq{\qquad}

\def\ex{\text{e}}

\def\ii{\text{i}}

\def\exp{\text{exp}}

\def\cM{\mathcal{M}}

\def\cU{\mathcal{U}}

\def\bs{\mathbf{s}}

\def\bn{\mathbf{n}}
\def\bm{\mathbf{m}}

\def\bC{\mathbb{C}}

\def\fc{\mathsf{c}}

\def\fq{\mathsf{q}}
\def\fq{{q}}

\def\sb{\mathsf{b}}
\def\sf{\mathsf{f}}
\def\sh{\mathsf{h}}

\def\sB{\mathsf{B}}

\def\sP{\mathsf{P}}
\def\sR{\mathsf{R}}

\def\sV{\mathsf{v}}

\newcommand{\mb}[1]{\mathbf{  {#1}}}

\makeatletter
\renewcommand*\env@matrix[1][*\c@MaxMatrixCols c]{%
  \hskip -\arraycolsep
  \let\@ifnextchar\new@ifnextchar
  \array{#1}}
\makeatother

\theoremstyle{plain}

\theoremstyle{remark}

\newcommand{\ot}{\otimes}
\newcommand{\ra}{\to}

\newcommand{\fsl}{{\mathfrak s}{\mathfrak l}}

\newcommand{\al}{\alpha}

\newcommand{\ga}{\gamma}

\newcommand{\de}{\delta}
\newcommand{\De}{\Delta}
\newcommand{\ep}{\epsilon}
\newcommand{\la}{\lambda}

\newcommand{\vf}{\varphi}

\newcommand{\pa}{\partial}
\newcommand{\CA}{{\mathcal A}}

\newcommand{\CC}{{\mathcal C}}

\newcommand{\CF}{{\mathcal F}}

\newcommand{\CH}{{\mathcal H}}
\newcommand{\CI}{{\mathcal I}}
\newcommand{\CL}{{\mathcal L}}
\newcommand{\CM}{{\mathcal M}}
\newcommand{\CN}{{\mathcal N}}
\newcommand{\CO}{{\mathcal O}}  
\newcommand{\CP}{{\mathcal P}}  
\newcommand{\CQ}{{\mathcal Q}}  
\newcommand{\CR}{{\mathcal R}}
\newcommand{\CS}{{\mathcal S}}

\newcommand{\CU}{{\mathcal U}}
\newcommand{\CV}{{\mathcal V}}
\newcommand{\CW}{{\mathcal W}}

\newcommand{\CY}{{\mathcal Y}}

\newcommand{\SD}{{\mathsf D}}

\newcommand{\SL}{{\mathsf L}}
\newcommand{\SM}{{\mathsf M}}

\newcommand{\SV}{{\mathsf V}}

\newcommand{\fe}{{\mathfrak e}}
\newcommand{\fg}{{\mathfrak g}}

\newcommand{\se}{{\mathsf e}}
\renewcommand{\sf}{{\mathsf f}}
\newcommand{\sk}{{\mathsf k}}

\newcommand{\sv}{{\mathsf v}}
\newcommand{\su}{{\mathsf u}}

\newcommand{\BR}{{\mathbb R}}

\newcommand{\BC}{{\mathbb C}}

\newcommand{\BZ}{{\mathbb Z}}

\newcommand{\rf}[1]{(\ref{#1})}

\begin{document}\thispagestyle{empty}
%\title{On the three-point conformal blocks in $\CW_3$ Toda CFT}
\title{Toda conformal blocks, quantum groups, and flat connections}
\author{Ioana Coman, Elli Pomoni, J\"org Teschner}
\address{Department of Mathematics, \\
University of Hamburg, \\
Bundesstrasse 55,\\
20146 Hamburg, Germany,\\[1ex]
and:\\[1ex]
DESY theory, \\
Notkestrasse 85,\\
20607 Hamburg,
Germany}
\maketitle
\begin{center}{\bf Abstract}\end{center}
\begin{quote}
{\small
This paper investigates the relations between the Toda conformal field theories, 
quantum group theory and the quantisation of moduli spaces of flat connections. 
We use the free field representation of the $\CW$-algebras to define natural bases for
spaces of conformal blocks of the Toda conformal field theory associated to the Lie algebra 
$\fsl_3$ on the three-punctured sphere
with representations of  generic type associated to the three punctures. 
The operator-valued monodromies of degenerate fields can be used to describe the
quantisation of the moduli spaces of flat $\mathrm{SL}(3)$-connections.
It is shown that the matrix elements of the  
monodromies can be  expressed as Laurent polynomials of 
more elementary operators which have a simple definition in the free field 
representation. These operators 
are identified as quantised counterparts of
natural higher rank analogs of the Fenchel-Nielsen 
coordinates from Teichm\"uller theory. Possible applications to the study of the 
non-Lagrangian SUSY field theories are briefly outlined.
}
\end{quote}

\section{Introduction}

Relations between conformal field theories (CFTs), quantum groups and 
the quantisation of moduli spaces of flat connections
have been investigated extensively in the past in connection with 
the Chern-Simons theory with compact gauge groups. It can be expected that this subject
has a somewhat richer counterpart in the cases where the gauge group becomes non-compact. 
Interest in the non-compact cases is motivated in particular by recent development in supersymmetric
field theory, as will be briefly outlined next.

\subsection{Motivation from $\CN=2$ supersymmetric field theory}

An important source of motivation for the study of the non-compact cases are
the relations between $\CN=2$ supersymmetric field theories in four dimensions
and two-dimensional conformal field theories discovered in
\cite{AGT}.  This development is deeply related to the study of the relations between $d=4$, $\CN=2$ 
supersymmetric field theories and quantum integrable models initiated in \cite{NS}.
An interesting class $d=4$, $\CN=2$ supersymmetric field theories, nowadays often referred to as class $\CS$, 
consists of theories 
$\mathfrak{X}(C,\mathfrak{g})$  labelled by a Riemann surface $C$ and a Lie algebra $\mathfrak{g}$ 
of ADE-type \cite{Ga,GMN}. The theory $\mathfrak{X}(C,\mathfrak{g})$ on a four-manifold $M^4$ 
is believed to be the  effective description for the six-dimensional $(2,0)$  theory associated to $\fg$ on 
$M^4\times C$ when the area of $C$ is small.
In the simplest case where $\mathfrak{g}=\mathfrak{sl}_2$
an essential role in this story is played by  the algebra $\CA_\mathfrak{X}$ of supersymmetric
Wilson and 't Hooft loops representing
an important sub-algebra of the algebra of all observables in $\mathfrak{X}(C,\mathfrak{g})$.
The algebra $\CA_\mathfrak{X}$ turns out to be isomorphic to the
algebra $\CA_V$ of Verlinde line operators in Liouville CFT \cite{AGGTV,DGOT,GOP,IOT}, 
which is furthermore isomorphic to the algebra $\CA_\CM$ of quantised functions
on the moduli space of flat $\mathrm{SL}(2)$-connections on Riemann surfaces \cite{TV13}, as summarised in
Table 1.
\begin{table}\label{Comparison1}
\begin{center}
\begin{tabular}{c|c|c}
Field theory  $\mathfrak{X}(C,\mathfrak{sl}_2)$ & Liouville theory & Moduli of flat connections on $C$\\ \hline\hline \\[-2ex]
Algebra $\CA_{\mathfrak{X}} $ generated by  & Algebra $\CA_V$ of Verlinde  & Quantised algebra $\CA_\CM$ of  \\
SUSY loop observables & line operators & functions on $\CM_{\rm flat}(\mathrm{SL}(2),C)$ \\[1ex] \hline \\[-2ex]
 Instanton partition & Liouville conformal  & Natural bases for  \\
  functions & blocks & modules  of $\CA_\CM$\\[1ex]\hline
 \end{tabular}
 \caption{\it Isomorphic algebras appear in different contexts.}
 \end{center}
\end{table}

One can argue \cite{TV13} that the isomorphisms between 
the algebras $\CA_\mathfrak{X}$, $\CA_V$ and $\CA_M$ imply the relations between
instanton partition functions for $\mathfrak{X}(C,\mathfrak{sl}_2)$ and conformal blocks
of Liouville CFT observed in \cite{AGT} and discussed in many subsequent works. 

In the cases where $\fg$ is a Lie algebra of higher rank one still expects to find 
a similar picture with Liouville CFT replaced by the conformal Toda CFTs associated
to finite-dimensional Lie algebras $\fg$. However, even in the next simplest case
$\mathfrak{g}=\mathfrak{sl}_3$ one encounters considerable additional difficulties 
obstructing direct generalisations of the results known for $\mathfrak{g}=\mathfrak{sl}_2$.
On the side of $\CN=2$ SUSY field theory one generically finds interacting field theories
not having a weak coupling limit with a useful Lagrangian description. 
It is not known how to calculate the partition 
functions in these cases. This difficulty has
a counterpart on the side of Toda CFT where the spaces of conformal blocks are known to 
be infinite dimensional generically, but otherwise poorly understood up to now.

This paper is a step towards the generalisation of the correspondences
between the theories $\mathfrak{X}(C,\mathfrak{g})$  and Toda CFT to the cases where 
no Lagrangian description is known. Motivated by the six-dimensional description of 
$\mathfrak{X}(C,\mathfrak{g})$, we take the quantised 
algebra of functions $\CA_\CM$ as an ansatz for the algebra 
$\CA_{\mathfrak{X}}$. This is perfectly consistent with previous work 
on the cases which do have a Lagrangian description \cite{Bu,GLF1,GLF2,LF}.
The relations between 
representations of  $\CA_\CM$ and the algebra of Verlinde line operators in 
Toda CFT investigated in \cite{CGT} suggest
that suitable Toda conformal blocks represent natural candidates for the
as yet unknown partition functions of the strongly coupled theories $\mathfrak{X}(C,\mathfrak{g})$.
This paper provides groundwork for such a  program in the  simplest nontrivial 
case $\fg=\mathfrak{sl}_3$. 

Our work is furthermore motivated by the papers \cite{BMPTY,MP,IMP}
where it was proposed that the conformal blocks of Toda CFT can be 
obtained in a certain limit from the partition functions of topological string theory
on certain toric Calabi-Yau manifolds used for the 
geometric engineering of $\mathfrak{X}(C,\mathfrak{g})$. 
With the help of the refined topological 
vertex one may represent the relevant
topological string partition 
functions as  infinite series admitting partial resummations. 
If the geometric engineering limit of these partition functions can be 
taken, it is expected to represent conformal blocks in Toda CFT. However, 
at the moment it is not known if this is the case and which basis for the conformal blocks in Toda CFT 
the topological string partition functions will correspond to.  
Only when this question is answered one  can extract  
predictions for the Toda three point functions from the results of 
\cite{BMPTY,MP,IMP}. We'll address the connections to topological string 
theory in subsequent papers.

\subsection{Context}  \label{Sec:Context}

One of our goals is to establish a precise relation between  conformal blocks in Toda CFT and specific 
states in the space of states obtained by quantising the moduli spaces of flat connections 
on Riemann surfaces $C$. The use of pants decompositions reduces the problem
to the basic case $C=C_{0,3}$. Some guidance is provided by the paradigm well-studied in the 
case of rational conformal field theories summarised in Table 2. 

\begin{table}\label{Comparison2}
\begin{tabular}{ccc}
\multicolumn{1}{c|}{Conformal field theory}   & \multicolumn{1}{c|}{Quantum group theory} & Moduli of flat connections \\ \hline\hline \\[-2ex]
\multicolumn{1}{c|}{Invariants in tensor products}  &  \multicolumn{1}{c|}{Invariants in tensor products of} & States \\
\multicolumn{1}{c|}{of $\CW$-algebra representations} & \multicolumn{1}{c|}{quantum group representations} & \\
\multicolumn{1}{c|}{(conformal blocks)} & \multicolumn{1}{c|}{} & 
\\[1ex] \hline
 \end{tabular}
 \caption{\it Comparison of modular functors from conformal field theory, quantum group theory (Reshetikhin-Turaev construction) and quantisation of the moduli spaces of flat connections on Riemann surfaces.}
\end{table}

However, the Toda CFTs of interest in this context are not expected to be rational CFTs. We'll therefore 
need a non-rational analog of the set of relations indicated in Table 2
generalising the situation encountered  in the case of Liouville theory, see \cite{TV13} and references therein.
The experiences made in this case suggest that 
a very useful role will be played in the  
non-rational cases by the Verlinde line operators, a natural family of operators $\SV_\ga$ on the spaces of conformal blocks % defined by means of the degenerate representations the chiral algebra
labelled by closed curves $\ga$ on the Riemann surface $C$.
The relevant spaces of conformal blocks can be abstractly 
{characterised} as modules for the algebra $\CA_V$
of Verlinde line operators \cite{TV13}. This means that these spaces of conformal blocks
have a natural scalar product making the Verlinde line operators self-adjoint. 
Natural representations of $\CA_V$ diagonalise maximal commutative 
sub-algebras of $\CA_V$. Such
sub-algebras of $\CA_V$ are labelled   by pants decompositions. 
Different pants decompositions give different representations of $\CA_V$ intertwined 
by unitary operators representing the fusion, braiding and modular transformations. 
%The  mapping class group is represented by operators which are unitary with respect to 
%this scalar product. 

A key step towards understanding the relation between conformal field theory and the quantum theory of 
flat connections is therefore to establish the isomorphism between
 the algebra of Verlinde line operators $\CA_V$ and the 
algebra of quantised functions on the corresponding moduli spaces of flat connections. In the case of 
Liouville theory this was established in \cite{TV13} using the explicit computations 
of Verlinde line operators performed in \cite{AGGTV,DGOT}. As discussed in \cite{TV13} 
one finds that the Verlinde line operators  generate a representation of the quantised algebra $\CA_\CM$
of functions on $\CM_{\rm flat}(\mathrm{PSL}(2,\BR))$. The generator of $\CA_\CM$ 
corresponding to the Verlinde line operator $\SV_{\ga}$ is the quantised counterpart $\SL_{\ga}$ of a 
trace function on $\CM_{\rm flat}(\mathrm{PSL}(2,\BR))$. The situation is schematically summarised
in Table 3.

An important feature of the representations associated to pants decompositions 
%calculated in \cite{AGGTV,DGOT} 
is the fact that the Verlinde line operators 
get expressed in terms of a set of more basic operators representing
quantised counterparts of certain coordinates on $\CM_{\rm flat}(\mathrm{PSL}(2,\BR))$
which are close relatives of the Fenchel-Nielsen coordinates 
for the Teichm\"uller spaces \cite{TV13}. The relevance of such coordinates had 
previously been emphasised in a related context in \cite{NRS}. The distinguished role
of these coordinates in our context is explained by the fact that they have two
important features: (i) They represent the algebraic structure of the moduli spaces 
of flat connections in the simplest possible way, and (ii) they are compatible with 
pants decompositions.

\begin{table}\label{Comparison3}
\begin{center}
%\begin{center}
\begin{tabular}{ccc}
& \multicolumn{1}{c|}{Conformal field theory}    & Moduli of flat connections on $C$\\ \hline\hline \\[-3ex]
\multicolumn{1}{c|}{Algebra} & \multicolumn{1}{c|}{$\CA_V$ of Verlinde} & Quantised algebra $\CA_\CM$ of  \\
\multicolumn{1}{c|}{}
& \multicolumn{1}{c|}{line operators} & functions on $\CM_{\rm flat}(\mathrm{SL}(2),C)$ \\ 
\multicolumn{1}{c|}{Generators} & \multicolumn{1}{c|}{Verlinde line operators $\SV_\ga$} & Quantised trace functions $\SL_\ga$\\ \hline \\[-3ex]
\multicolumn{1}{c|}{Module} & \multicolumn{1}{c|}{Spaces of conformal blocks}  & Spaces of states \\\hline
 \end{tabular}
 \caption{\it The vector spaces listed 
 in the row at the bottom can be abstractly characterised as modules of isomorphic algebras 
 $\CA_V\simeq \CA_\CM$.}
 \end{center}
\end{table}

\subsection{Contents of this paper}

The free field representation of the $\mathcal{W}$-algebras allows us to define
convenient bases for spaces of conformal blocks. Further developing the known 
connections between 
free field representations of conformal field theories and quantum group theory we 
will relate the spaces of conformal blocks to spaces of intertwining maps between tensor
products of quantum group representations and irreducible representations.  This reduces the
computation of monodromies to a problem in quantum group theory. The monodromy transformations 
get represented in terms of operator-valued matrices with matrix elements represented 
as finite-difference operators acting on the multiplicity labels for bases of conformal blocks. 
Our explicit computations allow us to  identify basic building blocks for the operator-valued monodromies
which can be identified with (quantised) coordinates for moduli spaces
of flat connections generalising the Fenchel-Nielsen coordinates mentioned above. 

Coordinates for $\CM_{\rm flat}(\mathrm{PSL}(3,\BR))$
satisfying the conditions (i), (ii) characterising coordinates of Fenchel-Nielsen type 
have been previously been introduced in \cite{Go,Ki}.
A systematic geometric approach to the definition of higher rank analogs of the 
Fenchel-Nielsen coordinates was proposed in \cite{HN,HK}. It would be very interesting 
to understand the relation between the coordinates defined in these references 
and the ones presented in this paper.

Some of the braiding matrices in Toda CFT have been calculated in \cite{GLF1,GLF2,LF}.
However, in all these cases the two representations associated to the punctures 
involved in the braiding
operations are degenerate or semi-degenerate, corresponding to theories 
$\mathfrak{X}(C, \fg)$ which have a Lagrangian description.

Section 2 introduces relevant background from conformal field theory and explains how 
to use conformal blocks with degenerate field insertions to define operator-valued
analogs of  monodromy 
matrices. A construction of conformal blocks for the $\CW_3$ algebra using the free field 
representation is introduced in Section 3. It is explained how this construction relates
bases for the spaces of conformal blocks on $C_{0,3}$ to bases in the space
of Clebsch-Gordan maps for the quantum group $\CU_q(\fsl_3)$. 
The connection to quantum group theory provided by the free field representation is used in
Section 4 to calculate basic cases of the monodromies of degenerate fields around generic chiral 
vertex operators in $\fsl_3$ Toda CFT. The results of this computation are then used in Section 5
in order to relate the basic building blocks of the operator valued monodromy matrices to 
coordinates of Fenchel-Nielsen type on the moduli spaces of flat $SL(3)$-connections.
There exist two different limits in Toda CFT corresponding to a classical limit 
for $\CA_V$, $b\ra 0$ and $b\ra i$, which both yield a 
deformation parameter $q^2=e^{-2\pi \mathrm{i}b^2}$ equal to $1$. We point out 
relations to the Yang's function for the Hitchin system, and to the Riemann-Hilbert 
problem for holomorphic $SL(3)$-connections on $C_{0,3}$ arising in 
the two limits, respectively. In the concluding Section 6 we discuss in particular the extension 
of our results to continuous families of conformal blocks as expected to be 
relevant for a full understanding of Toda CFT.

\section{Conformal blocks and quantum monodromies}\label{sec:confbl}
\setcounter{equation}{0}

The $A_{N-1}$ Toda CFTs are defined by the Lagrangian
\begin{equation}\label{LToda}
\mathcal{L}=\frac{1}{8\pi}(\partial_\nu\varphi, \partial_\nu\varphi)+\mu\sum_{i=1}^{N-1}e^{b(e_i,\varphi)}~,
\end{equation}
where $\vf$ is a $N-1$ component scalar field, the vectors 
$e_i$ are the simple roots of the Lie algebra $\mathfrak{sl}_N$, and $(.,.)$ denotes the standard scalar 
product in $\BR^{N-1}$.

\subsection{Chiral algebra $\mathcal{W}_N$ and its free field construction}

The chiral symmetry algebra of $A_{N-1}$ Toda CFT  is a $W_N$-algebra 
generated by holomorphic currents $W_d(z)$, $d=2,\dots,N$.
A quick way to introduce the algebras $\mathcal{W}_N$ starts from 
a collection of $N-1$ free chiral bosons, 
\begin{equation} 
\varphi_j(z)=q_j - \text{i} p_j \text{ln} z + \sum_{n\neq 0}\frac{\text{i}}{n}
a_n^jz^{-n} ~, \qquad j=1,\ldots , N-1 ~
\end{equation}
with modes satisfying the commutation relations 
\begin{equation}
[a_n^i,a_m^j]=n\delta_{ij}\delta_{n+m}~,\quad 
[p_j,q_k]=-i\delta_{jk},
\end{equation} 
and $(a_n^i)^\dagger=a_{-n}^i$.  
The currents $W_d(z)$ with mode expansions 
\begin{equation}  
W_d(z)=\sum_{n=-\infty}^{\infty} W_{d,n} z^{-n-d}~, \qquad 
j=2,\ldots ,N ~, 
\end{equation}
can then be defined through a deformed version of the Miura transformations, 
\begin{equation} \label{Miura}
:\prod_{i=0}^{N-1} \left( Q\partial + ( h_{N-i}, \partial\varphi(z)) 
\right) : 
= 
\sum_{k=0}^N W_{N-k}(z) (Q\partial)^k  
~,
\end{equation}
where $::$ denotes Wick ordering and 
$h_i = \omega_1 - \sum_{j=1}^{i-1} e_j $ are the weights 
of the  fundamental representation of $\mathfrak{sl}_N$ with highest 
weight $\omega_1$. The chiral algebra $\CW_N$ defined
in this way will be the chiral symmetry of Toda  CFT if the parameter $Q$ in 
\rf{Miura} is related to the parameter $b$ in the Lagrangian \rf{LToda} as 
$Q=b+b^{-1}$.

The currents $T(z)=W_2(z)$ generate a Virasoro sub-algebra within $\mathcal{W}_N$ with central charge
\begin{equation}
c=(N-1)\big(1+N(N+1)Q^2 \big).
\end{equation}

Highest weight representations of the algebras $\CW_N$ are generated from highest weight vectors 
$\fe_{\al}$ labelled by an $N-1$ component vector $\al$ satisfying 
\begin{equation}
W_{d,n}\fe_{\al}=0,\quad n>0,\qquad W_{d,0}\fe_{\al}=w_d(\al)\fe_{\al},\qquad d=2,\dots,N,
\end{equation}
where $w_d(\al)$ are polynomials of $\al$ of degree $d$, with 
\begin{equation}
w_2(\al)=\frac{1}{2}(2Q\rho_W^{}-\al,\al),
\end{equation}
being the conformal dimension. We are here using the notation
$\rho_W^{}$ for the Weyl vector of $\mathfrak{sl}_N$, which is equal to half the sum of all positive roots.

The space of states of the Toda CFTs will then be of the form 
$\CH=\int_{\mathbb{S}} d\al \,\CV_{\al}\otimes\bar{\CV}_{\al}$,
where $\bar{\CV}_{\al}$ is a  representation
for the algebra $\CW_N$ generated by the anti-holomorphic currents $\bar{W}_d(\bar{z})$.
The set of  labels $\mathbb{S}$ for the representations appearing in $\CH$ is 
expected to be the set of vectors $\al$ 
of the form $\al=\CQ+iP$, $P\in\mathbb{R}^{2}$, $\CQ=\rho_W(b+b^{-1})$, 
$\rho_W$ being the Weyl vector of $\mathfrak{sl}_N$. 

\subsection{Definition of conformal blocks}\label{defblocks}

One of the basic problems in Toda CFT originates from the fact that 
the spaces of conformal blocks corresponding to the three-punctured spheres are
infinite-dimensional, as will briefly be reviewed below. 
This implies that the Toda CFT three-point function can be represented 
in  the following form
%\begin{equation}\label{holofact3pt}\begin{aligned}
%\langle \,&V_{\al_3}(v_3\otimes w_3;z_3,\bar{z}_3) V_{\al_2}(v_2\otimes w_2;z_2,\bar{z}_2)V_{\al_1}(v_1\otimes w_1;z_1,\bar{z}_1)\,\rangle
%=\\
%&=\int dk d\bar{k}\;  C_{\mu}(k,\bar{k})
%\CF_{k}^{}(v_3(z_3) v_2(z_2)v_1(z_1))
%\CF_{\bar{k}}^{}(w_3(\bar{z}_3) w_2(\bar{z}_2)w_1(\bar{z}_1)).
%\end{aligned}\end{equation}
\begin{equation}\label{holofact3pt}\begin{aligned}
\langle \,&V_{\al_3}(v_3\otimes w_3;z_3,\bar{z}_3) V_{\al_2}(v_2\otimes w_2;z_2,\bar{z}_2)V_{\al_1}(v_1\otimes w_1;z_1,\bar{z}_1)\,\rangle
=\\
&=\int dk d\bar{k}\;  C_{\rho}(k,\bar{k})
f^{\rho}_{k}(v_3\otimes v_2\otimes v_1;z_3,z_2,z_1)
\bar{f}^{\rho}_{\bar{k}}(w_3\otimes w_2\otimes w_1;\bar{z}_3,\bar{z}_2,\bar{z}_1).
\end{aligned}\end{equation}
In \rf{holofact3pt} we are using the notation $V_{\al}(v\otimes w;z,\bar{z})$ for the vertex operator associated to 
a state $v\otimes w\in \CV_{\al}\otimes\CV_{\al}$ by the state-operator correspondence. 
It can be represented as  a normal ordered product of the primary field $V_{\al}(z,\bar{z})=
V_{\al}(\fe_\al\otimes \fe_\al;z,\bar{z})$ associated to the highest weight state $\fe_\al\otimes \fe_\al$
with differential polynomials in the fields representing the $\CW_3$-algebra.
The superscript $\rho$ refers to the triple of  labels $\rho=(\al_3,\al_2,\al_1)$ for the 
representations of the $\mathcal{W}$-algebra associated to the vertex operators appearing 
in the correlation function \rf{holofact3pt}. $f^{\rho}_k$ and $\bar{f}^{\rho}_{\bar{k}}$ 
are bases for the relevant (sub-)spaces of the spaces of conformal blocks on the three-punctured
spheres $C_{0,3}=\mathbb{P}^1\setminus\{z_1,z_2,z_3\}$ having $\CW$-algebra representations
with labels $\al_i$ assigned to the punctures $z_i$ for $i=1,2,3$, respectively. 

We will now review the basic definitions and results on the conformal blocks 
of $\CW$-algebras that will be used in the following. The presentation will be 
rather brief, the formulation being to a large extend analogous to the one 
presented in \cite{TLect} for the case of the Virasoro algebra.

The conformal blocks on $C_{0,n}=\mathbb{P}^1\setminus\{z_1,\dots z_n\}$,
are defined as the solutions to the Ward identities expressing the $\CW$-algebra symmetry of
$A_{N-1}$ Toda CFT under the $\CW_N$-algebra.
For any collection of meromorphic $(1-d)$-differentials $\xi_d=\xi_d(y)(dy)^{1-d}$, $d=2,\dots,N$, 
with poles only at $y=z_r$, $r=1,\dots,n$,
we may define operators $W_d[\xi_d]$ on the tensor product $\CR=\bigotimes_{i=1}^n \CV_{\al_n}$ by
setting 
\begin{equation}
W_d[\xi_d]:=\sum_{r=1}^n\sum_{k\in\BZ}\xi_{d,k}^{(r)}W_{d,k}^{}(z_r)\,,
\end{equation}
with $\xi_{d,k}^{(r)}$ being
the coefficients of the Laurent expansions of $\xi_d$ around $z_r$, $r=1,\dots,n$, %defined via
\begin{equation}
\xi_d(y)=\sum_{k\in\BZ}\xi_{d,k}^{(r)} (y-z_r)^{k+d-1}\,,
\end{equation}
and 
 $W_{d,k}^{}(z_r)$ representing  $W_{d,k}^{}$
 on the r-th tensor factor of $\CR$ associated to the puncture $z_r$,
\begin{equation}
W_{d,k}^{}(z_r)\,=\,{\rm id}\otimes\dots\otimes{\rm id}\otimes\underset{\text{r-th}}{W_{d,k}}\otimes{\rm id}
\otimes\dots\otimes{\rm id}.
\end{equation}
Using these notations we will define conformal blocks\footnote{This is the definition often 
given in the mathematical literature. The relation to the conformal Ward identities is discussed for 
the case $N=2$ e.g. in \cite{TLect}.}
as linear maps $f:\CR\ra \BC$ 
satisfying
\begin{equation}\label{CWIm}
f\big(W_d[\xi_d] v\big)\,=\,0\,,\qquad
\forall\;\,v\in \CR\,,\;\;
\forall\;\,d=2,\dots , N, \
\end{equation}
for all $\xi_d$ as introduced above.
The set of linear equations \rf{CWIm} defines a subspace $\mathrm{CB}(C,\CR)$
in the dual  $\CR'$ of the vector space
$\CR$ called
the space of conformal blocks.  

We will often find it convenient to assume that $z_n=\infty$ which does not restrict the 
generality in any serious way.
By using a part of the defining invariance conditions \rf{CWIm} one may %always 
reduce the computation of the linear functionals $f(v)$, $v\in\CR$, to the 
special values\footnote{See \cite[Section 1.7]{TLect} for the case $N=2$.} 
\begin{equation}\label{CWIm2}
g_f(v_{\infty})=f(\fe_{\al_1}\otimes\cdots\otimes \fe_{\al_{n-1}}\otimes v_{\infty}), \qquad v_{\infty}\in\CV_{\al_n}.
\end{equation}
There is a one-to-one correspondence between functionals $f:\CR\ra \BC$ 
satisfying \rf{CWIm}, and functionals $g:\CV_{\al_n}\ra \BC$ satisfying a 
reduced system of invariance conditions 
following from \rf{CWIm} and \rf{CWIm2}.  We will therefore also call such functionals $g$ 
conformal blocks. 

In this paper will mostly discuss the case $N=3$ corresponding to $\fsl_3$ Toda theory 
in the following, where we will use the simplified 
notations 
\begin{equation}
T(y)\equiv W_{2}(y)=\sum_{n\in\BZ}y^{-n-2}L_n,\qquad
W(y)=W_{3}(y)=\sum_{n\in\BZ}y^{-n-3}W_n.
\end{equation}
We expect to be able to reduce the case of an $n$-punctured sphere to the one for 
the three-punctured sphere by means of the gluing construction. 
For this case it is straightforward to show that the defining invariance property \rf{CWIm} allows one
to compute the values of 
$f(v)$ on arbitrary $v\in\CV_{\al_1}\otimes\CV_{\al_2}\otimes\CV_{\al_3}$ in terms 
of the particular values $F_{f,l}=f(\fe_{\al_1}\ot \fe_{\al_2}\ot(W_{-1})^l  \fe_{\al_3})$. 
This means that any conformal block $f$ on $C_{0,3}$ is uniquely defined by the
sequence of complex numbers $F_f=(F_{f,l})_{l\in\BZ_{\geq 0}}^{}$. 
As opposed to the case of the Virasoro algebra 
we therefore find an infinite-dimensional space of conformal blocks in the case of $\CW_3$.

The representation  of the $\CW_3$-algebra attached to the puncture $z_r$ 
generated by the operators $W_{d,n}(z_r)$ induces an action on the spaces of conformal blocks
$\mathrm{CB}(C,\CR)$ via 
\begin{equation}\label{W-act}
(W_{d,n}(z_r)f)(v)=f (W_{d,n}(z_r)v).
\end{equation}
The 
conformal block $W_{d,n}(z_r)f$ obtained by acting on $f$ with $W_{d,n}(z_r)$ 
is characterised by a sequence $(F_{W_{d,n}(z_r)f,l}^{})_{l\in\BZ_{\geq 0}}^{}$ which may be computed
by
specialising the definition \rf{W-act} to vectors of the form 
$v=\fe_{\al_1}\otimes \fe_{\al_2}\ot (W_{-1})^{l} \fe_{\al_3}$ 
and using \rf{CWIm} to express the right hand side of \rf{W-act} as a linear combination
of the $F_{f,l'}$, $l'\in\BZ_{\geq 0}$. 

Before one addresses the problem to compute the Toda three-point functions 
$C_{k,\bar{k}}(\al_1,\al_2,\al_3)$ using an expansion of the form \rf{holofact3pt}
one needs to find useful bases $\{f_k^\mu;k\in\CI\}$ for the space of conformal blocks 
appearing in \rf{holofact3pt}, with $\CI$ being a suitable index set.
According to the discussion above one might be tempted to consider the  
basis $\{f_l;l\in \BZ_{\geq 0}\}$  for $\mathrm{CB}(C,\CR)$ defined such that
$F_{f_l,n}=\de_{l,n}$. However, such a basis does not appear to be the most useful basis 
for the purpose to construct physical correlation functions in the form \rf{holofact3pt}.
Bases which appear to be more useful in this regard will be defined using the free 
field representation for the $\CW$-algebras.

It will also be useful to keep in mind the one-to-one correspondence between conformal blocks 
for three-punctured spheres $C_{0,3}=\mathbb{P}^1\setminus\{0,z_2,\infty\}$ 
and chiral vertex operators $V^\rho(v_2,z_2):\CV_{\al_1}\ra\CV_{\al_3}$, $\rho=(\al_3,\al_2,\al_1)$, $v_2\in\CV_{\al_2}$,
defined by the relation
\begin{equation}\label{bltoCVO}
f(v_1\otimes v_2\otimes v_3)=\langle v_3, V^\rho(v_2,z_2)v_1\rangle_{\CV_{\al_3}}^{},
\end{equation}
using an invariant bilinear form $\langle\, .\,,\,.\,\rangle_{\CV_{\al_3}}$ on $\CV_{\al_3}$. 
One may, more generally, relate conformal blocks $f_V$ 
to families of multi-local chiral vertex operators 
$V(w;{z}):\CV_{\al_1}\ra \CV_{\al_n}$, $w\in \CV_{\al_2}\otimes \dots \otimes\CV_{\al_{n-1}}$,
${z}=(z_1,\dots,z_{n-1})$,
defined such that
\begin{equation}\label{Vf}
f_V(v_1\otimes \dots \otimes v_n)=\langle v_n, V(w;{z})v_1\rangle_{\CV_{\al_n}}^{}.
\end{equation}
Such multi-local vertex operators can be constructed as compositions of vertex operators 
$V^\rho(v,z)$.
We may therefore construct large families of conformal blocks by 
constructing the chiral vertex operators 
$V^\rho(v,z)$, or equivalently  the conformal blocks associated to $C_{0,3}$. 
This will be our main goal here. 

%This will be done by means of the free field construction.

\subsection{Degenerate representations}\label{sec:degrep}

A special role is played by the vertex operators associated to degenerate representations
of the $\CW_3$-algebra. We will only need the examples where the representation label
$\al$ is equal to either $-b\omega_1$ or $-b\omega_{N-1}$, with $\omega_1$  and 
$\omega_{N-1}$ being the weights associated to the fundamental and anti-fundamental 
representation of $\mathfrak{sl}_N$, respectively. In these cases
one will find relations called null vectors 
among the vectors generated by the algebra $\CW_N$ 
from the highest weight vector $\fe_{\al}$. 
In the case $N=3$ one will find the following three null vectors  in the 
representation $\CV_{-b\omega_{1}}$:
\begin{align}
{\mathfrak{n}}_1:\quad & 0=(L_{-1}+\kappa_{1,1}W_{-1})\fe_{-b\omega_1}\,,\\
{\mathfrak{n}}_2:\quad & 0=(L_{-1}^2+\kappa_{2,1}^{} L_{-2}+\kappa_{2,2}W_{-2})\fe_{-b\omega_1}\,,\\
{\mathfrak{n}}_3:\quad & 0=(L_{-1}^3+\kappa_{3,1}^{} L_{-2}L_{-1}+\kappa_{3,2}^{}L_{-3}+\kappa_{3,3}^{}W_{-3})
\fe_{-b\omega_1}\,.
\end{align}
Explicit formulae for the coefficients can be found e.g. in \cite{FL}.
The relations in the  representation $\CV_{-b\omega_{N-1}}$
take a similar form, obtained by replacing $\kappa_{i,i}$ by $-\kappa_{i,i}$, $i=1,2,3$.

The null vectors in degenerate representations have well-known consequences for the 
vertex operators associated to such representations. If $\al_2=-b\omega_1$, for example, 
it follows from the relations above that $V^\rho(v_2,z_2):\CV_{\al_1}\ra\CV_{\al_3}$
can only be non-vanishing if $\al_3=\al_1-bh_\imath$ where $h_\imath$ are the weights of the fundamental representation of $\mathfrak{sl}_3$ for 
$\imath=1,2,3$. For $\al_2=-b\omega_{N-1}$ one will similarly find $\al_3=\al_1+bh_\imath$ 
for $\imath=1,2,3$. We may therefore abbreviate
the notations for the corresponding vertex operators to
$D_\imath(y)$ and $\bar{D}_{{\imath}}(y)$, respectively.  They map 
\begin{equation}
D_{\imath}:\CV_{\al}\ra \CV_{\al-bh_\imath},\qquad
\bar{D}_{\imath}: \CV_{\al}\ra \CV_{\al+bh_\imath},\qquad \imath=1,2,3.
\end{equation}

It is furthermore well-known that these relations imply differential equations satisfied by the 
conformal blocks if some of the tensor factors of $\CR=\bigotimes_{r=1}^n\CV_{\al_r}$ 
are degenerate in the sense above. 
To be specific we will in the following consider conformal blocks on a sphere 
with $n+2$ punctures $z_1=0,z_2,\dots,z_{n-1},y,y_0,\infty$, 
with representation $\CV_{-b\omega_{N-1}}$
associated to $y_0$ and representation $\CV_{-b\omega_{1}}$
associated to $y$. As in \rf{Vf} we may then consider the conformal blocks 
\begin{equation}
\CF_V(y,y_0;{z})_{\bar{\imath}\imath}^{}=\big\langle \fe_n, \bar{D}_{\bar{\imath}}(y_0) {D}_{\imath}(y)\, 
V(\fe_{\al_2}\otimes\dots\otimes\fe_{\al_{n-1}};{z})\,\fe_{\al_1}\big\rangle_{\CV_{\al_n}}^{}.
\end{equation}
The differential equations following from the null vectors include, in particular, 
\begin{align}\label{nvdecop}
\bigg(\frac{\pa^3}{\pa y^3}+\kappa_{3,1}^{}\,\mathsf{T}(y)\frac{\pa}{\pa y}
+\kappa_{3,2}\,\mathsf{T}'(y)+\kappa_{3,3}\,\mathsf{W}(y)\bigg)
\CF_V(y;y_0;z)_{\bar{\imath}\imath}^{}\,=\,0\,,
\end{align}
where 
\begin{align}
\mathsf{W}(y)=& \frac{w_3(-b\omega_1)}{(y-t_0)^3}+
\frac{\mathsf{W}_{-1}(y_0)}{(y-y_0)^2}+\frac{\mathsf{W}_{-2}(y_0)}{y-y_0}+\sum_{r=1}^{n-1}\left(
\frac{w_3(\al_r)}{(y-z_r)^3}+
\frac{\mathsf{W}_{-1}(z_r)}{(y-z_r)^2}+\frac{\mathsf{W}_{-2}(z_r)}{y-z_r}\right)  %\sum_{k=1}^M\left(
\,,\notag\\
\mathsf{T}(y)=& \frac{w_2(-b\omega_1)}{(y-y_0)^2}+\frac{1}{y-y_0}\frac{\pa}{\pa y_0}+\sum_{r=1}^{n-1}\left(
\frac{w_2(\al_r)}{(y-z_r)^2}+\frac{1}{y-z_r}\frac{\pa}{\pa z_r}\right)
%\sum_{k=1}^M\left(
\,.
\end{align}
Note that the null vectors ${\mathfrak{n}}_1$ allow us to express 
$\mathsf{W}_{-1}(y_0)$  in terms of 
$\mathsf{L}_{-1}(y_0)\sim\frac{\partial}{\partial y_0}$.
We may furthermore use the null vector ${\mathfrak{n}}_2$
together with the Ward identities to express $\mathsf{W}_{-2}(y_0)$ in terms of 
a differential operator $\mathcal{W}_{-2}(y_0)$ constructed out of 
derivatives $\frac{\partial}{\partial y_0}$ and $\mathsf{L}_{-1}(z_r)$. However, even in the
case $n=3$ one can not use the  invariance conditions defining the 
conformal blocks to express all
$ \mathsf{W}_{-i}(z_r)\CF_V(y;y_0;z)_{\bar{\imath}\imath}$, $i=1,2$ in terms 
of derivatives of $\CF_V(y;y_0;z)_{\bar{\imath}\imath}$, in general. 

Not having a closed 
system of differential equations represents an obstacle for the use of the differential
equations \rf{nvdecop}. Our goal is to define the analytic continuation
of $\CF_V(y;y_0;z)_{\bar{\imath}\imath}$ with respect to $y$
and compute
the resulting monodromies. Without a more explicit description of the action of
the algebra $\CW_3$ on spaces of conformal blocks one can not use
the differential equations discussed above to reach this goal. 
This is the main difference to the case where $N=2$. Other methods
are needed to reach our goal, as will be introduced next.

\subsection{Free field construction of conformal blocks}

We propose that useful bases for the spaces of conformal blocks can be defined using 
the free field representation of the algebra $\CW_3$. This section gives an outline of the
necessary constructions which is sufficient for the formulation of our main results. A 
more detailed description will be given in Section \ref{sec:freefield} below.

Key elements are the normal ordered exponentials
\begin{equation}\label{NOE}
V_{\al}(z)=e^{(\al,\vf(z))}.
\end{equation}
One of the most important properties for the following will be the exchange relations  
\bea \label{eq:exchangeRel1}
V_{\beta}(z') V_{\alpha}(z) = \ex^{-\pi\ii(\alpha ,\beta)} V_{\alpha}(z) V_{\beta}(z')~,
\eea
where the left hand side is understood as the analytic continuation from a region 
$|z'|>|z|$, and $z'$ moves in a counter-clockwise direction around $z$. 
The special fields $S_i(z)=V_{be_i}(z)$ will be referred to as 
the screening currents. From the screening currents $S_i(z)$ one can construct 
screening charges $Q_i(\ga)$ as
\begin{equation}\label{SC}
Q_i(\ga)=\int_{\ga} dz'\; e^{b(e_i,\vf(z'))},
\end{equation}
$\ga$ being a suitable  contour. Apart from the screening charges associated to the 
simple roots $e_i$, $i=1,2$, it will frequently be useful to consider the composite screening 
charge $Q_{12}$ defined as
\begin{equation}\label{composite}
Q_{12}(\ga)=\int_{\ga} dy\; S_{12}(y),\qquad S_{12}(y)=\int_{\ga_y}dy'\;S_1(y)S_2(y'),
\end{equation}
with $\ga_y$ being a suitable contour encircling the point $y$ which will be specified more precisely
below.

As will be discussed in more detail in Section \ref{sec:freefield} we may use these ingredients to 
construct conformal blocks on $C_{0,3}$ by expressions of the following form
\begin{align}\label{freeCVO}
g(v_{\infty})=
& \,\int_{\CC_{u_1}} du_1\dots \int_{\CC_{u_{m}}} du_{m}
\int_{\CC_{v_{1}}} dv_1\dots \int_{\CC_{v_{n}}} dv_{n} 
\int_{\CC_{w_{1}}} dw_1\dots \int_{\CC_{w_{p}}} dw_{p} \\
& \quad\qquad\big\langle v_{\infty}\,,\,
S_1(u_1)\dots S_1(u_m)  S_{12}(v_1)\dots S_{12}(v_n) S_2(w_1)\dots S_2(w_p)V_{\al_2}^{}(z_2)
\, \fe_{\al_1}\big\rangle .
\notag\end{align}
The precise definition of the expression on the right hand side of \rf{freeCVO} requires 
further explanations. The integrand is the multi-valued analytic function  obtained by 
standard normal ordering of the exponential fields appearing in the matrix element 
on the right hand side of \rf{freeCVO}.
Both the contours of integration and the precise choice of branch defining 
the integrand still need to be specified. 
All this will be discussed in more detail in Section \ref{sec:freefield} below.

The vertex operators appearing in \rf{freeCVO}
map from a  module $\CV_{\al_1}$ of the $\CW_3$ algebra
to the module $\CF_{\al_3}$, where $\al_3$ is 
related to $\al_2$, $\al_1$ and $\mathbf{n}=(m,n,p)$
by the conservation of the 
free field momentum,
\begin{equation}\label{cons}
\al_3=\al_1+\al_2+b(s_1e_1+s_2e_2),\qquad
\begin{aligned}
s_1&:= n+m,\\
s_2&:=n+p.
\end{aligned}
\end{equation}
Fixing the triple $\rho=(\al_3,\al_2,\al_1)$ of representation labels
only fixes the two combinations $s_1=m+n$ and $s_2=n+p$ of the three parameters $\mathbf{n}=(m,n,p)$.
It is therefore natural to parameterise $\mathbf{n}$ in terms of $s_1$, $s_2$ and an additional 
multiplicity label $k$ such that
\begin{equation}
\begin{aligned}
&m= s_1-k,\\
&p=s_2-k,
\end{aligned} 
\qquad n=k.
\end{equation}
For fixed external parameters $\rho=(\al_3,\al_2,\al_1)$ one may therefore 
use \rf{freeCVO}
to construct a one-parameter family of 
conformal blocks $F_k^\rho$.

\subsection{Quantum monodromies} \label{sec:quantmono}

A useful probe for the conformal blocks ${F}_{k}^\rho$ 
will be the $3\times 3$ matrix   of conformal blocks on $C_{0,3+2}$
having matrix elements denoted by 
$\mathcal{F}_{k}^\rho(y;y_0)_{\bar{\imath}\imath}^{}$, 
$\imath,\bar{\imath}=1,2,3$, defined as 
\begin{equation}\label{3+2blocks}
\mathcal{F}_{k}^\rho(y;y_0)_{\bar{\imath}\imath}^{}=
\langle \,{e}_{\al_3}^{}\,,\, \bar{D}_{\bar{\imath}}^{}(y_0) D_\imath^{}(y)\,
V_k^\rho(z)\,\fe_{\al_1}^{}\rangle.
\end{equation}
We will regard $y_0$ as a base-point, typically
assumed to be near $\infty$ on $C_{0,3}$ and sometimes use the abbreviated notation 
$\mathcal{F}_{k}^\rho(y)_{\bar{\imath}\imath}^{}\equiv
\mathcal{F}_{k}^\rho(y;y_0)_{\bar{\imath}\imath}^{}$.
We are going to demonstrate in Section \ref{sec:monocomp} below that 
$\mathcal{F}_k^{\rho}(y)_{\bar{\imath}\imath}^{}$ has an analytic continuation with respect to 
$y$ generating a representation of the fundamental group $\pi_1(C_{0,3})$ of the form
\begin{equation}\label{monod}
\mathcal{F}_{k}^\rho(\ga.y)_{\bar{\imath}\imath}^{}=
\sum_{\jmath}\sum_{l}(M_\ga)_{\imath,k}^{\jmath,l}\, \mathcal{F}_{l}^{\rho}(y)_{\bar{\imath}\jmath}^{} .
\end{equation}
We will see that only finitely many terms are nonzero in the summation over $l$.  
It will be useful to regard $\mathcal{F}_k^{\rho}(y)_{\bar{\imath}\imath}^{}$ as the components of 
vectors $\mathbf{F}^{\rho}_{}(y)_{\bar{\imath}\imath}^{}$ in a vector space with respect to a basis labelled
by  $k$, and accordingly 
the coefficients  $(M_\ga)_{\imath,l}^{\jmath,k}$ as matrix elements of operators 
$(\SM_\ga)_{\imath}{}^{\jmath}$, $\imath,\jmath=1,2,3$,  allowing us to write \rf{monod} in the form
\begin{equation}\label{monod'}
\mathbf{F}^{\rho}(\ga.y)_{\bar{\imath}\imath}^{}=
\sum_{\jmath}(\SM_\ga)_{\imath}{}^{\jmath}\cdot \mathbf{F}^{\rho}(y)_{\bar{\imath}\jmath}^{} .
\end{equation}
We may anticipate at this point that the 
matrices $(\SM_\ga)_{\imath}{}^{\jmath}$ will turn out to be
``quantum'' analogs of the monodromy matrices
of a flat $\mathrm{SL}(3)$-connection on $C_{0,3}$. 

The derivation of the relations \rf{monod'} described below will allow us to show
that the elements $(\SM_\ga)_{\imath}{}^{\jmath}$ of the quantum monodromy matrices 
can be expressed as functions $M_\ga(\su,\sv)_{\imath}{}^{\jmath}$ of the two operators
 \begin{equation}
\su\, \CF_{k}^\rho = q^{-2k} \CF_{k}^\rho,\qquad
\sv\, \CF_{k}^\rho = q^{-k} \CF_{k+1}^{\rho},
\end{equation}
generating a Weyl algebra 
$\CW_\hbar$ characterised  by the commutation relation 
$\su\sv=q^{-2}\sv\su$,
using the notation $q=e^{-\mathrm{i}\pi b^2}$. 
It will turn out that the $M_\ga(\su,\sv)_{\imath}{}^{\jmath}$ are Laurent polynomials 
of low orders in $\su$, $\sv$.  This property will be
essential for the interpretation of $\su$ and $\sv$ as quantised analogs of 
Fenchel-Nielsen type coordinates on moduli spaces of flat
connections to be discussed in Section \ref{Sec:flatconn}.

%It is straightforward to generalise our methods used to derive 
%the relations \rf{monod'} to the cases where $C_{0,3}$ is replaced 
%by $C_{0,n}$, with $V_k^\rho$ replaced by any multi-local vertex
%operator $V$ constructed as composition of vertex operators 
%$V_k^\rho$. 

\section{Free field construction of chiral vertex operators}\label{sec:freefield}
\setcounter{equation}{0}

We will now describe how to define the chiral vertex operators (CVOs) precisely, 
and how to compute the monodromies resulting from the analytic continuation of the location of
a degenerate chiral  vertex operator around the support of a generic CVO. 
%In the cases where $n_1$, $n_{12}$ and $n_2$ are integers o
One needs to 
specify a suitable set of contours for the integration of the screening currents
in order to complete the construction of the chiral vertex operators. Monodromies may
then be calculated by deforming the contours appropriately. 
The main goal of the following two sections will be to derive the claims on the 
structure of the quantum monodromy matrices formulated at the end of the previous 
section. The consequences of these results concerning the relations to the moduli 
spaces of flat connections will be explained in Section \ref{Sec:flatconn} below.

The approach described below is inspired by the works \cite{GS,FW} where 
the connection between the free field representation of the Virasoro algebra and 
the quantum group $\CU_q(\mathfrak{sl}_2)$ was investigated. 
Relations between free field representations of other conformal field theories and 
quantum groups of higher ranks have been studied in \cite{RRR,BMP,BFS}. 
However, the aims and scope of these works are different from ours, making it
difficult to extract the results relevant for us from the references above. 
We therefore give a self-contained treatment below.

\subsection{Basic definitions}

\begin{figure}[t]
\centering
\includegraphics[width=0.7\textwidth]{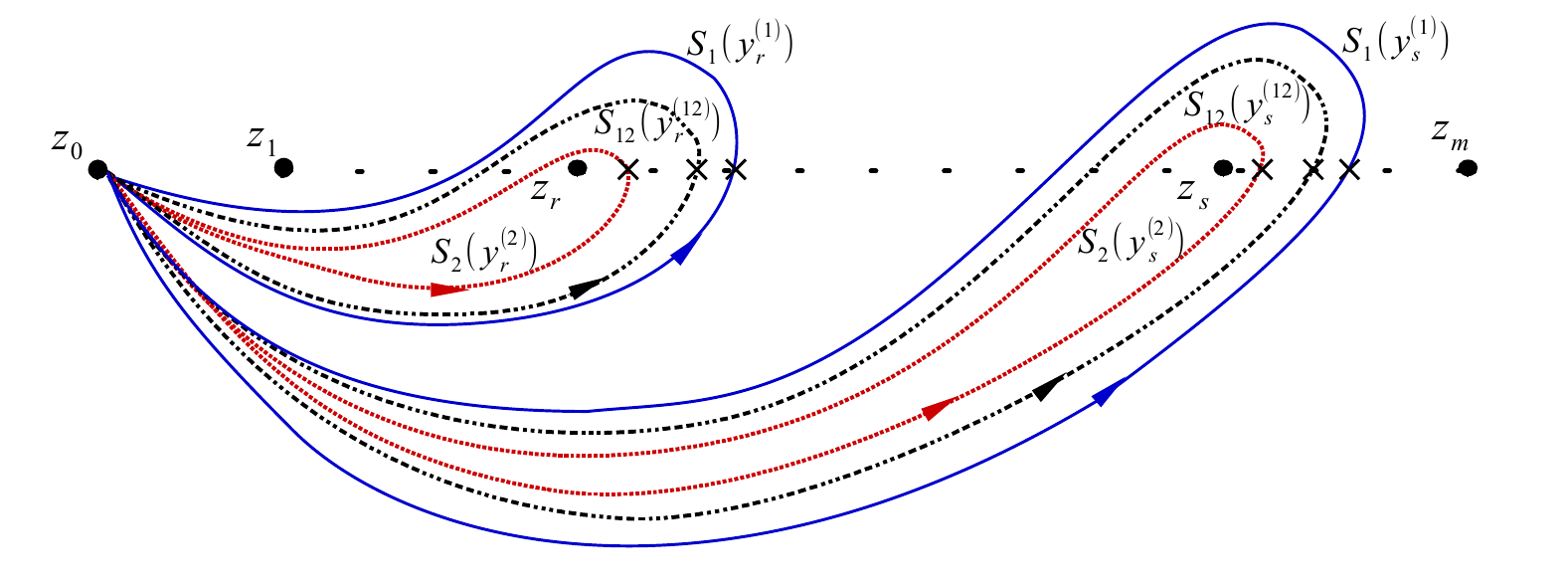}
\caption{{\it Nested non-intersecting and non-self-intersecting multi-contours for integration 
over screening currents associated with different vertex operators. 
The multi-contours around a puncture and which are associated 
with a particular screening charge are depicted collectively 
by one loop. }}
\label{Contours1}
\end{figure}

For both tasks it will be useful to represent  conformal blocks
as linear combinations of a family of auxiliary matrix elements of the following form
\begin{equation}\label{ansatz}
\big\langle v_{\infty}\,,\,V^{\rho_m}_{k_m}(z_m)\cdots V^{\rho_2}_{k_2}(z_2)\fe_{\al_1}\big\rangle=
\!\!\sum_{\mb{n}_m , \ldots , \mb{n}_1}\!\!
C\big(\begin{smallmatrix} \rho_m, \ldots, \rho_2 \\ 
k_m, \dots ,k_2 
\end{smallmatrix}\big)_
{\mb{n}_m , \ldots , \mb{n}_1}
{W}_{\mb{n}_m , \ldots , \mb{n}_1}^{\alpha_m , \ldots , \alpha_1} (z_m , \ldots ,z_2, 0) ,
\end{equation}
where $\rho_l=(\beta_{l},\al_l,\beta_{l-1})$ for $l=2,\dots,m$, $\beta_m=\al_\infty$
and $v_\infty\in\mathcal{F}_{\alpha_\infty}$. The functions
${W}_{\mb{n}_m , \ldots , \mb{n}_1}^{\alpha_m , \ldots , \alpha_1} (z_m , \ldots ,z_1)$
in \rf{ansatz} 
are defined  as multiple integrals
\begin{align}\label{SVO1}
 {W}_\mb{n_m , \ldots , n_1}^{\alpha_m , \ldots , \alpha_1} &(z_m , \ldots , z_1)  = \\ 
 &=\int_{\check{\Gamma}} 
d \mb{y}_m  \ldots d \mb{y}_1  \;\big\langle \,v_{\infty}\,,\,
\mb{S}^{\mb{n}_m}_m(\mb{y}_m) V_{\alpha_m} (z_m) \ldots 
\mb{S}^{\mb{n}_1}_1(\mb{y}_1) V_{\alpha_1} (z_1)\fe_0\big\rangle,
\notag\end{align}
over the
contours $\check{\Gamma}=\Gamma^{\mb{n}_m}_m \times \ldots \times 
\Gamma^{\mb{n}_1}_1$ depicted in Figure \ref{Contours1}. The crosses in Figure \ref{Contours1} 
indicate normalisation points on the contour $\check{\Gamma}$
where the multi-valued integrand in \rf{SVO1} is real if $z_r\in\BR^+$, $r=2,\dots,m$, and
$0<z_1<\dots<z_m$.
We are using the notations
$\mb{n}_l=(n_{l,1},n_{l,12},n_{l,2})$ for
$l=1,\dots,m$, and 
\begin{equation}
\mb{S}^{\mb{n}_l}_l(\mb{y}_l):=
S_1(y_{1,1}^{(l)})\dots S_1(y_{1,n_{l,1}}^{(l)})  S_{12}(y_{12,1}^{(l)})\dots 
S_{12}(y_{12,n_{l,12}}^{(l)}) S_2(y_{2,1}^{(l)})\dots S_2(y_{2,n_{l,2}}^{(l)}),
\label{Scurrents}
\end{equation}
and assume that
\begin{equation}
\alpha_\infty=\sum_{l=1}^m (\alpha_i+b(n_{l,1}+n_{l,12})e_1+b(n_{l,2}+n_{l,12})e_2).
\end{equation}
The contours used to define the composite screening currents $S_{12}(y)$ in \rf{composite} should be chosen such
that the only singular point of the integrand in \rf{SVO1} encircled by $\ga_y$ is the point $y$.  

The functions ${W}_\mb{n_m , \ldots , n_1}^{\alpha_m , \ldots , \alpha_1}$ do not by themselves 
represent conformal blocks,
as is easily seen as follows. The commutators between screening currents $S_i(z)$ and the generators
of the $\mathcal{W}_3$-algebra are total derivatives with respect to the variable $z$. When the 
screening currents $S_i(z)$ are integrated over open contours as depicted in Figure \ref{Contours1}, 
one will therefore find boundary terms  in the commutators between
screening charges and the generators
of the $\mathcal{W}_3$-algebra 
supported at the base-point $z_0$. The boundary terms will spoil the validity of the Ward identities
for the $\mathcal{W}_3$-algebra, in general. 

However, we will see that the boundary terms all cancel for specific choices
of the coefficients 
$C$ in (\ref{ansatz}).
Representing compositions of chiral vertex operators in the form (\ref{ansatz})
is useful since  both the problem to find the coefficients 
$C$ in (\ref{ansatz})
and the calculation of monodromies will be shown to reduce to purely algebraic problems 
in quantum group theory. In order to explain this relation we shall begin by 
collecting the definitions and notations concerning the quantum group
$\cU_q(\fsl_3)$ relevant for our goals.

\subsection{Quantum group background}

%%%%%%%%%%%%%%%%%%%%%%%%%%%%%%%%%%%%%%%%%%%%%%%%%%%%%%%%%
%%%%%%%%%%%%%%%%%%%%%%%%%%%%%%%%%%%%%%%%%%%%%%%%%%%%%%%%%

%[91] V. Chari and A. Pressley, A guide to quantum groups. Cambridge University Press, 1994.

%A Hopf algebra is a collection 
%$(\cU,m,\eta,\Delta,\epsilon,S)$, where the various 
%elements are the unit $\eta : \bC (q) \rightarrow \cU$, 
%product $m : \cU\otimes\cU \rightarrow \cU$, 
%coproduct $\Delta : \cU \rightarrow \cU\otimes\cU$, 
%counit $\epsilon : \cU \rightarrow \bC (q)$ and an 
%algebra antiautomorphism called antipode 
%$S : \cU \rightarrow \cU$. The triple $(\cU,m,\eta)$ is a 
%unital associative algebra and $(\cU,\Delta,\epsilon)$ 
%is a counital associative coalgebra. $\Delta$ and $
%\epsilon$ are unital algebra homomorphisms and satisfy 
%the following relations 
%\bea
%(\Delta\otimes1)\circ\Delta &=& (1
%\otimes\Delta)\circ \Delta \qqq 
%\text{(co-associativity) } \nn\\
%(\epsilon\otimes1)\circ\Delta &=& (1
%\otimes\epsilon)\circ \Delta  = 1 
%\nn\\ 
%m\circ(S\otimes1)\circ \Delta &=&  m\circ(1\otimes S)\circ \Delta \qqq 
%\text{(pentagon)} ~ . 
%\eea 
The quantum group $\cU = \cU_q(\fsl_N)$ is 
a Hopf algebra, generated as an associative algebra over 
$\bC (q)$ by $\se_i$, $\sf_i$ and $\sk^\pm_i$ for $i=1,\ldots , N-1$ which satisfy the relations 
\begin{equation}
[\se_i , \sf_j] = \delta_{ij} \frac{\sk_i-\sk_i^{-1}}{\fq-\fq^{-1}} ,\qquad
\begin{aligned}
&\sk_i^\pm \se_j = \fq^{\pm\kappa_{ij}} \se_j \sk_i^\pm ~, 
\\
&\sk_i^\pm \sf_j = \fq^{\mp\kappa_{ij}} \sf_j \sk_i^\pm ~ 
\end{aligned}
\end{equation}
for $i=1,\ldots , N-1$ and where $\kappa$ is the $\fsl_N$ Cartan matrix
$ 
\kappa_{ij} = 2\delta_{ij}-\delta_{i+1,j}-\delta_{i,j+1} ~.
$
The Serre relations are 
\be 
\se_i^2 \se_j - (\fq+\fq^{-1}) \se_i \se_j \se_i + \se_j \se_i^2 = 0 ~, \quad
\sf_i^2 \sf_j - (\fq+\fq^{-1}) \sf_i \sf_j \sf_i + \sf_j \sf_i^2 = 0 ~,
\ee
with $|i-j|= 1$. Let us now restrict to the case $N=3$. Defining 
\be \label{def:qgp-e12f12}
\se_{12} = \se_1 \se_2 - \fq \se_2 \se_1 ~,\qqq 
\sf_{12} = \sf_1 \sf_2 - \fq \sf_2 \sf_1 ~,
\ee
it follows from the Serre relations that
%$\sk_i \sf_{12} = \fq^{-1} \sf_{12} \sk_i$ 
%and similarly for $\se_{12}$ 
%and 
\bea \label{eq:qgp-e12f12}
\se_1 \se_{12} &=& \fq^{-1} \se_{12} \se_1 ~, \qqq 
\se_2 \se_{12} = \fq \se_{12} \se_2 ~,  \\
%\sk_i \sf_{12} = \fq^{-1} \sf_{12} \sk_i ~, \qqq
\sf_1 \sf_{12} &=& \fq^{-1} \sf_{12} \sf_1 ~, \qqq 
\sf_2 \sf_{12} = \fq \sf_{12} \sf_2~ .  
\eea
Using
these relations it is possible to show that arbitrary monomials formed out of the 
basic generators $\sf_i$, $i=1,2$ can be represented as linear combinations of the 
ordered monomials $\sf_1^{n_1}\sf_{12}^{n_{12}}\sf_2^{n_2}$ with 
$(n_1,n_{12},n_2)\in\mathbb{N}^3$.

\subsubsection{Representations}

The irreducible highest weight representation of $\cU_q(\fsl_3)$ with highest weight $\lambda$
will be denoted by $\mathcal{R}_{\lambda}$. The highest weight vector $v_\lambda\in\mathcal{R}_{\lambda}$ 
is annihilated by the 
generators $\se_i$, $\se_i v_\lambda = 0$. The action of $\sk_i$ is 
\be 
\sk_i v_\lambda = \fq^{(e_i , \lambda)} v_\lambda ~.
\ee
The representations $\mathcal{R}_{\lambda}$ are irreducible for generic weights $\lambda$. 
We call a weight $\la$ generic if it is not the highest weight of a finite-dimensional highest weight representation
of $\fsl_3$. 
We will use a  basis  for the $\cU_q(\fsl_3)$ Verma module $\mathcal{R}_{\lambda}$
which consists of the  following vectors
\be \label{basis}
e_{\mb{n}}^\lambda = 
\sf_1^{n_1}\sf_{12}^{n}\sf_2^{n_2}v_\lambda ~,\qquad \mb{n}=(n_1,n,n_2)~.
\ee
The vector $e_{\mb{n}}^\lambda$ has weight 
\be \label{weight}
\nu=\nu(\lambda,{\mb{n}}) = \lambda - (n+n_1) e_1 - (n+n_2) e_2 ~.
\ee
It will sometimes be useful to parameterise the labels $\mb{n}=(n_1,n,n_2)$
in terms of another pair of data $\mb{n}=[\delta,n]$ 
consisting of the shift of weights $\delta= - s_1 e_1 - s_2 e_2$
together with a multiplicity label $n$ such that $s_1=n_1+n$ and $s_2=n_2+n$. %and $n=m$.
For a fixed weight $\nu=\la+\de$ parameterised by positive integers $s_1$ and $s_2$, one can only 
have finitely many values of $n$ between $0$ and $d_\nu-1=\mathrm{min}(s_1,s_2)$. 
The multiplicity space of a weight $\nu$ has dimension $d_{\nu}$ 
and will be denoted by $\cM^{\nu}_\la$.

The action of the generators $x$ of $\cU_q(\fsl_3)$  is represented in the basis (\ref{basis}) 
by the matrices 
$R_{(x),\bn}^{\lambda , \bn'}$ defined by
\begin{equation}
\pi_\lambda(x)e_{\mb{n}}^\lambda = \sum_{\mb{n}'}
R_{(x),\bn}^{\lambda , \bn'}e_{\mb{n}'}^\lambda.
\end{equation}
Tables for the matrix elements of  $R_{(x),\bn}^{\lambda , \bn'}$ for $x=\se_i , \sf_i$
can be found  in appendix \ref{app:UsefulRel}. 
Here we will only note the following important feature concerning the dependence on
the multiplicity labels $k$. The action of $\pi_\lambda(x)$ on the basis vector $e_{\mb{n}}^\lambda$
can for $x=\se_i , \sf_i$, $i=1,2$, be represented in the following form
\begin{equation}\label{UVrep}
\begin{aligned}
\pi_\lambda(\se_i)e_{[\de,k]}^\lambda &= E_i(\lambda,\de,u_k)\,e_{[\de+e_i,k]}^\lambda+E_i'(\lambda,\de,u_k)
\,\mathsf{v}^{-1}\,e_{[\de+e_i,k]}^\lambda~, \\
\pi_\lambda(\sf_i)e_{[\de,k]}^\lambda &= F_i(\lambda,\de,u_k)\,e_{[\de-e_i,k]}^\lambda+F_i'(\lambda,\de,u_k)
\,\mathsf{v}^{+1}\,e_{[\de-e_i,k]}^\lambda~,
\end{aligned}
\end{equation}
where $u_k=q^{-2k}$,  $\mathsf{v} e_{[\de,k]}^\lambda= q^{-k}
e_{[\de,k+1]}^\lambda$, and the coefficient functions $E_{i}$, $E_i'$, $F_i$ and $F_i'$ are Laurent polynomials
in the variable $u_k$. This simple feature will play a crucial role for us in the following.

Apart from the representations associated to generic weights $\la$ we will be interested in 
particular in the fundamental representation  generated from 
the  highest weight vector $v_{\omega_1}$ and having a basis 
$\{v_{\omega_1}, \sf_1 v_{\omega_1}, \sf_{12} v_{\omega_1}\}$.

\subsubsection{Tensor products of representations}\label{tensor}

The coproduct $\Delta$ of $\cU_q(\fsl_3)$  is defined by
\be \label{def:coproduct:ei-fi}
\Delta (\se_i) = \sk_i \otimes \se_i + \se_i \otimes 1 ~, \quad
\Delta (\sf_i) = \sf_i \otimes \sk_i^{-1} + 1 \otimes \sf_i ~, \quad
\Delta(\sk_i) =\sk_i \otimes\sk_i.
\ee
The coproduct is an algebra homomorphism, so equations 
\eqref{def:coproduct:ei-fi} imply, for example,
\bea \label{def:coproducte12}
\Delta (\se_{12}) &=& 
\sk_1 \sk_2 \otimes \se_{12} + \se_{12} \otimes 1 
+ (q^{-1}-q) \se_2 \sk_1 \otimes \se_1  ~
\nn\\
\Delta (\sf_{12}) &=& 
\sf_{12} \otimes \sk_1^{-1} \sk_2^{-1}  + 
1 \otimes \se_{12} 
+ (q^{-1}-q) \sf_1 \otimes \sf_2 \sk_1^{-1}  ~.
\eea
The Clebsch-Gordan (CG) maps describe the embedding of  irreducible representations into 
tensor products of representations. In the case of the highest
weight representations introduced above, one may represent the CG maps $\mathsf{C}^{\lambda_3}_{\lambda_1\lambda_2}:\mathcal{R}_{\lambda_3}\rightarrow
\mathcal{R}_{\lambda_1}\otimes\mathcal{R}_{\lambda_2}$ in 
the form 
\begin{equation}
e^{\lambda_3}_{\mb{n}_3}(\lambda_2,\lambda_1)=
\sum_{\substack{\mb{n}_2,\mb{n}_1 \\ \nu_3=\nu_2+\nu_1}}
\left(\begin{smallmatrix}
\lambda_{3}  \\   \mb{n}_3
\end{smallmatrix}\middle|
\begin{smallmatrix}
 \lambda_2 & \lambda_1 \\  \mb{n}_2 & \mb{n}_1
\end{smallmatrix}\right) 
e^{\lambda_1}_{\mb{n}_1} \otimes e^{\lambda_2}_{\mb{n}_2}~,
\end{equation} 
Here and below we use the notation $\nu_i=\nu_i(\la_i,\mb{n}_i)$ with $\nu(\lambda,{\mb{n}})$
defined in \rf{weight}.

A  set of equations for the Clebsch-Gordan 
coefficients (CGC) $\left(\begin{smallmatrix}
\lambda_{3}  \\   \mb{n}_3
\end{smallmatrix}\middle|
\begin{smallmatrix}
 \lambda_2 & \lambda_1 \\  \mb{n}_2 & \mb{n}_1
\end{smallmatrix}\right) $ follows from the intertwining property
\be \label{intertw}
\Delta(x) e_{\bn_3}^{\lambda_3}(\lambda_2,\lambda_1) = \sum_{\bn'_3}R_{(x),\bn_3}^{\lambda_3 , \bn'_3} 
e_{\bn_3'}^{\lambda_3}(\lambda_2,\lambda_1)~,
%\sum_{\substack{
%\bn_2,\bn_1 \\ \nu_3'=\nu_2+\nu_1}} 
 %\left(\begin{smallmatrix}
%\lambda_{3}  \\   \bn'_3
%\end{smallmatrix}\middle|
%\begin{smallmatrix}
 %\lambda_2 & \lambda_1 \\  \mb{n}_2^{\phantom '} & \mb{n}_1
%\end{smallmatrix}\right) 
%e^{\lambda_1}_{\mb{n}_1} \otimes e^{\lambda_2}_{\mb{n}_2} ~,
\ee
for any element $x\in \cU_q(\fsl_3)$.
When $x=\se_i\in\{\se_1,\se_2\}$ and $\bn_3=0$, this 
equation implies that for every fixed pair $(\bn_1,\bn_2)$ 
\be \label{intertw2}
0= \sum_{\mb{n}'_2} \big(\begin{smallmatrix}
\lambda_3  \\ 0 
\end{smallmatrix}\big|
\begin{smallmatrix}
 \lambda_2 & \lambda_1 \\  \mb{n}'_2 & \mb{n}_1
\end{smallmatrix}\big)
\fq^{(e_i,\nu_1)}  
 R_{(\se_i),\mb{n}'_2}^{\lambda_2 , \mb{n}_2}  + 
\sum_{\mb{n}'_1} \big(\begin{smallmatrix}
\lambda_3  \\ 0 
\end{smallmatrix}\big|
\begin{smallmatrix}
 \lambda_2 & \lambda_1 \\  \mb{n}_2 & \mb{n}'_1
\end{smallmatrix}\big) 
R_{(\se_i),\mb{n}'_1}^{\lambda_1 , \mb{n}_1}  
\ee
Equations
(\ref{intertw2}) can be regarded as a set of recursion relations.
For generic values of $\lambda_i$, $i=1,2,3$ one can show that these equations determine 
$\left(\begin{smallmatrix}
\lambda_{3}  \\  0
\end{smallmatrix}\middle|
\begin{smallmatrix}
 \lambda_2 & \lambda_1 \\  \mb{n}_2 & \mb{n}_1
\end{smallmatrix}\right)$
uniquely in terms of 
$\left(\begin{smallmatrix}
\lambda_{3}  \\  0
\end{smallmatrix}\middle|
\begin{smallmatrix}
 \lambda_2 & \lambda_1 \\  \mb{n}_2 & 0
\end{smallmatrix}\right)$. 
Indeed, it is not hard to see (see Appendix B for further details) that  one 
can use equations (\ref{intertw2}) to derive a recursion relation expressing 
$\left(\begin{smallmatrix}
\lambda_{3}  \\  0
\end{smallmatrix}\middle|
\begin{smallmatrix}
 \lambda_2 & \lambda_1 \\  \mb{n}_2 & \mb{n}_1
\end{smallmatrix}\right)$
with $\mb{n}_1=[-s_1e_1-s_2e_2,k_1]$ in terms of  the CGC
$\left(\begin{smallmatrix}
\lambda_{3}  \\  0
\end{smallmatrix}\middle|
\begin{smallmatrix}
 \lambda_2 & \lambda_1 \\  \mb{m}_2 & \mb{m}_1
\end{smallmatrix}\right)$
having $\mb{m}_1=[-s_1'e_1-s_2'e_2,k_1']$ with $s_1'+s_2'=s_1+s_2-1$.

The value of $\left(\begin{smallmatrix}
\lambda_{3}  \\  0
\end{smallmatrix}\middle|
\begin{smallmatrix}
 \lambda_2 & \lambda_1 \\  \mb{n}_2 & 0
\end{smallmatrix}\right)$ is not further constrained by (\ref{intertw2}).
A basis for the space of  CG maps $\mathsf{C}^{\lambda_3}_{\lambda_1\lambda_2}$ is therefore obtained from the CGC
 $\left(\begin{smallmatrix}
\lambda_{3}  \\   \mb{n}_3
\end{smallmatrix}\middle|
\begin{smallmatrix}
 \lambda_2 & \lambda_1 \\  \mb{n}_2 & \mb{n}_1
\end{smallmatrix}\right)_k$
characterised by the property that 
\begin{equation}\label{normcond}
\left(\begin{smallmatrix}
\lambda_{3}  \\   0
\end{smallmatrix}\middle|
\begin{smallmatrix}
 \lambda_2 & \lambda_1 \\  \mb{n}_2 & 0
\end{smallmatrix}\right)_k = \delta_{k,k_2}, \quad \text{if}\quad \mb{n}_2=[\delta_2,k_2],\quad
\delta_2=\lambda_3-\lambda_1-\lambda_2.
\end{equation}
%with $\delta_2$ fixed by the condition $\delta_2=\lambda_3-\lambda_1-\lambda_2$.  
As $k_2$ can only take a finite set of values
determined by  $\delta_2$ we see that the  coefficients 
$\left(\begin{smallmatrix}
\lambda_{3}  \\   \mb{n}_3
\end{smallmatrix}\middle|
\begin{smallmatrix}
 \lambda_2 & \lambda_1 \\  \mb{n}_2 & \mb{n}_1
\end{smallmatrix}\right)_k$ define a basis for the space of CG maps. 
For generic weights $\la_1$, $\la_2$, $\la_3$ there is an isomorphism 
between the space of CG maps and the subspace with weight $\lambda_3-\lambda_1$ within
$\mathcal{R}_{\lambda_2}$, or equivalently the 
subspace with weight $\lambda_3-\lambda_2$ within
$\mathcal{R}_{\lambda_1}$.

It will be important to keep in mind that the  $\left(\begin{smallmatrix}
\lambda_{3}  \\  0
\end{smallmatrix}\middle|
\begin{smallmatrix}
 \lambda_2 & \lambda_1 \\  \mb{n}_2 & \mb{n}_1
\end{smallmatrix}\right)_k$ defined using \rf{normcond}
depend on the multiplicity labels only through the matrix elements
$R_{(\se_i),\mb{n}'}^{\lambda , \mb{n}}$. The features concerning the 
dependence of $R_{(\se_i),\mb{n}'}^{\lambda , \mb{n}}$
on the multiplicity labels observed in (\ref{UVrep}) will
be inherited. 

Note furthermore that the recursive procedure determining the CGC 
$\left(\begin{smallmatrix}
\lambda_{3}  \\  0
\end{smallmatrix}\middle|
\begin{smallmatrix}
 \lambda_2 & \lambda_1 \\  \mb{n}_2 & \mb{n}_1
\end{smallmatrix}\right)_k$
proceeds
by induction in $s=s_1+s_2$ if $\mb{n}_1=[-s_1e_1-s_2e_2,k_1]$. In each step one generically increases the 
range of values for the multiplicity label $k_2$ in $\mb{n}_2=[\de_2,k_2]$ 
with non-vanishing $\left(\begin{smallmatrix}
\lambda_{3}  \\  0
\end{smallmatrix}\middle|
\begin{smallmatrix}
 \lambda_2 & \lambda_1 \\  \mb{n}_2 & \mb{n}_1
\end{smallmatrix}\right)_k$ by one unit.
In the case  $\mb{n}_1=[-s_1e_1-s_2e_2,k_1]$
with $s_1+s_2\leq 2$, for example, we find non-vanishing $\left(\begin{smallmatrix}
\lambda_{3}  \\  0
\end{smallmatrix}\middle|
\begin{smallmatrix}
 \lambda_2 & \lambda_1 \\  \mb{n}_2 & \mb{n}_1
\end{smallmatrix}\right)_k$  
only for $k-2\leq k_2\leq k$. Further details are given in Appendix \ref{sec:app:CGcoeff}. 

\subsubsection{Iterated Clebsch-Gordan maps}

%As usual, we may relate invariants in $n$-fold tensor products $\bigotimes_{l=1}^n\mathcal{R}_l$ 
%to intertwining maps $\mathsf{C}:\bigotimes_{l=1}^{n-1}\mathcal{R}_l\rightarrow \bar{\mathcal{R}}_n$
%with $\bar{\mathcal{R}}_n$ being the conjugate of ${\mathcal{R}}_n$ by introducing 
%the invariant 
%$\mathsf{B}_{\mathcal{R}_n}:\bar{\mathcal{R}}_n\otimes {\mathcal{R}}_n\rightarrow \mathbb{C}$, 
%and
%\begin{equation}
%I(v_{n-1}\otimes v_{n}):=\mathsf B_{\mathcal{R}_n}^{}(\mathsf{C}(v_{n-1})\otimes v_n)~.
%\end{equation}
Families of intertwining maps $\mathsf{C}:\bigotimes_{l=1}^{n-1}\mathcal{R}_l\rightarrow \bar{\mathcal{R}}_n$
can be constructed as compositions of the Clebsch-Gordan maps introduced in Section \ref{tensor}. 
For $m=3$, for example, one may consider  linear combinations of the form
\begin{align}\label{fourpt-qgrp}
&e^{\lambda_4}_{\mb{n}_4}(\lambda_3,\lambda_2,\lambda_1)
=\sum_{\substack{\bn_3, \bn_{2}, \bn_{1} \\ 
\nu_4=\nu_{1}+\nu_{2}+\nu_3}}
%F_{k_3,k_2}^{\lambda_{12}}
\left(\begin{smallmatrix}
\lambda_4  \\ \mb{n}_4
\end{smallmatrix}\middle|
\begin{smallmatrix}
 \lambda_3 & \lambda_{2} &\lambda_1  \\  \mb{n}_3 & \mb{n}_{2} & \mb{n}_{1}
\end{smallmatrix} \right)
e_{\bn_1}^{\lambda_1} \otimes e_{\bn_2}^{\lambda_2} \otimes 
e_{\bn_3}^{\lambda_3},
\end{align}
One may argue in the same way as before that the coefficients in (\ref{fourpt-qgrp}) are uniquely determined
in terms of the particular values for $\mb{n}_4=0$ and $\mb{n}_1=0$. The space 
$\mathrm{CG}(\lambda_4|\lambda_3\lambda_2\lambda_1)$ of Clebsch-Gordan maps 
is therefore isomorphic to the subspace denoted by $\mathcal{R}_{\lambda_3,\lambda_2}^{\lambda_4-\lambda_1}$
in $\mathcal{R}_{\lambda_3}\otimes\mathcal{R}_{\lambda_2}$ with total weight 
$\lambda_4-\lambda_1$. It is then easy to see that a basis for this space is provided by the composition of 
Clebsch-Gordan maps, 
\begin{equation}\label{compCG}
\left(\begin{smallmatrix}
\lambda_4  \\ 0 
\end{smallmatrix}\middle|
\begin{smallmatrix}
 \lambda_3 & \lambda_{2} &\lambda_1  \\  \mb{n}_3 & \mb{n}_{2} & 0
\end{smallmatrix} \right)_{k_3,k_2}^{\lambda_{12}}=
\sum_{\substack{\bn_{12} \\ 
\nu_{12}=\lambda_1+\nu_2}}
\left(
\begin{smallmatrix}
\lambda_4  \\ 0 
\end{smallmatrix}\middle|
\begin{smallmatrix}
 \lambda_3 & \lambda_{12} \\  \mb{n}_3 & \mb{n}_{12}
\end{smallmatrix}\right)_{k_3}  
%\lambda_{12}=\lambda_{1} + \lambda_{2,\bn_2}
\big(\begin{smallmatrix}
\lambda_{12}  \\ \mb{n}_{12}
\end{smallmatrix}\big|
\begin{smallmatrix}
 \lambda_2 & \lambda_1 \\  \mb{n}_2 & 0
\end{smallmatrix} \big)_{k_2}~.
\end{equation}
%In order to see this, let us note that the allowed values of $\lambda_{12}$ are
%of the form $\lambda_{12}=\lambda_1+\lambda_2-\delta_{12}$.  
The coefficients (\ref{compCG}) represent the matrix elements of  
a map $\mathsf{D}^{\lambda_4}_{\lambda_3\lambda_2\lambda_1}$
from $\mathrm{CG}(\lambda_4|\lambda_3\lambda_2\lambda_1)$ into 
$\mathcal{R}_{\lambda_3,\lambda_2}^{\lambda_4-\lambda_1}$.

\subsubsection{R-matrix}

The universal R-matrix is an invertible element $\sR\in \cU_q(\fsl_N) \otimes \cU_q(\fsl_N)$ 
with the property 
\be \label{defpropR}
\sR \Delta(x) \sR^{-1} = (\sP\circ\Delta)(x) ~, \qqq x\in \cU_q(\fsl_N) ~,
\ee
where $\sP\in \cU_q(\fsl_N) \otimes \cU_q(\fsl_N)$ is the permutation  
that acts as $\sP(x\otimes y) = y\otimes x$. It satisfies the 
Yang-Baxter equation 
$\sR_{12}\sR_{13}\sR_{23} = \sR_{23}\sR_{13}\sR_{12}$ 
%\be 
%\sR_{12}\sR_{13}\sR_{23} = \sR_{23}\sR_{13}\sR_{12}
%\ee 
and the action of the coproduct on it is 
\be 
(\Delta \otimes 1) \sR = \sR_{13} \sR_{23} ~, \qqq 
(1 \otimes \Delta) \sR = \sR_{13} \sR_{12}  ~.
\ee
It can be shown that the conditions \rf{defpropR} have a unique solution of the form $\sR=q^{t}\bar{\sR}$ with
$t=\sum_{ij} (\kappa_{ij}^{-1})\sh_i\otimes\sh_j$, with $\kappa_{ij}$ being the Cartan matrix, 
if $\sk_i=\fq^{\sh_i}$, and
%\be 
%\sR = \fq^{\sum_{ij} (\kappa_{ij}^{-1}) 
%\sh_i\otimes\sh_j} ~
%\ex_{\fq^{-2}}^{(\fq-\fq^{-1})\se_2\otimes\sf_2} ~
%\ex_{\fq^{-2}}^{(\fq^{-2}-1)\se_{12}\otimes\sf_{12}} ~
%\ex_{\fq^{-2}}^{(\fq-\fq^{-1})\se_1\otimes\sf_1} ~,
%\ee
\bea \label{eq:universalR-matrix} 
\bar{\sR} = 1+ (\fq-\fq^{-1})(\se_1\otimes\sf_1 + 
\se_2\otimes\sf_2 - \fq^{-1} \se_{12}\otimes\sf_{12} )
+ (\fq -\fq^{-1}) ^2\se_2\se_1\otimes\sf_2\sf_1
+ \dots ~~~~~
\eea 
The omitted terms in the expression \rf{eq:universalR-matrix}   are of higher order in $\se_i$, $i=1,2$. 

%with exponentials 
%
%\be 
%\ex_{\fq}^x = \sum_{n=0}^\infty \frac{x^n}{[n]_\fq !}~, 
%\qqq [n]_\fq = \frac{1-\fq^{2n}}{1-\fq^2}  ~.
%\ee
The universal R-matrix is related to the braid matrices $\sB_{\la_1\la_2}:\CR_{\la_1}\otimes\CR_{\la_2}\ra
\CR_{\la_2}\otimes\CR_{\la_1}$ through 
\begin{equation}
\sB_{\la_1\la_2}=\sP\circ\mathsf{r}_{\la_1\la_2},\quad \text{where}\quad
\mathsf{r}_{\la_1\la_2}=(\pi_{\la_1}\otimes\pi_{\la_2})(\sR)
\end{equation} 
is the evaluation of the universal R-matrix in the tensor product of representations $\CR_{\la_1}\otimes\CR_{\la_2}$
and  $P$
is the permutation of tensor factors. Our notation for the  matrix elements of $\sB_{\la_1\la_2}$ is defined such 
that
%%
%%\be 
%%\ex_{\fq}^x = \sum_{n=0}^\infty \frac{x^n}{[n]_\fq !}~, 
%%\qqq [n]_\fq = \frac{1-\fq^{2n}}{1-\fq^2}  ~.
%%\ee
%%The R-matrix is related to the braid matrix through $\sB=\sP\circ\sR$, 
\begin{equation}\label{braid1}
\mathsf{B}_{\la_1\la_2}\, e_{\mb{n}_1}^{\lambda_1}\otimes e_{\mb{n}_2}^{\lambda_2}=
\sum_{\bm_1,\bm_2}(B_{\lambda_1\lambda_2})_{\bn_1}^{\bm_1}{}^{\bm_2}_{\bn_2}
e_{\mb{m}_2}^{\lambda_2}\otimes e_{\mb{m}_1}^{\lambda_1}
~.
%W_{\bm_2  \bm_1  }^{\alpha_2   \alpha_1} ~.
\end{equation}

%%%%%%%%%%%%%%%%%%%%%%%%%%%%%%%%%%%%%%%%%%%%%%%%%%%%%%%%%
%%%%%%%%%%%%%%%%%%%%%%%%%%%%%%%%%%%%%%%%%%%%%%%%%%%%%%%%%

\subsection{Braiding of screened vertex operators}
\label{sec:braidingSVO1}

The spaces of functions 
generated by the functions 
${W}_\mb{n_m , \ldots , n_1}^{\alpha_m , \ldots , \alpha_1} (z_m , \ldots , z_1)$ introduced in %(\ref{def:basisSVO1})
\eqref{SVO1}
carry a natural action 
of the braid group which turns out to be related to the representation 
of the braid group on tensor products of quantum group representations. 
We will now briefly explain this first crucial link between quantum group theory and free field representation to be 
used in this paper. 

\begin{figure}[h!]
\centering
\includegraphics[width=0.5\textwidth]{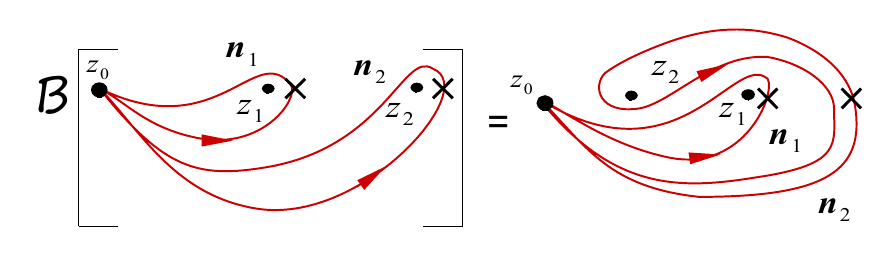}
\caption{{\it Braiding of two generic multi-contours. The collections of contours associated with general products of screening charges are depicted here as simple loops around the punctures. }}
\label{Braidop}
\end{figure}

Let us consider 
${W}_\mb{n_2 n_1}^{\alpha_2  \alpha_1} (z_2, z_1) $ defined using 
the contours from Figure \ref{Contours1}.
The braiding operations are defined using the analytic continuation of this function with respect 
to $z_{2}$ and $z_1$ along a path exchanging their positions. This analytic continuation 
can be defined by integrating along suitably deformed contours as indicated in Figure \ref{Braidop}.
By means of contour deformations one can represent the result of this operation as 
a linear combination of integrals over the contours introduced in Figure \ref{Contours1}.

\begin{figure}[h!]
\centering
\includegraphics[width=0.3\textwidth]{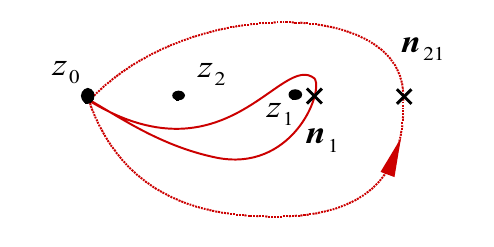}
\caption{{\it A set of auxiliary contours for the braiding calculation.}}
\label{Braid1}
\end{figure}

This can be done by the following recursive procedure. In a first sequence of steps, 
performed in an induction over $\mb{n}_2=(n_{2,1},n_{2,21},n_{2,2})$ one may 
deform the contours on the right  of Figure \ref{Braidop} into linear combinations 
of the contours depicted in Figure \ref{Braid1} . 

\begin{figure}[h!]
\centering
\includegraphics[width=1.0\textwidth]{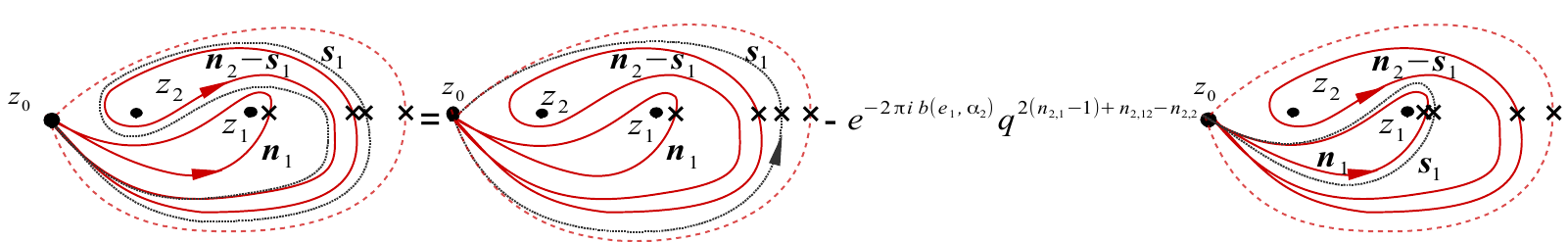}
\caption{{\it Deformation of the outermost of the contours encircling $z_2$ representing a step in the passage to linear combinations of the contours depicted in Figure \ref{Braid1}, where $\mathbf{s}_1=(1,0,0)$.
}}
\label{Braid2}
\end{figure}

The induction step is indicated in Figure \ref{Braid2}.

In a second sequence of steps one may then deform the contours surrounding both $z_1$ and $z_2$
in Figure \ref{Braid1}
into a linear combination of contours surrounding only $z_1$ or $z_2$ at a time. 
The result of this procedure  can be represented as linear combination of the original contours
introduced in Figure \ref{Contours1}, leading to a relation of the form
\begin{equation}\label{braid2}
\mathcal{B}
\big(W_{\bn_2  \bn_1  }^{\alpha_2   \alpha_1}  
(z_2,z_1)\big)=\sum_{\bm_1,\bm_2}(b_{\alpha_1\alpha_2})_{\bn_1}^{\bm_1}{}^{\bm_2}_{\bn_2}
W_{\bm_2  \bm_1  }^{\alpha_2   \alpha_1}  
(z_1,z_2)
\end{equation}
We will explain in more detail in Appendix \ref{sec:app:braiding2} below how such computations can be done
and calculate the matrix elements of $b_{\alpha_1\alpha_2}$ for the 
important special case where $\alpha_2=-b\omega_1$ with $\alpha_1$ generic.  
The result can be compared to the action of the braid group on tensor products 
of quantum group representations defined in (\ref{braid1}).
It turns out that the matrix $b_{\alpha_1\alpha_2}$ appearing in 
(\ref{braid2}) coincides with the matrix $B_{\lambda_1\lambda_2}$ in (\ref{braid1}) provided that
$\lambda_l = -\alpha_l / b$ for $l=1,2$. This result can easily be generalised to 
the cases where $\alpha_2=-b\lambda_2$ with $\lambda_2$ being the 
weight of any finite-dimensional representation of $\cU_q(\fsl_3)$ by noting that any
finite-dimensional representation of $\cU_q(\fsl_3)$ appears in the iterated tensor 
products of fundamental representations, and that the co-product of the quantum group 
has a simple representation in terms of the free field representation, as indicated in Figure \ref{Co-product}
below.
\begin{figure}[H]
\centering
\includegraphics[width=1.0\textwidth]{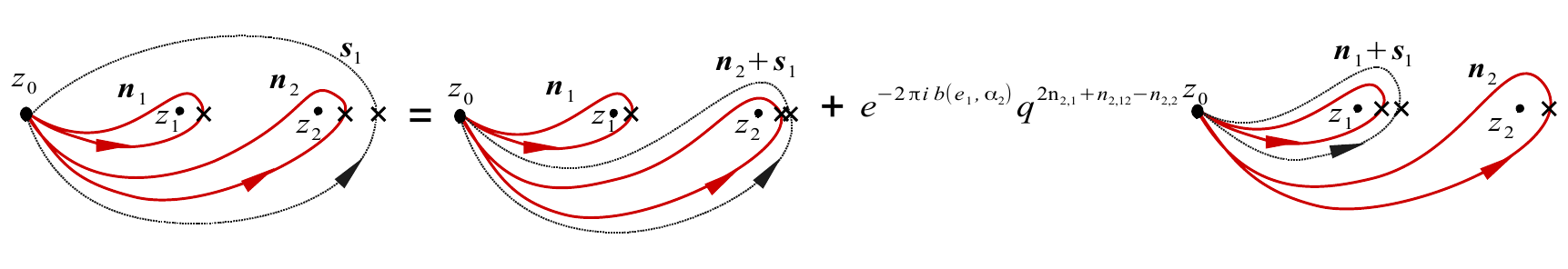}
\caption{{\it Decomposition of a simple contour associated 
with the root $e_1$ around two punctures into contours 
enclosing only one of the punctures. This decomposition represents
the co-product $\De(\sf_1)=1\otimes \sf_i+\sf_1\otimes \sk_1^{-1}$.
%The factor $f_{21}$ is given in equation \eqref{eq:monodromy-NP1}.
}}
\label{Co-product}
\end{figure}

As representations of the braid group one may therefore identify the vector spaces
generated by the functions ${W}_{\mb{n}_m , \ldots , \mb{n}_1}^{\alpha_m , \ldots , \alpha_1} (z_m , \ldots , z_1)$
with the tensor product of quantum group representations $\bigotimes_{l=1}^m\mathcal{R}_{\lambda_l}$ 
assuming that the weights are related as 
$\lambda_l = -\alpha_l / b$, $l=1,\dots,m$.

\subsection{Construction of conformal blocks}

We had noted above that the 
functions ${W}_\mb{n_m , \ldots , n_1}^{\alpha_m , \ldots , \alpha_1} (z_m , \ldots , z_1) $
are not the objects we are ultimately interested in. They do not
satisfy the Ward identities for the  $\mathcal{W}_3$-algebra characterising the 
conformal blocks. 
It turns out, however, that there exist suitable linear combinations
of the form (\ref{ansatz}) which will indeed represent conformal blocks. We are now going to argue 
that 
this will be the case if the coefficients 
$C$ in (\ref{ansatz}) are
the matrix elements of the Clebsch-Gordan maps describing the embedding of the
irreducible representation with weight $\lambda_{\infty}=-b^{-1}\alpha_{\infty}$ into the 
tensor product $\bigotimes_{l=1}^m\mathcal{R}_{\lambda_l}$.
For $m=3$, for example, one may get expressions of the following form
\begin{align}\label{fourpt-gen}
&\langle v_{\infty},V^{\rho_3}_{k_3}(z_3) V^{\rho_2}_{k_2}(z_2) \fe_{\al_1}\rangle
=\\
&\qquad\qquad=\sum_{\substack{\bn_3, \bn_{2}, \bn_{1}, \bn_{12} \\ 
\lambda_4=\nu_{12}+\nu_3\\ 
\nu_{12}=\nu_1+\nu_2}}
\left(\begin{smallmatrix}
\lambda_4  \\ 0 
\end{smallmatrix}\middle|
\begin{smallmatrix}
 \lambda_3 & \lambda_{12} \\  \mb{n}_3 & \mb{n}_{12}
\end{smallmatrix}\right)_{k_3}  
%\lambda_{12}=\lambda_{1} + \lambda_{2,\bn_2}
\left(\begin{smallmatrix}
\lambda_{12}  \\ \mb{n}_{12}
\end{smallmatrix}\middle|
\begin{smallmatrix}
 \lambda_2 & \lambda_1 \\  \mb{n}_2 & \mb{n}_1
\end{smallmatrix}\right)_{k_2} 
{W}_{\mb{n}_3 , \mb{n}_2 , \mb{n}_1}^{\alpha_3, \alpha_2 , \alpha_1} (z_3 , z_2 , 0),
\notag\end{align}
where  $\rho_2=(\alpha_{12},\alpha_2,\alpha_1)$ and 
$\rho_{3}=(\alpha_\infty,\alpha_3,\alpha_{12})$ with $\alpha_{12}=-b\lambda_{12}$.
Note that the summations in (\ref{fourpt-gen}) include summations over the multiplicity labels
of $\mb{n}_3$, $\mb{n}_{2}$, $\mb{n}_{1}$  and $\mb{n}_{12}$ together with a summation over two out  
the three weights $\nu_1$, $\nu_2$ and $\nu_3$.

\subsubsection{Relation to quantum group theory}\label{sec:W-comm-qgrp}

In order to see the relation between the problem to find linear combinations of the 
form (\ref{ansatz})  representing conformal blocks
and the Clebsch-Gordan problem one needs to notice that the 
boundary terms appearing in the commutators of the screening charges with the 
generators of the algebra $\mathcal{W}_3$ turn out to be related to the action of the 
generators $\se_i$ on tensor products of  quantum group representations defined 
by means of the co-product. In order to formulate a more precise 
statement, let us introduce some useful notations.
Let 
\begin{equation}\label{def:SVoperator}
{V}_{\mb{n}_m , \ldots , \mb{n}_1}^{\alpha_m , \ldots , \alpha_1} (z_m , \ldots , z_1)
=\int_{\check{\Gamma}} 
d \mb{y}_m  \ldots d \mb{y}_1  \;
\mb{S}^{\mb{n}_m}_m(\mb{y}_m) V_{\alpha_m} (z_m) \ldots 
\mb{S}^{\mb{n}_1}_1(\mb{y}_1) V_{\alpha_1} (z_1)
\end{equation}
be the operator appearing in the matrix elements (\ref{SVO1})
defining the functions 
${W}_\mb{n_m , \ldots , n_1}^{\alpha_m , \ldots , \alpha_1}$. 
 The correspondence with vectors in tensor products of quantum group representations
 observed above can be schematically represented as 
\be \label{block-module}
{V}_{\mb{n}_m , \ldots , \mb{n}_1}^{\alpha_m , \ldots , \alpha_1} \quad\leftrightarrow  \quad 
e_{\bn_1}^{\lambda_1} \otimes \ldots \otimes 
e_{\bn_m}^{\lambda_m} ~,
\ee
allowing us to 
introduce the notation 
\begin{equation}
\big(\Delta(x)
{V}_{\mb{n}_m , \ldots , \mb{n}_1}^{\alpha_m , \ldots , \alpha_1} \big)(z_m , \ldots , z_1)
\end{equation}
for the vertex operator associated to 
$\Delta(x)e_{\bn_1}^{\lambda_1} \otimes \ldots \otimes 
e_{\bn_m}^{\lambda_m}$ for $x\in\cU_q(\fsl_3)$ via the correspondence (\ref{block-module}). 
In the computation of the commutator of the vertex operators 
${V}_{\mb{n}_m , \ldots , \mb{n}_1}^{\alpha_m , \ldots , \alpha_1} $ with the 
generators of the 
algebra $\mathcal{W}_3$ one may distinguish terms from the commutators 
of the screening currents from the  terms coming from commutators with other exponential
fields. The contribution of the former is then found to be a linear combination of 
terms which are proportional to 
$(\Delta(\se_i)
{V}_{\mb{n}_m , \ldots , \mb{n}_1}^{\alpha_m , \ldots , \alpha_1} )(z_m , \ldots , z_1)$.
Considering, for example, the commutator with $L_{-1}$ we find
\begin{equation}
\big[\,L_{-1}\,,\, {V}_{\mb{n}_m , \ldots , \mb{n}_1}^{\alpha_m , \ldots , \alpha_1} (z_m , \ldots , z_1)\,\big]
=\sum_{l=1}^m\partial_{z_l}
{V}_{\mb{n}_m , \ldots , \mb{n}_1}^{\alpha_m , \ldots , \alpha_1} (z_m , \ldots , z_1)
+(\text{boundary terms}),
\end{equation} 
with boundary terms proportional to 
$(\Delta(x)
{V}_{\mb{n}_m , \ldots , \mb{n}_1}^{\alpha_m , \ldots , \alpha_1} )(z_m , \ldots , z_1)$ for 
\begin{equation}
x=(q-q^{-1})(\sk_1^{-1}\se_1+\sk_2^{-1}\se_2).
\end{equation}
The derivation of these statements is outlined in 
Appendix \ref{sec:app:eaction}. We conclude that the boundary terms in the commutators with generators
of the algebra $\mathcal{W}_3$ will vanish if the coefficients $C$ in (\ref{ansatz}) are taken to be the 
expansion coefficients of the images of highest weight vectors under the CG maps.

\subsubsection{Independence of the choice of base-point}

The linear combinations (\ref{ansatz}) representing conformal blocks  do not depend on 
the position of the base point $z_0$ indicated in Figure \ref{Contours1}. Indeed, the derivative of ${W}_\mb{n_m , \ldots , n_1}^{\alpha_m , \ldots , \alpha_1}$ with respect to $z_0$ yields boundary terms which cancel in the linear combinations (\ref{ansatz}) 
by the same arguments as used above. 

\begin{figure}[t]
\centering
\includegraphics[width=1.0\textwidth]{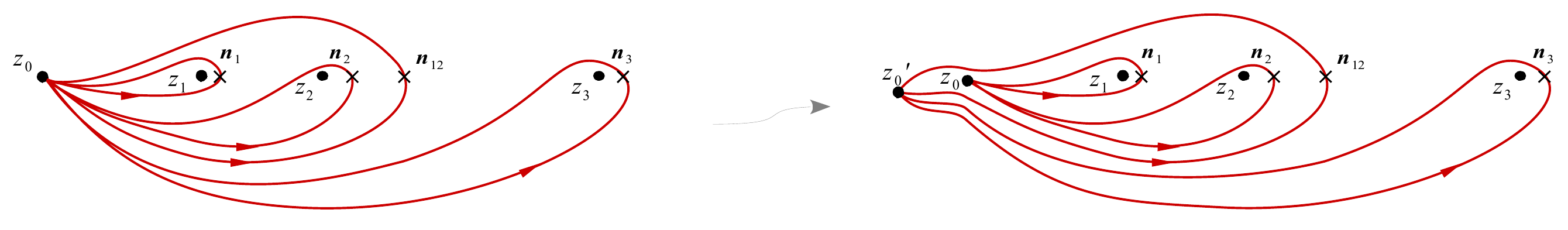}
\caption{{\it Different base points $z_0$ and $z_0'$ associated with different integration contours.
}}
\label{Different-base-points}
\end{figure}

The independence of $z_0$
can be used for our advantage in  two ways.
One may note, on the one hand, that the expressions for conformal blocks resulting from 
(\ref{ansatz}) can be replaced by expressions where different base points $z_0$ are associated
to 
each individual CG coefficient appearing in expressions like (\ref{fourpt-gen}). In (\ref{fourpt-gen}), for 
example,
one could have a base-point $z_0$ for the integrations with screening numbers $\bn_1$ and $\bn_2$, 
and another base-point $z_0'$ for those with screening numbers $\bn_{12}$ and $\bn_3$, as 
depicted in Figure \ref{Different-base-points}. In this form it 
is manifest that the conformal blocks defined via (\ref{fourpt-gen}) factorise into three-point conformal blocks, 
as they should. 
This representation furthermore exhibits 
clearly the relation between the intermediate representation $\mathcal{V}_{\alpha_{12}}$ of the algebra $\mathcal{W}_3$
appearing in the conformal blocks on the left side of (\ref{fourpt-gen})
%appearing in  (\ref{fourpt-gen}) 
and the intermediate representation $\mathcal{R}_{\lambda_{12}}$ appearing in the 
Clebsch-Gordan coefficients on the right side of (\ref{fourpt-gen}),
with representation labels related by $\alpha_{12}=-b\lambda_{12}$.

If, furthermore, 
the components of $\alpha_1$ are sufficiently negative, one may simplify the expression (\ref{fourpt-gen})
by taking the limit where the base-point $z_0$ used in the definition of the contours approaches
$z_1$. %This is related to the projection onto the highest weight vector in the tensor factor $\mathcal{R}_{\lambda_1}$.
Taking into account (\ref{normcond}) the expression (\ref{fourpt-gen}) gets simplified to 
\begin{align}\label{fourpt-sim}
&\langle v_{\infty},V^{\rho_3}_{k_3}(z_3) V^{\rho_2}_{k_2}(z_2) \fe_{\al_1}\rangle
%=\\&\qquad
=\sum_{\substack{\bn_3 \\ 
\nu_{3}+\nu_2=\lambda_4-\lambda_1}}
D_{k_3k_2}^{\lambda_{12}}\left(\begin{smallmatrix}
\lambda_4  \\ 0 
\end{smallmatrix}\middle|
\begin{smallmatrix}
 \lambda_3 & \lambda_{2} &\lambda_1  \\  \mb{n}_3 & \mb{n}_{2} & 0
\end{smallmatrix} \right)
{W}_{\mb{n}_3 , \mb{n}_2 , 0}^{\alpha_3, \alpha_2 , \alpha_1} (z_3 , z_2 , 0),
\end{align}
where $\mb{n}_2=[\lambda_{12}-\lambda_1-\lambda_2,k_2]$,
$\rho_2=(\alpha_{12},\alpha_2,\alpha_1)$ and 
$\rho_{3}=(\alpha_\infty,\alpha_3,\alpha_{12})$ with $\alpha_{12}=-b\lambda_{12}$. 
The expression for the coefficients $D_{k_3,k_2}^{\lambda_{12}}$ in \rf{fourpt-sim}
gets simplified to 
\begin{equation}\label{Dsimple}
D_{k_3,k_2}^{\lambda_{12}}\left(\begin{smallmatrix}
\lambda_4  \\ 0 
\end{smallmatrix}\middle|
\begin{smallmatrix}
 \lambda_3 & \lambda_{2} &\lambda_1  \\  \mb{n}_3 & \mb{n}_{2} & 0
\end{smallmatrix} \right)=
\sum_{\substack{\bn_{12} \\ 
\nu_{12}=\lambda_4-\nu_3}}
\left(
\begin{smallmatrix}
\lambda_4  \\ 0 
\end{smallmatrix}\middle|
\begin{smallmatrix}
 \lambda_3 & \lambda_{12} \\  \mb{n}_3 & \mb{n}_{12}
\end{smallmatrix}\right)_{k_3}  ~.
%\lambda_{12}=\lambda_{1} + \lambda_{2,\bn_2}
%\big(\begin{smallmatrix}
%\lambda_{12}  \\ \mb{n}_{12}
%\end{smallmatrix}\big|
%\begin{smallmatrix}
 %\lambda_2 & \lambda_1 \\  \mb{n}_2 & 0
%\end{smallmatrix} \big)_{k_2}~,
\end{equation}
We see that sending $z_0$ to $z_1$ is related to the projection of the tensor product %of representations
$\bigotimes_{l=1}^3\mathcal{R}_{\lambda_l}$ onto the subspace 
$v_{\lambda_1}\otimes\bigotimes_{l=2}^3\mathcal{R}_{\lambda_l}$. This projection obviously commutes with
the braiding operation applied to the second and third tensor factors, allowing us to simplify the 
computations.

It can also be useful to consider the limit where $z_0\ra z_2$. In this case one may
use contours supported near the circles around the origin with radius $|z_2|$, allowing us 
represent the CVOs as composite operators 
of the form $Q_1^{n_1}(\ga)Q_{12}^{n_{12}}(\ga)Q_2^{n_2}(\ga) V_{\al_2}(z_2)$, 
with screening charges $Q_i(z)=\int_{\ga}S_i(y)$ being defined by integration over 
$\ga=\{y\in\BC;|y|=|z_2|\}$ for $i=1,12,2$.

%\newpage

%%%%%%%%%%%%%%%%%%%%%%%%%%%%%%%%%%%%%%%%%%%%%%%%%%%%%%%%

\section{Computation of monodromies}\label{sec:monocomp}
\setcounter{equation}{0}

We will now explain how the connection between  free field representation
and quantum group theory helps us to compute the representation 
of monodromies on spaces of conformal blocks. Recall that we are interested in 
calculating the monodromies of the conformal blocks
\begin{equation}\label{3+2blocks'}
\mathcal{F}_{k}^\rho(y;y_0)_{\bar{\imath}\imath}^{}=
\langle \,{\fe}_{\al_3}^{}\,,\, \bar{D}_{\bar{\imath}}^{}(y_0) D_\imath^{}(y)\,
V_k^\rho(z)\,\fe_{\al_1}^{}\rangle,
\end{equation}
regarded as function of $y$. The task can be simplified slightly by sending the base-point $y_0$
to infinity, reducing the problem to the computation of the monodronies of the four-point conformal blocks
\begin{equation}\label{3+2blocks''}
\mathcal{G}_{k}^\rho(y)_{\imath}^{}=
\langle \,v_{\infty}\,,\,  D_\imath^{}(y)\,
V_k^\rho(z)\,\fe_{\al_1}^{}\rangle,
\end{equation}
where $v_{\infty}\in\CV_{\al_4}$ with $\al_4$ being related to the variables $\al_3$ and $\bar{\imath}$ 
in \rf{3+2blocks'} as $\al_4=\al_3-bh_{\bar{\imath}}$.

%\subsection{Form of the braid matrix in the general case}

We will therefore consider the  space of conformal blocks on the four-punctured sphere 
$\mathbb{P}^1\setminus\{z_1=0,z_2=z,z_3=y,z_4=\infty\}$ 
with representation $\CV_{\al_r}$ assigned to $z_r$, $r=1,\dots,4$. The free field 
representation 
identifies this space with
the subspace of $\bigotimes_{l=1}^{3}\mathcal{R}_{\lambda_l}$ with weight $\lambda_4-\lambda_1$,
assuming that $\al_r=-b\la_r$ for $r=1,\dots,4$.

The fundamental group of $C_{0,3}=\mathbb{P}^1\setminus\{0,z,\infty\}$ is generated by the loops
$\ga_0$, $\ga_z$, $\ga_\infty$ around $0$, $z$ and $\infty$. Since the loop $\ga_0\circ\ga_z\circ\ga_\infty$
is contractible, it suffices to compute the monodromies along two out of the three loops. The monodromy
of the conformal blocks $\mathcal{G}_{k}^\rho(y)_{\imath}^{}$
introduced in \rf{3+2blocks''} around $\infty$ is a simple diagonal matrix. 
It will therefore suffice to compute  the monodromy of $\mathcal{G}_{k}^\rho(y)_{\imath}^{}$
%corresponding to the analytic continuation of $y$ 
along a loop surrounding $z_2=z$. The free field representation introduced above relates this monodromy
to 
\begin{equation}\label{Dyntwist}
(\mathsf{D}^{\lambda_4}_{\lambda_1\lambda_2\lambda_3})^{-1}\cdot
\mathsf{B}_{\lambda_3\lambda_2}\cdot
\mathsf{B}_{\lambda_2\lambda_3}\cdot
\mathsf{D}^{\lambda_4}_{\lambda_1\lambda_2\lambda_3}~,
\end{equation}
where $\mathsf{D}$ is the linear operator defined in \rf{fourpt-gen}, and 
$\mathsf{B}_{\lambda_3\lambda_2}:\CR_{\la_3}\otimes\CR_{\la_2}\ra 
\CR_{\la_2}\otimes\CR_{\la_3}$ is the braiding operation.

As explained previously, we will mainly be interested in the case where 
the representation $\mathcal{R}_{\lambda_3}$ assigned to $z_3=y$ is the fundamental representation,
$\lambda_3=\omega_1$. 
The simplifications resulting from $\la_3=\omega_1$ will be discussed next.

\subsection{Conformal blocks with degenerate fields}

In the case $\la_3=\omega_1$ the weight
$\lambda_{12}$ in (\ref{fourpt-sim}) can only take the values 
$\lambda_\imath=\lambda_4-h_\imath$, $\imath=1,2,3$.  
 The expansion  (\ref{fourpt-sim}) further simplifies to
\begin{align}\label{fourpt-deg}
&\langle v_{\infty},D_{\imath}(y) V^{\rho_2}_{k}(z)  \fe_{\al_1}\rangle
%=\\&\qquad\qquad
=\sum_{\jmath=1}^3\sum_{\substack{\bn_2 \\ 
\lambda_4=h_\jmath+\nu_2+\lambda_1}}\!\!
D_{k,0}^{\lambda_{\imath}}\big(\begin{smallmatrix}
\lambda_4  \\ 0 
\end{smallmatrix}\big|
\begin{smallmatrix}
 \omega_1 & \lambda_{2} &\lambda_1  \\   \mb{d}_\jmath & \mb{n}_2 & 0
\end{smallmatrix} \big)
{W}_{\mb{d}_\jmath ,\mb{n}_2 ,  0}^{-b\omega_1, \alpha_2 , \alpha_1} (y , z, 0),
%\notag
\end{align}
where $\mb{d}_i=[\omega_1-h_i,0]$, $i=1,2$, and $\mb{d}_3=[\omega_1-h_3,1]$.
%The weights $\nu_2$ are fixed by the constraint $\lambda_4=h_\jmath+\nu_2+\lambda_1$. 
One should also remember that $\mb{n}_2$ is fixed as $\mb{n}_2=[\lambda_{12}-\lambda_1-\lambda_2,k]$ through 
equation \rf{normcond}. 
Taking this constraint into account, we will  in the following replace the 
notation ${W}_{\mb{d}_\jmath,\mb{n}_2 , 0}^{-b\omega_1, \alpha_2 , \alpha_1} $
by ${W}_{\jmath, k}^{\alpha_4, \alpha_2 , \alpha_1} $. 
Noting that $\nu_{12}=\lambda_4-h_\imath$ one finds that the summation over $\mb{n}_{12}$ 
in \rf{Dsimple} can take only the
four possible values $\mb{s}_{i}=[-e_i,0]$, $i=1,2$  and $[-e_{12},\ep]$, $e_{12}=e_1+e_2$, $\ep\in\{0,1\}$.
One may then write (\ref{fourpt-deg}) a bit more concisely in the form
\begin{equation}
\label{fourpt-deg'}
\langle v_{\infty},D_{\imath}^{}(y) {V}^{\rho}_k(z)\fe_{\al_1}\rangle
=\sum_{\jmath=1}^3
{D}_{\imath\jmath}(\la_4)
 {W}_{\jmath, k}^{\alpha_4, \alpha_2 , \alpha_1}  (z_3 , z_2 , z_1).
\end{equation}
The remaining coefficients in (\ref{fourpt-deg'}) are 
\be 
D_{ii}  = 1 ~, \quad
D_{i+1 ~ i} = 
\big(\begin{smallmatrix}
\lambda_4 \\ 0 
\end{smallmatrix}\big|
\begin{smallmatrix}
 \omega_1 & \lambda_4 - h_{i+1} \\  \mb{d}_i & \bs_i
\end{smallmatrix}\big)
~, \quad
D_{31} = \sum_{\ep=0}^1
\big(\begin{smallmatrix}
\lambda_4 \\ 0 
\end{smallmatrix}\big|
\begin{smallmatrix}
 \omega_1 & \lambda_4 - h_3 \\  0 & [-e_{12},\ep]
\end{smallmatrix}\big)
~.
\ee
Using the notation
\be [n] = \frac{1-\fq^{2n}}{1-\fq^2} ~. \ee
the non vanishing matrix elements of $D$ can be represented explicitly as
\begin{align}
&
\begin{aligned}
& D_{21} = \fq^{-1} [-(e_1,\lambda_4)]^{-1} ~,\\
& D_{32} = \fq^{2(e_2,\lambda_4)-2} 
\big([1-(e_2,\lambda_4)]-[1+(e_1,\lambda_4)]\big) 
\big([(e_2,\lambda_4)-1]\fc^{(2)}_{\lambda_4}\big)^{-1}~,\\
& 
{D}_{31} = \fq^{2(e_{12},\lambda_4)-1}
\left(1-[1-(e_2,\lambda_4)] \right)
 \big([(e_2,\lambda_4)-1]\fc^{(2)}_{\lambda_4}\big)^{-1}~,
\end{aligned}\\
%\end{align}
&\fc^{(2)}_{\lambda_4} = %\fq^{1-(e_{12},\lambda_4)} 
\fq^{2+2(e_1,\lambda_4)}
(1+[1-(e_{12},\lambda_4)]) - (1-\fq^2)
([1-(e_2,\lambda_4)]-[1+(e_1,\lambda_4)]).\nn
\end{align}
The derivation of these expressions is outlined in Appendix \ref{simpleClebsch}.

%We can similarly write
%\begin{equation}
%\label{fourpt-deg2'}
%\langle v_{\infty},\mb{V}^{\rho_3}(z_3) D_{\imath}^{}(z_2)  V^{\rho_1}(z_1)v_0\rangle
%=\sum_{\jmath=1}^3
%\mathsf{C}\big[{}_{\alpha_4}^{\alpha_3}{}^{-b\omega_1}_{~~\alpha_1}\big]^{}_{\imath\jmath}
%\mb{W}_{\delta_\jmath}\big[{}_{\alpha_4}^{\alpha_3}{}^{-b\omega_1}_{~~\alpha_1}\big]  (z_3 , z_2 , z_1),
%\end{equation}
%where the non-vanishing matrix coefficients are 
%\begin{align}
%& \sC_{\imath\imath} (k) = 1
%~, \quad 
%\sC_{12} (k) = 
%\big(\begin{smallmatrix}
%\lambda_4 \\ 0 
%\end{smallmatrix}\big|
%\begin{smallmatrix}
% \lambda_3 & \lambda_1 +h_1 \\ \bn_3 (k) -\bs_1 & \bs_1
%\end{smallmatrix}\big)_k
%~, \quad 
%\sC_{23} (k) = 
%\big(\begin{smallmatrix}
%\lambda_4 \\ 0 
%\end{smallmatrix}\big|
%\begin{smallmatrix}
 %\lambda_3 & \lambda_1 + h_2 \\ \bn_3 (k) -\bs_3 & \bs_2
%\end{smallmatrix}\big)_k\\
%&\sC_{13} (k) = \sum_{\bn_{21}\in\{(1,0,1),(0,1,0)\}}
%\big(\begin{smallmatrix}
%\lambda_4 \\ 0 
%\end{smallmatrix}\big|
%\begin{smallmatrix}
 %\lambda_3 & \lambda_1+h_1 \\ \mb{n}_3 (k) -\bs_3 & \mb{n}_{21}
%\end{smallmatrix}\big)_k + 
%\big(\begin{smallmatrix}
%\lambda_4 \\ 0 
%\end{smallmatrix}\big|
%\begin{smallmatrix}
 %\lambda_3 & \lambda_1 +\omega_1 \\ \mb{n}_3 (k-1) -\bs_3 & \mb{n}_{21}
%\end{smallmatrix}\big)_k~
%\end{align}
%and $\mb{n}_3 (k)$ is the vector defining the weight 
%$\lambda_1 + \lambda_3 - \lambda_4 + \omega_1 $ and 
%which has multiplicity $k$
%\be 
%\mb{n}_3 (k) = 
%[\lambda_1 + \lambda_3 - \lambda_4 + \omega_1 , 
%k ] ~.
%\ee

%%%%%%%%%%%%%%%%%%%%%%%%%%%%%%

\subsection{Braiding with fundamental representation}

The form of the braid relations also simplifies when  $\la_2=\omega_1$ or $\la_3=\omega_1$. 
Let us first consider the case $\la_2=\omega_1$.
The braid matrix preserves weight subspaces in the tensor product. Recalling that the only allowed 
weights in the fundamental representation are $h_\imath$, $\imath=1,2,3$ one notes that the 
subspace in $\mathcal{R}_{\omega_1}\otimes\mathcal{R}_{\lambda}$ having fixed total weight $\nu$
is spanned by vectors of the form $e^{\omega_1}_{\delta_\imath}\otimes e^{\lambda}_{\mb{n}}$ 
with weight of $e^{\lambda}_{\mb{n}}$ being $\nu-h_\imath$ for $\imath=1,2,3$. For fixed $\imath$ one gets 
a subspace of $\mathcal{R}_{\omega_1}\otimes\mathcal{R}_{\lambda}$  isomorphic to the 
multiplicity space $\mathcal{M}^{\nu-h_\imath}_{\lambda}$. The projection of the braiding operator 
$\mathsf{B}_{\omega_1\la}:\mathcal{R}_{\omega_1}\otimes\mathcal{R}_{\lambda}\rightarrow
\mathcal{R}_{\lambda}\otimes\mathcal{R}_{\omega_1}$ onto a subspace of total weight $\nu$
can be 
represented by an upper triangular matrix $\mathsf{b}^{+}\equiv \sb^+(\la,\nu)$ 
of operators $\mathsf{b}^{+}_{\imath\jmath}$ 
mapping from $\mathcal{M}^{\nu-h_\imath}_\lambda$ to $\mathcal{M}^{\nu-h_\jmath}_\lambda$
for $\imath,\jmath=1,2,3$. 

The operators $\sb_{\imath \jmath}^+$ can be conveniently represented as follows.  
The basis \rf{basis} allows us to introduce 
a basis for the multiplicity spaces $\mathcal{M}^{\nu}_\la$ given
by the vectors $\mathfrak{m}_{\la,k}^\nu=\sf_1^{s_1-k}\sf_{12}^{k}\sf_2^{s_2-k}v_\lambda$ 
with $k=0,\dots,\mathrm{min}(s_1,s_2)$ and $\nu=\la- s_1 e_1 - s_2 e_2$.
$\mathcal{M}^{\nu}_\la$  can be embedded
into
the infinite-dimensional vector space $\mathcal{M}$ with basis $\mathfrak{m}_k$, $k\in\BZ$,
by mapping $\mathfrak{m}_{\la,k}^\nu$ to $\mathfrak{m}_{k}$.
On $\mathcal{M}$ one can define
the operators
$\sv^{-1}$ and $\sk$ as
\begin{equation}
\sv^{-1}\, \mathfrak{m}_{k} = 
q^{k} \, \mathfrak{m}_{k-1}, 
\qquad \sk\, \mathfrak{m}_{k} = k\,  \mathfrak{m}_{k}.
\end{equation} 
Out of these operators one may then construct 
the operators $\hat{\sb}_{\imath \jmath}^+$  as
\begin{align} \label{eq:braiding-Vo10MT}
&\hat{\mathsf{b}}_{11}^+ = 
\fq^{(h_1, \nu )} ~, \qqq 
\hat{\mathsf{b}}_{22}^+ = 
\fq^{(h_2, \nu + e_1 )} ~, \qqq 
\hat{\mathsf{b}}_{33}^+ = 
\fq^{(h_3, \nu + e_{12} )}
\\
&\hat{\mathsf{b}}_{12}^+ = 
\fq^{(h_2, \nu + e_1 )} 
(\fq-\fq^{-1})\fq^{-s_2-(e_1,\lambda)}
\left(
\fq^{1+2\sk} [s_1-\sk][1-s_1+s_2-\sk+(e_1,\lambda)] + 
\sV^{-1} [\sk]
\right)
\nn\\
&\hat{\mathsf{b}}_{23}^+ = 
\fq^{(h_3 , \nu + e_{12} )} 
(\fq^{-1}-\fq) \left(
[s_2-\sk][1-s_2+\sk+(e_2,\lambda)]\fq^{-(e_2,\lambda)}
- \sV^{-1} [\sk]\fq^{1-2 s_2 + (e_2,\lambda)}
\right)
\nn\\
& \begin{aligned} \hat{\mathsf{b}}_{13}^+ =
\fq^{(h_3 ,\nu + e_{12} )}
&(1-\fq^2)
\Big(
 [s_1-\sk][s_2-\sk][s_2-1-\sk-(e_2,\lambda)]
\fq^{1+4\sk-3s_2+(e_2-e_1,\lambda)}\\  
& +\sV^{-1}[\sk] %\times
\big([s_2-2-(e_{12},\lambda)]\fq^{2-2\sk-s_2+(e_{12},\lambda)}
-
[s_2-\sk]\fq^{-2-s_2-(e_{12},\lambda)} \nn\\
 & \qquad\qquad+
[s_1-\sk]\fq^{2\sk-3 s_2+(e_2-e_1,\lambda)}  
+ [s_2-\sk+1]\fq^{2\sk-2-3 s_2+(e_2-e_1,\lambda)}
\big)\nn \\
 & -\fq^{-1} \sV^{-2} [\sk][\sk-1]\fq^{-3s_2-(e_2-e_1,\lambda)}
\Big).
\nn\end{aligned}
\end{align}
It can easily be checked that the image of the restriction of $\hat{\sb}_{\imath \jmath}^+$
to $\mathcal{M}^{\nu-h_\imath}_\la$ is contained in 
the subspace of $\CM$ isomorphic to $\mathcal{M}^{\nu-h_\jmath}_\la$.
Using these definitions one may represent our results for the operators ${\sb}_{\imath \jmath}^+$
as the projections of $\hat{\sb}_{\imath \jmath}^+$ to 
operators from $\mathcal{M}^{\nu-h_\imath}_\la$ to $\mathcal{M}^{\nu-h_\jmath}_\la$.

In the second case $\la_3=\omega_1$ one may similarly represent the projection of
$\mathsf{B}_{\la\omega_1}$ onto the subspace of weight $\nu$
by a matrix $\sb_-\equiv \sb_-(\nu)$ having
matrix elements representable through
\bea \label{eq:braiding-Vo11MT}
 \mathsf{b}_{11}^- &=& \fq^{(h_1, \nu )} ~, \qqq 
 \mathsf{b}_{22}^- = 
\fq^{(h_2, \nu + e_1 )} ~, \qqq 
\mathsf{b}_{33}^- = 
\fq^{(h_3 , \nu + e_{12} )}
\\
 \mathsf{b}_{21}^- &=& 
\fq^{(h_1, \nu )} (\fq-\fq^{-1}) ~, \qqq 
\mathsf{b}_{32}^- = 
\fq^{(h_2, \nu + e_1 )}
(\fq^2-1)\fq^{-s_1+1}\left(\sV\fq^{2\sk} [s_1-1-\sk] - \fq^{2\sk} \right)
\nn\\
\mathsf{b}_{31}^- &=& 
\fq^{(h_1, \nu )}
(\fq-\fq^{-1}) \sV \fq^{s_1-1} ~.
\qqq\qqq\qqq\qqq\qqq\qqq\qqq\qqq\qqq\qqq\qqq\qqq \nn
\eea
A feature of particular importance for us is the fact that the braid matrix can be represented 
as a matrix having elements which are finite difference operators of second order 
in the multiplicity label. This is a direct consequence of \rf{UVrep} given the relation 
between braiding and quantum group R-matrix, as is easily seen using
\rf{eq:universalR-matrix}.

\subsection{Form of the monodromy matrix}

With both of the braid matrices $\sb_+$ and $\sb_-$ 
explicitly derived, we now find that the matrix $M_2$ representing the monodromy around 
$z_2$ takes the form $M_2=D^{-1}\sb_-\sb_+D$
with $D$ being a lower triangular  matrix with  matrix elements acting as pure multiplication operators on the
multiplicity spaces. 
It may be useful to note that 
\begin{align*}
\sb_- = \left(\begin{array}{ccc}
\fq^{(h_1,\nu)} & 0 & 0 \\ 
\sb^-_{21} & \fq^{(h_2,\nu + e_1)} & 0  \\ 
\sb^-_{31} & \sb^-_{32} & \fq^{(h_3,\nu + e_{12})}
  \end{array} \right) ,\qquad
\sb_+ = \left(\begin{array}{ccc}
\fq^{(h_1,\nu)} & 
\sb_{12}^+ & 
\sb_{13}^+ \\ 
0 & \fq^{(h_2,\nu + e_1)} &
\sb^+_{23}  \\ 
0 & 0 & \fq^{(h_3,\nu + e_{12})}
  \end{array} \right), \nn
\end{align*}
We may represent $\sb_-\sb_+$ as
\begin{align*} 
\sb_-\sb_+ = \left(\begin{array}{ccc}
\fq^{2(h_1,\nu)} & \fq^{(h_1,\nu)}\sb_{12}^+
%[\sV^{-1},\sU^{-1},s_{1,2},k] 
& 
\fq^{(h_1,\nu)}\sb_{13}^+ \\ 
\fq^{(h_1,\nu)}\sb^-_{21} & 
\fq^{2(h_2,\nu+e_1)} + 
\sb^-_{21} \sb_{12}^+%[\sV^{-1},\sU^{-1},s_{1,2},k] 
& \sb_{21}^- \sb_{13}^+ + \fq^{(h_2,\nu + e_1)}
\sb^+_{23}
\\ 
\fq^{(h_1,\nu)} \sb^-_{31} %[\sV,s_1] 
& 
\sb^-_{31} %[\sV,s_1] 
\sb_{12}^+%[\sV^{-1},\sU^{-1},s_{1,2},k]  
+  
\fq^{(h_2,\nu + e_1)} 
\sb^-_{32}%[\sV,\sU^{-1},s_1,k] 
& \fq^{2(h_3,\nu+e_{12})}  + 
 \sb^-_{32}
\sb^+_{23}+ \sb_{31}^-\sb^+_{13}
  \end{array}\right) ~,
\end{align*}   
It has now become straightforward to derive our main claim concerning the form
of the monodromy matrix: It has matrix elements which are finite difference operators 
in the multiplicity labels of low order.

%%%%%%%%%%%%%%%%%%%%%%%%%%%%%%%%%%%%%%%%%%%%%%%%%%%%%%%%%
%%%%%%%%%%%%%%%%%%%%%%%%%%%%%%%%%%%%%%%%%%%%%%%%%%%%%%%%%

\section{Relations to the quantisation of moduli spaces of 
flat connections}\label{Sec:flatconn}

\setcounter{equation}{0}

Our next goal will be to explain how the ``quantum monodromy'' relations
\rf{monod'} can be used to define a quantisation of the moduli spaces of 
flat $\mathfrak{sl}_3$-connections on $C=C_{0,3}$, and more generally on 
punctured Riemann spheres $C_{0,n}$. After recalling the description of 
these moduli spaces as algebraic varieties, we will recall the definition of  
the Verlinde line operators $\SV_{\ga}$, a natural 
family of operators acting on the spaces of conformal blocks labelled
by closed curves $\ga$ on $C$. 
It was shown in \cite{CGT} that the Verlinde line operators
generate representations of the quantised algebras of functions on the 
moduli spaces of flat connections with $\SV_\ga$ representing 
the quantised counterpart of a trace function $L_\ga$ associated to $\ga$.
Our main goal in this section is to explain why the operators $\su$, $\sv$
representing elementary building blocks of the operator-valued 
monodromy matrices $M_\ga(\su,\sv)_{\imath}{}^{\jmath}$ 
are closely related to 
quantum counterparts of a natural higher rank generalisation of the 
Fenchel-Nielsen coordinates.

\subsection{Moduli spaces of flat $\mathrm{SL}(3)$-connections}

Let us denote the space of holomorphic connections on $\mathbb{P}^1\setminus\{z_1,\dots,z_n\}$ 
of the 
form
\begin{equation}
\nabla'=\pa_y+A(y), \qquad A(y)=\sum_{r=1}^n\frac{A_r}{y-z_r},\qquad
\sum_{r=1}^n{A_r}=0,
\end{equation}
with $A_r\in\mathfrak{sl}_N$ modulo overall conjugation by elements of $\mathrm{SL}(N)$
by $\mathcal{M}_{dR}(\mathrm{SL}(N),C_{0,n})$.
Computing the holonomy defines a map
$\mathcal{M}_{dR}(\mathfrak{sl}_N,C_{0,n})\ra\mathcal{M}_B(\mathrm{SL}(N),C_{0,n})$, 
with $\mathcal{M}_B(\mathrm{SL}(N),C_{0,n})$ being the 
{character variety, the space of representations $\rho:\pi_1(C)\ra \mathrm{SL}(N)$,
modulo overall conjugation.
$\CM_{B}(\mathrm{SL}(3),C_{0,n})$ has the structure of an algebraic variety.  
A convenient set of regular 
functions is  provided by the traces of holonomies
$L_{\ga}=\mathrm{tr}(M_{\ga})$, where 
$M_\ga$ is the holonomy along a closed curve $\ga$. 

The description can be 
reduced to the case $n=3$ by pants decompositions.
As a useful set of generators for $\CM_{B}(\mathrm{SL}(3),C_{0,3})$
one can choose the following combinations of trace functions \cite{L1,L2}, see also \cite{CGT}:
Let $M_i$, $i=1,2,3$ be the monodromies around the three punctures of $C_{0,3}$, let $A_i=\mathrm{tr}(M_i)$,
$\bar{A}_i=\mathrm{tr}(M_i^{-1})$ for $i=1,2,3$ and 
\begin{equation}
\begin{aligned}
&N=\mathrm{tr}(M_1^{}M_3^{-1})-A_1\bar{A}_3,\\
&\bar{N}=\mathrm{tr}(M_1^{-1}M_3^{})-\bar{A}_1A_3,
\end{aligned}\quad
\begin{aligned}
&W=\mathrm{tr}(M_3M_2M_1)-A_1\bar{A}_1-A_2\bar{A}_2-A_3\bar{A}_3,\\
&\bar{W}=\mathrm{tr}((M_3M_2M_1)^{-1})-A_1\bar{A}_1-A_2\bar{A}_2-A_3\bar{A}_3.
\end{aligned}
\end{equation}
Other trace functions $L_\ga$ can be expressed in terms of these generators using the skein
relations.

It was shown in \cite{L1,L2,CGT} that the  generators introduced above
satisfy two algebraic relations of the form 
\begin{equation}\label{classrel}
\CP_i\big(\{L_{\ga},\ga\in\pi_1(C_{0,3})\}\big)=0,\qquad i=1,2.
%\CP_2(N,\bar{N},W,\bar{W};\mathbf{A})=0.
\end{equation}
The explicit formulae for the polynomials $\CP_1$ and $\CP_2$ can be found in the references cited above.

The algebra $\mathcal{A}_0=\mathrm{Fun}(\CM_{B}(\mathrm{SL}(3),C_{0,3}))$ is a Poisson-algebra 
with Poisson bracket defined using the Atiyah-Bott symplectic structure. 
The Poisson structure
is algebraic, it closes among the polynomial functions of $(N,\bar{N},W,\bar{W})$ and 
$\mathbf{A}=(A_i, \bar{A}_i)_{i=1,2,3}$ \cite{L1,L2,CGT}.
The functions 
$(A_i, \bar{A}_i)_{i=1,2,3}$ generate the center of this Poisson algebra. 
%\jtr{To be completed; introduce notion of Fenchel-Nielsen type coordinates here.}
%\begin{equation}
%\begin{aligned}
%& \{N,\bar{N}\}=\frac{\partial \CP_1}{\partial W} \frac{\partial \CP_2}{\partial \bar{W}}-
%                                     \frac{\partial \CP_1}{\partial \bar{W}} \frac{\partial \CP_2}{\partial {W}} \\
%& \{N,W\}=\frac{\partial \CP}{\partial \bar{W}_1} \frac{\partial \CP_2}{\partial \bar{N}}-
%                                      \frac{\partial \CP_1}{\partial \bar{N}} \frac{\partial \CP_2}{\partial \bar{W}}
%\end{aligned}\qquad
%\begin{aligned}
%& \{N,\bar{W}\}=\frac{\partial \CP_1}{\partial \bar{N}} \frac{\partial \CP_2}{\partial {W}}-
%                                      \frac{\partial \CP_1}{\partial {W}} \frac{\partial \CP_2}{\partial \bar{N}}\\
%& \{W,\bar{W}\}=\frac{\partial \CP_1}{\partial N} \frac{\partial \CP_2}{\partial \bar{N}}-
%                                      \frac{\partial \CP_1}{\partial \bar{N}} \frac{\partial \CP_2}{\partial {N}}
%\end{aligned}
%\end{equation}

%\subsection{Verlinde line operators -- II}

\subsection{Quantisation of the moduli spaces of flat connections from CFT}

Spaces of conformal blocks carry two natural module structures. 
One type of module structure is canonically associated to the definition of conformal blocks as solution
to the conformal and $\CW$-algebra Ward identities, as defined  
in Section \ref{sec:confbl}. Of particular interest for us is
another type of module structure defined in terms of the so-called Verlinde line operators
and their generalisations. This subsection will present the definition of the Verlinde line
operators, and explain why the algebra of Verlinde line operators is a quantum 
deformation of the algebra of regular functions on  
$\mathrm{Fun}(\CM_{B}(\mathrm{SL}(3),C_{0,3}))$ \cite{CGT}.

In order to define the Verlinde line operators let us note that 
the identity operator appears in the operator product expansion
\begin{equation}\label{OPE-1} 
\begin{aligned} \bar{D}_{\bar{\imath}}^{}(y_0) 
D_\imath^{}(y)&=\,\de_{\bar{\imath},\imath}\,P_{\imath}\,(y_0-y)^{\lambda_0}({\rm id}+Y(y_0-y))\\
&\;\;\,+\sum_{l=2}^3 C_{\bar{\imath},\imath}^{l}\,(y_0-y)^{\lambda_l}(V_{l}(y)+\CO(y_0-y)),
\end{aligned} \end{equation} 
where $Y(y_0-y)=\sum_{n=1}^\infty (y_0-y)^nY_n(y)$, with $Y_n(y)$ being combinations
of the generators of the $\CW_3$ algebra, and  $P_{\imath}\in\BC$, $\imath=1,2,3$.  
The chiral vertex operators $V_{l}(y)$ correspond to the 
other finite-dimensional representations appearing in the tensor product of fundamental and 
anti-fundamental representation of $\fsl_3$.
%Let us note that the expectation values 
%$\langle \bar{e}_{\al_3}^{},Y(y_0-y) V_{\al}^\nu(z)e_{\al_1}^{}\rangle$ 
%are defined through the definition of the conformal blocks $\CF_\mu^\nu$. 
This  shows that  a canonical
projection $\mathbb{P}$ from the space of conformal blocks 
on $C_{0,3+2}$ spanned by the $\mathcal{F}_{k}^\rho(y)_{\bar{\imath}\imath}^{}$ to the
space of conformal blocks on $C_{0,3}$ spanned by ${F}_{k}^\rho$ can be defined by setting
$\mathbb{P}(\mathcal{F}^{\rho}_k(y)_{\bar{\imath}\imath}^{})=
\de_{\bar{\imath}\imath} P_{\imath}{F}_{k}^{\rho}$. 

By taking linear combinations of \rf{OPE-1} one may eliminate the terms containing 
the chiral vertex operators $V_l(y)$, leading to a relation of the form
\begin{equation}\label{OPE-2} 
 \sum_{\imath=1}^3 E^{\imath}\,\bar{D}_{{\imath}}^{}(y_0) 
D_\imath^{}(y)=(y_0-y)^{\lambda_0}({\rm id}+Y(y_0-y)).
\end{equation} 
This allows us to define an embedding $\mathbb{E}$ of the space of conformal blocks 
on $C_{0,3}$ into the space of conformal blocks on $C_{0,3+2}$ by setting
$\mathbb{E}(F_{k}^\rho)=\sum_{\imath=1}^3 E^{\imath}\,\mathcal{F}_{k}^\rho(y)_{{\imath}\imath}$.

As before we find it convenient to regard the functions $\mathcal{F}_{k}^\rho$
as components of a vector $\mathbf{F}_{}^\rho$ with respect to a basis labelled by $k$.
We had observed in Section \ref{sec:confbl} that 
 the analytic continuation of 
$\mathbf{F}^{\rho}(y)_{\bar{\imath},\imath}$ along 
$\ga\in\pi_1(C_{0,3})$ defines an operator $\SM_{\ga}$ 
on the space of conformal blocks on $C_{0,3+2}$,
\begin{equation}
\SM_{\ga}\big( \mathbf{F}^{\rho}(y)_{{\bar{\imath}}\imath}^{}\big)\equiv
\mathbf{F}^{\rho}(\ga.y)_{{\bar{\imath}}\imath}^{}=
\sum_{\jmath}(\SM_\ga)_{\imath}{}^{\jmath}\cdot \mathbf{F}^{\rho}(y)_{{\bar{\imath}}\jmath}^{}.
\end{equation}
The composition $\mathbb{P}\circ \SM_{\ga}\circ \mathbb{E}$ will then be an operator
from the space of conformal blocks on $C_{0,3}$ to itself which may be represented in the 
form
\begin{equation}\label{Verlop}
(\mathbb{P}\circ \SM_{\ga}\circ \mathbb{E})(F^{\rho})=
%\sum_{\imath,\bar{\imath}=1}^3 E_{\imath} \mathbb{P}((\mathbf{F}_{\mu})_{{\imath},\imath}^{}(\ga.y))=
\SV_{\ga}\cdot F^{\rho}, \qquad \SV_{\ga}= \sum_{\imath=1}^3 E^{\imath}\,(\SM_\ga)_{\imath}{}^{\imath}\,
P_{\imath}.
\end{equation}
The operators $\SV_{\ga}$ will be called Verlinde line operators. They can be regarded as 
``quantum'' versions of the traces of the monodromy matrices $M_{\ga}$ of a flat $\mathrm{SL}(3)$-connection.

Assuming the validity of the mondromy relations \rf{monod'} it was shown in \cite{CGT} that
the algebra generated by the operators $\SV_{\ga}$ is
isomorphic to the algebra $\mathcal{A}_\hbar$ of quantised functions on $\CM_{\rm flat}(\mathrm{SL}(3),C_{0,3})$, 
the moduli space of flat $\mathrm{SL}(3)$-connections on $C_{0,3}$.
This means in particular that
the operators $\SV_{\ga}$ satisfy deformed versions of the relations \rf{classrel}
\begin{equation}\label{quantrel}
\CP_i^\hbar\big(\{\SV_{\ga},\ga\in\pi_1(C_{0,3})\}\big)=0,\qquad i=1,2,
%\CP_2(N,\bar{N},W,\bar{W};\mathbf{A})=0.
\end{equation}
with polynomials $\CP_i^\hbar$ differing from the polynomials $\CP_i$ in \rf{classrel} by fixing an operator 
ordering and $\hbar$-dependent deformations of the coefficients. The algebra generated by the Verlinde line operators 
can be identified with the skein algebra  defined from  quantum group theory 
using the Reshetikhin-Turaev construction \cite{CGT}.\footnote{An alternative proof of this result  can be 
based on the observation that the operator $\SD$ appearing in \rf{Dyntwist} represents the 
dynamical twist used in \cite{CGT}.}
It follows that the
 operators $\SV_{\ga}$  satisfy q-commutation relations 
reproducing the Poisson brackets of the corresponding 
trace functions to leading order in $\hbar$. 
%It follows that the Poisson algebra $\mathcal{A}_0$
%defined as the classical limit of $\mathcal{A}_\hbar$ is the algebra of regular 
%functions on the character variety $\CM_{B}(\mathrm{SL}(3),C_{0,3})$ with Poisson bracket 
%defined using the Atiyah-Bott symplectic form.

We had observed in Section \ref{sec:quantmono} that the elements of the 
quantum monodromy matrices can be expressed as functions of two operators
$\su$ and $\sv$. It follows from \rf{Verlop} that the same is true for the 
Verlinde line operators $\SV_{\ga}$. We are now going to 
argue that the classical limits of $\su$ and $\sv$ are closely related to 
natural higher rank analogs of the Fenchel-Nielsen coordinates.

\subsection{The notion of Fenchel-Nielsen type coordinates}

As a preparation let us first discuss defining properties of 
a class of coordinate system which  will be called 
coordinates of Fenchel-Nielsen type. In order to motivate our 
proposal we will briefly review the
case of flat $\mathfrak{sl}_2$ connections.

Recall that useful coordinates 
on $\mathcal{M}_B(\mathrm{SL}(2),C_{0,4})$ are the trace coordinates, associating 
to a closed curve $\ga$ the trace of the holonomy along $\ga$. 
%The trace functions
%can be used to describe $\mathcal{M}_B(\mathrm{SL}(N),C_{0,n})$ as algebraic variety.
For $N=2$, $C=C_{0,4}=\mathbb{P}^1\setminus\{z_1,z_2,z_3,z_4\}$
one may, for example,  introduce the holonomies $M_i$ around the 
punctures $z_i$ and define  the following set of trace functions:
\begin{align}
&L_i=\operatorname{Tr} M_i=2\cos2\pi m_i,\qquad i=1,\ldots,4,\\
&L_s=\operatorname{Tr} M_1 M_2,\qquad L_t=\operatorname{Tr} M_1 M_3,\qquad L_u=\operatorname{Tr} M_2 M_3,
\end{align}
%\begin{figure}[t]
%\epsfxsize12.5cm
%\centerline{\epsfbox{c04a_v2new.eps}}
%\caption{Left: Generators for the fundamentall group on $C_{0,4}$. Right}
%\label{c04}\vspace{.3cm}
%\end{figure}
In order to describe  $\mathcal{M}_B(\mathrm{SL}(2),C_{0,4})$
as an algebraic Poisson-variety let us introduce the polynomial
\begin{subequations}
\begin{align}
 \label{JFR}
\mathcal{P}_{0,4}(L_s,L_t,L_u):&=L_1L_2L_3L_4+L_sL_tL_u+L_s^2+L_t^2+L_u^2+L_1^2+L_2^2+L_3^2+L_4^2\\
 &-\left(L_1L_2+L_3L_4\right)L_s-\left(L_1L_3+L_2L_4\right)L_t
-\left(L_2L_3+L_1L_4\right)L_u-4.
\notag
\end{align}
The polynomial $\CP_{0,4}$ allows us to write both the equation describing 
$\mathcal{M}_B(\mathrm{SL}(2),C_{0,4})$ as an algebraic variety and the Poisson brackets 
among the regular functions $L_s$, $L_t$ and $L_u$ in the following form:
\begin{align} & \text{(i)}\quad \mathcal{P}(L_s,L_t,L_u)=0,\\
&\text{(ii)} \quad \{L_s,L_u\}=\frac{\partial}{\partial L_t}\mathcal{P}(L_s,L_t,L_u)\;\; \text{and cyclic}.
\end{align}
\end{subequations}
A set of 
Darboux coordinates $\lambda,\kappa$ 
for $\mathcal{M}_{B}(\mathrm{SL}(2),C_{0,4})$ can then be defined 
by parameterising $L_s$, $L_t$ and $L_u$ in terms of coordinates $u,v$ as
\begin{align*}
&\begin{aligned}
& (u-u^{-1})^2\,L_t=
C_t^+(u)\,v+C_t^0(u)+C_t^-(u)\,v^{-1}\,,\\
&  (u-u^{-1})^2\,L_u=
C_u^+(u)\,v+C_u^0(u)+C_u^-(u)\,v^{-1}\,,
\end{aligned}\quad L_s=u+u^{-1},
\end{align*}
with coefficient functions $C_t^{\ep}(u)$, $C_u^{\ep}(u)$, $\ep=-1,0,1$ explicitly given as
\begin{align*}
&C_u^0(u)=L_s(L_1L_3+L_2L_4)-2(L_2L_3+L_1L_4)\, ,\\
&C_t^0(u)=L_s(L_2L_3+L_1L_4)-2(L_1L_3+L_2L_4)\, ,\\
&C_u^{\pm}(u)= (u+u^{-1}-2\cos 2\pi (m_1\mp m_2))
(u+u^{-1}-2\cos 2\pi (m_3\mp m_4))\,,\\
& C_t^{\pm}(u)=-u^{\pm 1}C_u^{\pm}(u)\,.
%  \sin\pi(a+s(m_1\mp m_2))\sin\pi(a+s(m_3\mp m_4))\,,
%\\ C_{-}(\la)&=-4\prod_{s=\pm 1} \sin\pi(\la+s(m_1+m_2))\sin\pi(\la+s(m_3+m_4))\,.
\end{align*}
These coordinates are relatives of the coordinates previously introduced in \cite{Ji,NRS}. 
The coordinates defined by representing $u$, $v$ as
$u=e^{\mathrm{i}\la}$, $v=e^{\mathrm{i}\kappa}$ 
are closely related to the Fenchel-Nielsen coordinates
from Teichm\"uller theory.  Restricting  $\la$ and $\kappa$ to the  
flat $\mathrm{PSL}(2,\BR)$-connections coming from the uniformisation of Riemann
surfaces one finds purely imaginary values of these coordinates. 
$2\mathrm{i}\lambda$ will coincide 
with the geodesic length along $\ga_s$, and $\mathrm{i}\kappa$ will 
differ from the Fenchel-Nielsen twist parameter $\tau$ only by a simple
function of $\la$, $\mathrm{i}\kappa=\tau+\eta(\la)$. 
Compared to the Fenchel-Nielsen coordinates
the parametrisation introduced above has the advantage
that  the expressions for the 
trace functions in terms of $\lambda, \kappa$ have 
more favourable analytic properties.
In the case of the Fenchel-Nielsen coordinates one would find 
square-roots in these expressions.

This motivates  the following  definition.
Darboux coordinates 
$(\mu,\nu)$, $\mu=(\mu_1,\dots,\mu_d)$, $\nu=(\nu_1,\dots,\nu_d)$
for $\CM_{B}(C)$ 
are of ``Fenchel-Nielsen type'' 
if
\[
\begin{aligned}
\text{(R)}\quad &\text{the expressions for $L_{\ga}(\mu,\nu)$ have the best possible analytic structure, being}\\
& \text{rational functions of 
                $u_a=e^{\mathrm{i}\mu_a}$, $v_a=e^{\mathrm{i}\nu_a}$ or even Laurent polynomials,}\\
\text{(P)}\quad &\text{the coordinates are compatible with a pants decomposition of $C$.}\\
\end{aligned}
\]

Requirement (R)
means that coordinates of Fenchel-Nielsen type are reflecting the Poisson-algebraic 
structure of $\CM_B(C)$ in an optimal way. Note that this property  is shared by the Fock-Goncharov
coordinates \cite{FG1} for $\CM_B(C)$. The main feature referred to in the terminology 
``Fenchel-Nielsen type'' is (P), shared by the Fenchel-Nielsen coordinates
for the Teichm\"uller spaces. Compatibility with a pants decomposition means
that appropriate subsets of the coordinates would represent coordinate systems
for the surfaces obtained by cutting the surface $C$ along any simple closed curve
defining the pants decomposition.
The Fock-Goncharov coordinates 
 for $\CM_B(C)$ do not have this property.

%\begin{claim}\label{claim1a}
%For each closed curve $\ga$ on $C_{0,3}$ there exists an operator $\SV_\ga:
%\CS_{\mu}\ra\CS_{\mu}$. 
%\end{claim}

\subsection{Quantum Fenchel-Nielsen type coordinates}

In order to relate the operators $\su$ and $\sv$ to quantum counterparts of  
Fenchel-Nielsen type coordinates for $\mathcal{M}_B(\mathrm{SL}(3),C_{0,3})$
one mainly needs to observe the following simple consequence of 
our calculation of quantum monodromy matrices. 
\[
\text{(L)}\quad\left\{
\begin{aligned}
&\text{The operators $\SV_\ga$ generating a representation of $\mathcal{A}_\hbar$, $\hbar=b^2$, 
on the spaces of}\\
&\text{ conformal blocks  can be  represented as  Laurent polynomials $\CV_{\ga}^\hbar(\su,\sv)$ 
of $\su$  and $\sv$.}
\end{aligned}\right\}
\]

%The coefficients will be calculated (semi-)explicitly below. 

%This structure is naturally coming from the description of the conformal blocks in terms 
%of screening charges given above.

One may represent $\su$ and $\sv$ as $\su=e^{\mathrm{i}\hat{\mu}}$ and 
$\sv=e^{\mathrm{i}\hat{\nu}}$, with $\hat{\mu}\,\CF_{\mu,k}=2\pi b^2 k\CF_{\mu,k}$ 
and $\hat{\nu}\,\CF_{\mu,k}=-\mathrm{i}\frac{\pa}{\pa k}\CF_{\mu,k}$, 
respectively.
We will interpret our observation (L) as the statement 
that $\hat{\mu}$ and $\hat{\nu}$ represent  the quantised counterparts of 
Fenchel-Nielsen type coordinates for $\CM_{B}(\mathrm{SL}(3),C_{0,3})$.

In order to support this interpretation let us consider
the Poisson-algebra $\mathcal{R}_0$ generated by rational functions of two variables 
$u,v$ with Poisson-bracket $\{u,v\}=-2\pi \,uv$.
The classical limit  $\mathcal{A}_{0}$ of the algebras $\mathcal{A}_\hbar$ considered above 
is a Poisson-algebra which can be identified with the  sub-algebra of ${\mathcal{R}}_0$
generated by the Laurent polynomials $\CV_{\ga}^0(u,v)$
% of the form 
%\begin{equation}
%\CV_{\ga}(u,v)=\sum_{n\in \BZ}C_{\ga,n}^0(u)\,v^n.
%\end{equation} 
The logarithmic coordinates $\mu=-\mathrm{i}\log u$, $\nu=-\mathrm{i}\log v$ 
have Poisson bracket $\{\mu,\nu\}=2\pi$, identifying
them as Darboux-coordinates for $\mathcal{A}_{0}$. It follows from  
our observations above that mapping $\CV_\ga^0$ to $\CL_{\ga}$ defines an 
isomorphism of Poisson algebras. This means in particular that
the functions $\CV_{\ga}^0(u,v)$ satisfy the relations \rf{quantrel},
and have Poisson brackets that reproduce the Poisson brackets of the 
trace functions $L_{\ga}$ when $L_\ga$ is replaced by $\CV_\ga^0$.

These properties are completely 
analogous to the description of $\CM_{B}(\mathrm{SL}(2),C_{0,4})$, motivating 
us to call the coordinates $\mu$, $\nu$ Fenchel-Nielsen type coordinates for $\CM_{B}(\mathrm{SL}(3),C_{0,3})$.

\subsection{Yang's functions and solutions to the Riemann-Hilbert problem from classical limits}

We will now point out further
relations between Toda CFT and the theory of flat connections on 
$C_{0,3}$ arise in two different limits.
The standard classical limit of Toda field theory corresponds to $b\ra 0$.
Assuming that the vectors $\al_i$ are of the form $\al_i=-b^{-1}\eta_i$ for $i=1,2,3$
one may expect that the conformal blocks behave in the limit $b\ra 0$, $\al_i=-\eta_i/b$, 
$k=\mu/2\pi b^2$, with $\eta_i$, $i=1,2,3$,  and $\mu$ finite, as follows:
\begin{equation}\label{classlim}
\mathcal{F}_{k}^{\rho}(y_0,y)_{{\bar{\imath}}\imath}^{}\sim e^{-\frac{1}{b^2}\CY(\eta,\mu)}
\bar{\chi}_{\bar{\imath}}^{}(y_0)\chi_{\imath}^{}(y),
\end{equation}
where $\chi_{\imath}^{}(y)$, and $\bar{\chi}_{\bar\imath}^{}(y)$, $\imath,{\bar\imath}=1,2,3$, are three linearly 
independent solutions of the differential equations
\begin{align}\label{nvclass}
&\bigg(-\frac{\pa^3}{\pa y^3}+{T}(y)\frac{\pa}{\pa y}
+\frac{1}{2}{T}'(y)+{W}(y)\bigg)\chi_{\imath}^{}(y)\,=\,0\,,\\
&\bigg(-\frac{\pa^3}{\pa y^3}+{T}(y)\frac{\pa}{\pa y}
+\frac{1}{2}{T}'(y)-{W}(y)\bigg)\bar{\chi}_{\bar\imath}^{}(y)\,=\,0\,,
\end{align}
Differential equations of this form are closely related to a special class of 
flat $\mathfrak{sl}_3$-connections on $C_{0,3}$ called {\it opers}. The action of the 
operator-valued matrices $M_\ga(\su,\sv)$ 
on $\mathbf{F}_{}^\rho(y_0,y)_{{\bar\imath}\imath}^{}$ turns into the 
multiplication with the ordinary matrices
$M_\ga(u,v)$ with $u=e^{ \mathrm{i} \mu}$ and 
$v=e^{\mathrm{i}\nu}$  in the limit \rf{classlim}, where
\begin{equation}\label{genfctopers}
\nu=\frac{\mu}{2}+2\pi\frac{b^2}{\mathrm{i}}\frac{\pa}{\pa \mu}\CY(\rho,\mu).
\end{equation} 
The matrices 
$M_\ga(u,v)$ represent the monodromy of % the opers corresponding to 
the differential equation \rf{nvclass}. Equation \rf{genfctopers}
describes the image of $\mathrm{Op}_{\mathfrak{sl}_3}(C_{0,3})$ 
within $\CM_B(\mathrm{SL}(3),C_{0,3})$ under the holonomy map. It therefore 
represents a higher rank analog of the generating function of the variety of
$\mathfrak{sl}_2$-opers proposed to represent the Yang's function 
for the quantised Hitchin system
in \cite{T10,NRS,TLect}.

Another interesting limit is $b\ra i$, where $c=2$.  In this case one finds $q^2=1$, 
so that the operators $\su$ and $\sv$ commute with each other. The transformation
\begin{equation}
\mathcal{G}_{}^{\rho}(y_0,y;\kappa,\la)_{\bar{\imath}\imath}^{}= \sum_{n\in\BZ} e^{\pi\mathrm{i}\,n\la} 
e^{-\frac{\pi}{2}\mathrm{i}\,n^2}
\mathcal{F}_{\kappa+n}^{\rho}(y_0,y)_{\bar{\imath}\imath}^{}
\end{equation}
diagonalises $\su$ and $\sv$ simultaneously with 
eigenvalues $u=e^{-2\pi\mathrm{i}\,\kappa}$ and $v=\mathrm{i}e^{-\pi\mathrm{i}(\kappa+\lambda)}$, respectively.
It follows that the operator-valued monodromy matrices $\SM_\ga=M_{\ga}(\su,\sv)$ are transformed into 
ordinary matrices  $M_{\ga}(u,v)$. The conformal blocks $\mathcal{G}_{}^{\rho}$ represent 
solutions to the Riemann-Hilbert problem to construct multi-valued analytic functions
on $C_{0,3}$ with monodromies fixed by the data $\lambda,\kappa$.  
Comparison with the main result of \cite{ILT} suggests
that \begin{equation}
\mathcal{G}_{}^{\rho}(\kappa,\la)= \sum_{n\in\BZ} e^{\pi\mathrm{i}\,n\la} 
e^{-\frac{\pi}{2}\mathrm{i}\,n^2}
\mathcal{F}_{\kappa+n}^{\rho}
\end{equation}
may serve as a partial replacement for the isomondromic tau function in this context.

\section{Conclusions and outlook}
\setcounter{equation}{0}

Our main result is a remarkably simple relation between the screening charges of the 
free field representation for the $\CW_3$ algebra and the Fenchel-Nielsen type coordinates 
$\CM_{B}(\mathrm{SL}(3),C_{0,3})$. This direct link between the free field realisation of CFTs and
the (quantum) geometry of flat connections on Riemann surfaces seems to offer a key towards a more 
direct understanding of the relations between these two subjects. 
We believe that these relations deserve to be investigated much more 
extensively.

\subsection{More punctures, higher rank}

The generalisation to conformal blocks on surfaces with higher number of punctures should be straightforward. 
A simple counting of 
variables indicates that Fenchel-Nielsen type coordinates for surfaces $C_{0,n}$ 
with arbitrary $n$ can be obtained from the Fenchel-Nielsen coordinates of 
all subsurfaces of type $C_{0,3}$ and $C_{0,4}$ appearing in a pants decomposition 
of $C_{0,n}$.
%It will therefore be sufficient to introduce Fenchel-Nielsen type coordinates
%in the case $C=C_{0,4}$. 
This paper has introduced all ingredients necessary
to compute the monodromies of
degenerate fields on any punctured Riemann sphere. 
Using these ingredients within the set-up used in \cite{ILT} it is 
not hard to see that the operator-valued monodromy matrices can be 
represented as Laurent polynomials of basic shift and multiplication operators
which can be related to Fenchel-Nielsen type coordinates in the same way as before.

In the case of Toda CFTs associated to Lie algebras of higher ranks 
an interesting feature not discussed here will play a more
important role. Instead of only three types of  screening charges,
here denoted $Q_{1}$, $Q_{2}$ and $Q_{12}$ one will in the case of 
$\fsl_N$ Toda CFT have $\frac{1}{2}(N-1)N$  screening charges $Q_{i,j}$, 
$j>i$. These screening charges
are in one-to-one correspondence with the positive roots of $\fsl_N$, and 
can be  constructed as multiple q-commutators of the simple screening charges $Q_{i,i+1}$
associated to the simple roots. This implies that different ordering prescriptions
for the positive roots of $\fsl_N$ will correspond to different bases in the space
of conformal blocks on $C_{0,3}$. The physical correlation functions in Toda CFT 
can not depend on such choices. This invariance may represent a 
supplement to the crossing symmetry conditions exploited in the 
conformal bootstrap approach which may be potentially useful.

\subsection{Continuous bases for spaces of conformal blocks in Toda field theories}

The conformal blocks defined above generate a subspace of the space of 
conformal blocks which is sufficient to describe the holomorphic factorisation 
for the subset of the physical correlation 
functions in $\fsl_3$ Toda CFT admitting multiple integral representations of Dotsenko-Fateev type.
For the case of Liouville theory it is known that one
needs to consider continuous families of 
conformal blocks in order to construct generic correlation functions in a holomorphically  
factorised form.  
The same qualitative feature is expected to hold in Toda CFTs associated to higher
rank Lie algebras.
Based on the experiences from Liouville theory one may anticipate some essential
features of the generalisation from the discrete families of conformal blocks studied in this
paper to the cases relevant for generic Toda correlation functions. 

\subsubsection{Continuation in screening numbers}

Recall 
that the conformal blocks provided by the free field representation 
can be labelled by the so-called screening numbers, the
numbers of screening currents integrated over in the multiple-integral
representations. We conjecture that conformal blocks appearing in 
generic Toda correlation functions are analytic functions of a set of 
parameters which restrict to the conformal blocks coming from the 
free field construction when the parameters are specialised to a
discrete set of values labelled by the screening numbers. 

In order to support this conjecture let us offer the following observations.
A variant of the free field construction can be used to construct
{\it continuous} families of conformal blocks.
For real and sufficiently small values of the parameter $b$
one may 
define, following \cite{T01},  operators $Q_i(\ga)$, $i=1,2$ by
integrating along contours $\ga$  supported on the unit circle.
Following the discussion of the Virasoro case in \cite{T03} one may argue that
the operators $Q_i(\ga)$
are densely defined and can be made {\it positive} self-adjoint
with respect to the natural scalar product on the direct sum of unitary 
Fock modules $\int_{\mathbb{S}} d\beta \,\CV_{\beta}$,
with $\mathbb{S}$ being the set of vectors $\beta$ 
of the form $\beta=\CQ+iP$, $P\in\mathbb{R}^{2}$, $\CQ=\rho_W(b+b^{-1})$, 
$\rho_W$ being the Weyl vector of $\mathfrak{sl}_3$.
For purely imaginary values of $n_1$, $n_{12}$ and $n_2$ one  may therefore define
{\it unitary} operators denoted by $Q_1^{n_2}(\ga)$,  $Q_{12}^{n_{12}}(\ga)$, and $Q_2^{n_1}(\ga)$
by simply taking $Q_i^{n_i}(\ga)=\exp(n_i\log(Q_i(\ga))$, for $i=1,12,2$. We conjecture that 
$Q_i^{n_i}(\ga)$ can be defined for even more
general values of $n_i$ by analytic continuation.

This means in particular that the definition of the conformal blocks $F_{k}^\rho$
given in Section \ref{sec:quantmono} can be generalised
to provide  families of conformal blocks for $C_{0,3}$ labelled
by a set of {continuous} parameters $\rho=(\al_3,\al_2,\al_1)$ and $k$.
We conjecture  that the dependence of $F_{k}^\rho$
on $\rho$ and $k$ is meromorphic. This is consistent with, and supported by the fact that
the operator-valued monodromy matrices calculated in this paper 
have an obvious analytic continuation which is entire analytic in 
the parameter $k$ and meromorphic in $\rho$. If the 
conformal blocks on $C_{0,3}$ have an analytic continuation in these parameters as
conjectured, the quantum monodromies defined from the analytically continued 
conformal blocks must coincide with the obvious analytic continuation of the monodromy 
matrices calculated in this paper.

\subsubsection{Relation to quantum group theory and higher Teichm\"uller theory}

There are further encouraging hints that the description of the continuous families 
of Virasoro conformal blocks relevant for Liouville theory admits a 
higher rank generalisation representing the analytic continuation of the 
conformal blocks considered in this paper. Let us recall that
the conformal blocks needed for generic correlation functions come in families
associated to a continuous series of Virasoro representations \cite{ZZ}. The monodromies of conformal 
blocks can be represented in term of the $6j$-symbols associated to a continuous family
of representations of the non-compact quantum group $\CU_q(\mathrm{SL}(2,\BR))$ \cite{T01}.
The relevant family of quantum group representations \cite{PT1,PT2} is distinguished by remarkable
positivity properties closely related \cite{BT} to the phenomenon of modular duality \cite{Fa1,PT1,Fa2}.
 All this is 
closely related to the quantisation of the Teichm\"uller spaces \cite{CF,Ka1}:
It turns out that the $6j$-symbols 
of $\CU_q(\mathrm{SL}(2,\BR))$ mentioned above describe the change of pants decompositions 
on the four-holed sphere in quantum Teichm\"uller theory \cite{T03,NT}. 
The situation is summarised
in Table 4. 

\begin{table}\label{Comparison4}
\begin{center}
\begin{tabular}{c|c|c}
{Conformal field theory}   &{Quantum group theory} & Moduli space \\ \hline\hline \\[-3ex]
{Liouville theory}  &  {Modular double} & 
$\mathrm{PSL}(2,\BR)$-connections, \\ 
&of $\CU_q(\mathrm{SL}(2,\BR))$ & Fuchsian component
\\ \hline
 \end{tabular}
 \caption{\it Relations between conformal field theory, quantum group theory
 and moduli spaces of flat connections relevant for the case of Liouville theory.
 The Fuchsian component 
in the moduli space of flat $\mathrm{PSL}(2,\BR)$-connections is the connected 
component formed by all connections having holonomies $\rho:\pi_1(C)\ra\mathrm{PSL}(2,\BR)$
such that $\mathbb{H}/\rho$ is a Riemann surface of the same topological type as $C$.}
 \end{center}
\end{table}

The mathematical structures appearing in Liouville theory have generalisations associated to 
Lie algebras of higher ranks. The higher Teichm\"uller theories \cite{FG1} can be quantised \cite{FG2}. A key feature
facilitating the quantisation 
of the higher Teichm\"uller spaces is the positivity of the coordinates introduced in \cite{FG1} 
when restricted to the so-called Teichm\"uller component, a connected component in
the moduli spaces of flat  $\mathrm{PSL}(N,\BR)$-connections generalising the 
Fuchsian component isomorphic to the Teichm\"uller spaces for $N=2$ \cite{Hi}.
This 
generalises similar properties of the  ordinary Teichm\"uller spaces when described 
in terms of the coordinates going back to \cite{Pe}.

The main new feature in the higher rank cases is the appearance of non-trivial
spaces of three-point conformal blocks. We expect these spaces to be 
isomorphic to the multiplicity spaces of  Clebsch-Gordan maps for the 
positive representations of $\CU_q(\mathrm{SL}(N,\BR))$.
The study of the positive representations  of
$\CU_q(\mathrm{SL}(N,\BR))$, $N>2$, was initiated in 
\cite{FI}. 
The Clebsch-Gordan maps for this class of representations 
of $\CU_q(\mathrm{SL}(N,\BR))$ have recently been  constructed in \cite{SS2}. 
For the case $N=3$ one gets bases for the space of Clebsch-Gordan maps
labelled by one continuous parameter.
The constructions in \cite{SS1,SS2} reveal a direct connection 
between the positive representations of
$\CU_q(\mathrm{SL}(N,\BR))$  and  the quantised higher Teichm\"uller theories
generalising what was found for $N=2$ in 
\cite{K,NT}. 

It should be possible to establish a link between the continuous families of 
conformal blocks introduced above and 
a suitable basis for the space of Clebsch-Gordan maps between positive
representations of $\CU_q(\mathrm{SL}(3,\BR))$ by generalising the constructions
in \cite{T01} for the case of Liouville theory.  The positivity of the screening charges 
together with the direct relation between Clebsch-Gordan maps and free field 
representation established in the paper suggests that there is a scalar product on the 
space of conformal blocks on $C_{0,3}$ with $\rho=(\al_3,\al_2,\al_1)$, $\al_i=\CQ+iP_i$
 for which the conformal blocks $F_{k}^\rho$
with $k\in\mathrm{i}\BR$  generate a basis.

%\jtr{To be completed; point out relation to isomonodromic tau-function.}

%%%%%%%%%%%%%%%%%%%%%%%%%%%%%%%%%%%%%%%%%%%%%%%

%\newpage

%%%%%%%%%%%%%%%%%%%%%%%%%%%%%%%%%%%%%%%%%%%%%%%%%%%%%%%%%

%\newpage

{\bf Acknowledgements.} 
The authors would like to thank L. Hollands, A. Neitzke, G. Schrader and 
A. Shapiro for interesting discussions on related topics.

This work was supported by the Deutsche Forschungsgemeinschaft (DFG) through the 
collaborative Research Centre SFB 676 ``Particles, Strings and the Early Universe'', project 
A10. E. Pomoni's work is supported by the German Research Foundation (DFG) via the Emmy
Noether program ``Exact results in Gauge theories''.

\newpage

%%%%%%%%%%%%%%%%%%%%%%%%%%%%%%%%%%%%%%%%%%%%%%%%%%%%%%%%%%

\appendix

\section{Useful relations}\label{app:UsefulRel}
\setcounter{equation}{0}
%%%%%%%%%%%%%%%%%%%%%%%%%%%%%%%%%%%%%%%%%%%%%%%%%%%%%%%%%%%%
%%%%%%%%%%%%%%%%%%%%%%%%%%%%%%%%%%%%%%%%%%%%%%%%%%%%%%%%%%%%

We wish to compute the action of the raising operators $\se_i$ on 
some generic representation of $\cU_q(\fsl_3)$, which is 
constructed from a highest weight vector $v_\lambda$, that is 
$\se_i e_\bn^\lambda = \se_i \sf_1^{n_1} \sf_{12}^{n} \sf_2^{n_2} v_\lambda$. 
To do so we use the notations $\bn = (n_1,n,n_2) = (s_1 -k , k, s_2-k) = \ss_k$ and also 
$\bs_1=(1,0,0)$, $\bs_2=(0,0,1)$, 
$\bs_3=(1,0,1)$ and $\bs_{12}=(0,1,0)$, as well as the relations  
\begin{align} 
& [\se_1 , \sf_{12}] = \sf_2 \sk_1^{-1} ~, \qqq 
[\se_2 , \sf_{12}] = - \fq \sf_1 \sk_2 \\
& [\se_{12},\sf_1] = -\fq \sk_1 \se_2 ~, \qqq 
[\se_{12},\sf_2] = \sk_2^{-1} \se_1 \\
&
[\se_{12},\sf_{12}] = \frac{\fq}{\fq-\fq^{-1}} (\sk_1^{-1}\sk_2^{-1} - \sk_1 \sk_2) \\
& 
\se_1 \sf_1^{m} = \sf_1^{m}\se_1 + \frac{[m]}{\fq-\fq^{-1}} 
\sf_1^{m - 1} (\fq^{2(1-m)}\sk_1-\sk_1^{-1}) \\
& 
\se_2 \sf_2^m =\sf_2^m \se_2 + \frac{[m]}{\fq-\fq^{-1}} 
\sf_2^{m - 1} (\fq^{2(1-m)}\sk_2-\sk_2^{-1}) \\
&
\se_1 \sf_{12}^m = \sf_{12}^m \se_1 + [m] \sf_{12}^{m-1} \sf_2 \sk_1^{-1} \\
& 
\se_2 \sf_{12}^m = \sf_{12}^m \se_2 - [m] \fq^{2-m} \sf_1 \sf_{12}^{m-1} \sk_2 \\
& 
\se_{12} \sf_1^m = \sf_1^m \se_{12} - [m] \fq^{-2m+3} \sf_1^{m-1} \sk_1 \se_2 \\
&  
\se_{12} \sf_{12}^m = \sf_{12}^m \se_{12} - 
\frac{ [m]}{1-\fq^{-2}} 
\sf_{12}^{m-1} (\fq^{2(1-m)} \sk_1 \sk_2 - \sk_1^{-1} \sk_2^{-1})
\\
& 
\se_{12} \sf_2^m = \sf_2^m \se_{12} + [m] \sf_2^{m-1} \sk_2^{-1} \se_1~.
\end{align}
The matrix elements which give actions of generators 
$\se_i$ on representations of $\cU_q(\fsl_3)$ are 
\bea \label{eq:action-ei}
\se_1 e_{\ss_k}^\lambda &=& 
\sV^{-1}[k] \fq^{-s_2-(e_1,\lambda)}   
e_{\ss_k - \bs_1}^\lambda
+
[s_1-k] [1-s_1-k+s_2+(e_1,\lambda)] 
\fq^{1+2k-s_2-(e_1,\lambda)}  
e_{\ss_k - \bs_1 }^\lambda
\nn\\
\se_2 e_{\ss_k}^\lambda &=& 
-\sV^{-1} [k] \fq^{2-2s_2+(e_2,\lambda)} 
%\sf_1^{n_1+1} \sf_{12}^{n-1} \sf_2^{n_2} v_\lambda    
e_{\ss_k - \bs_2}^\lambda
+
[s_2-k] [1-s_2+k+(e_2,\lambda)] \fq^{1-(e_2,\lambda)} 
e_{\ss_k - \bs_2}^\lambda
\nn\\
\se_{12} e_{\ss_k}^\lambda &=& 
- 
[s_1-k][s_2-k] [1-s_2+k+(e_2,\lambda)] 
\fq^{3-2s_1+s_2+(e_1-e_2,\lambda)}     
e_{\ss_k - \bs_3}^\lambda
\nn\\
&~& + 
\sV^{-1} [k] \fq^{4-2k-s_2+(e_{12},\lambda)} 
(\fq^{-2(s_1-k)} [s_1 - k] + [s_2-1-(e_{12},\lambda)]) 
e_{\ss_k - \bs_3 }^\lambda ~,
\eea 
denoting the non-simple root $e_{12}= e_1+ e_2$.

The operator $\sV$ acts on representations $e_{\ss_k}$ through 
$\sV e_{\ss_k}^\lambda = \fq^{-k} e_{\ss_{k+1}}^\lambda$ 
and therefore shifts the value of the multiplicity label $k$. 
One may note however that $k$ is bounded and when it is 
zero, any term which includes $\sV^{-1}$ in equations 
\eqref{eq:action-ei} vanishes due to the presence of the factor $[k]$. 
Thus from equations \eqref{eq:action-ei}, for some generic 
fixed $\ss_k$, 
the matrix elements representing the generators $\se$ 
are non-zero only for particular $\ss_k'$
\bea 
R_{(\se_1),\ss_k'}^{\lambda , \ss_k}  && \neq 0 \quad \mbox{iff} 
\quad 
\ss_k' \in \{ \ss_k + (1,0,0) ~,~\ss_{k+1} + (1,0,0) \}
 ~, \\
R_{(\se_2),\ss_k'}^{\lambda , \ss_k}  && \neq 0 \quad \mbox{iff} 
\quad 
\ss_k' \in \{ \ss_k + (0,0,1) ~,~\ss_{k+1} + (0,0,1) \}
 ~, \\
R_{(\se_{12}),\ss_k'}^{\lambda , \ss_k}  && \neq 0 \quad \mbox{iff} 
\quad 
\ss_k' \in \{ \ss_k + (1,0,1) ~,~\ss_{k+1} + (1,0,1) \} 
 ~.
\eea
When $\ss_k=0$, the values of $\ss_k'$ are more restricted, 
since each component of $\ss_k'$ is non-negative
\bea 
R_{(\se_i),\ss_k'}^{\lambda , 0}  \neq 0 \quad \mbox{iff} 
\quad 
\ss_k' = \{ (1,0,0) \delta_{i,1} , (0,0,1) \delta_{i,2} , (1,0,1) \delta_{i,12}, (0,1,0) \delta_{i,12}\}
\eea
%If we hold $s_i$ fixed, then equations \eqref{eq:action-ei} set the representations of the $\se_i$ generators to be   
From equations \rf{eq:action-ei}, we read off the following matrix elements
\bea \label{explicit-matrix-coefficients}
R_{(\se_1),\ss_k}^{\lambda , \ss_{k-1}-\bs_1}  &=&
 [k] \fq^{-s_2 + k -(e_1,\lambda)} 
\\
R_{(\se_1),\ss_k}^{\lambda , \ss_{k}-\bs_1}  &=& 
[s_1 - k] [1-s_1-k+s_2+(e_1,\lambda)] \fq^{1+2k-s_2-(e_1,\lambda)}
\nn\\
R_{(\se_2),\ss_k}^{\lambda , \ss_{k-1}-\bs_2}  &=& 
-[k] \fq^{2-2s_2+k+(e_2,\lambda)}
\nn\\
R_{(\se_2),\ss_k}^{\lambda , \ss_{k}-\bs_2}  &=& 
[s_2-k] [1-s_2+k+(e_2,\lambda)] \fq^{1-(e_2,\lambda)}
\nn\\
R_{(\se_{12}),\ss_k}^{\lambda , \ss_{k-1}-\bs_3}  &=& 
[k] \fq^{4-k-s_2+(e_{12},\lambda)} ( \fq^{-2(s_1-k)}[s_1-k] + 
[s_2-1-(e_{12},\lambda)] ) ~
\nn\\
R_{(\se_{12}),\ss_k}^{\lambda , \ss_{k}-\bs_3}  &=& 
- [s_1-k] [s_2-k] 
[1-s_2+k+(e_2,\lambda)] \fq^{3 - 2s_1 + s_2 +(e_1-e_2,\lambda)} ~. \nn
\eea
The $\se_i$ label can be omitted since it is redundant, being implied 
by the difference of weights. 
%The equations for the matrix elements representing the actions of the 
%generators $\se_1$ and $\se_2$ imply those corresponding to $\se_{12}$. 
Similarly, the action of generators $\sf_i$ is represented by
\be 
R_{\ss_k}^{\lambda , \ss_k + \bs_1} = 1 ~, ~ 
R_{\ss_k}^{\lambda , \ss_{k+1} + \bs_3} = \fq^{s_1 - k} ~, ~ 
R_{\ss_k}^{\lambda , \ss_k + \bs_2} = \fq^{2k - s_1} ~, ~ 
R_{\ss_k}^{\lambda , \ss_{k+1} + \bs_2} = - \fq^{k - s_1} [s_1 - k]~.
\ee

%%%%%%%%%%%%%%%%%%%%%%%%%%%%%%%%%%%%%%%%%%%%%%%%%%%%%%%%%
%%%%%%%%%%%%%%%%%%%%%%%%%%%%%%%%%%%%%%%%%%%%%%%%%%%%%%%%%

\section{Clebsch-Gordan coefficients} 
\label{sec:app:CGcoeff}
\setcounter{equation}{0}
%%%%%%%%%%%%%%%%%%%%%%%%%%%%%%%%%%%%%%%%%%%%%%%%%%%%%%%%%
%%%%%%%%%%%%%%%%%%%%%%%%%%%%%%%%%%%%%%%%%%%%%%%%%%%%%%%%%

Here we sketch how to compute the Clebsch-Gordan 
coefficients which enter the construction 
of chiral vertex operators in section \ref{tensor}, starting from equation 
\eqref{intertw2} reproduced below 
\be \label{eq:generalCGrel}
0= \sum_{\mb{n}'_2} \left(\begin{smallmatrix}
\lambda_3  \\ 0 
\end{smallmatrix}\middle|
\begin{smallmatrix}
 \lambda_2 & \lambda_1 \\  \mb{n}'_2 & \mb{n}_1
\end{smallmatrix}\right)
\fq^{(e_i,\nu_1)}  
 R_{%(\se_i),
 \mb{n}'_2}^{\lambda_2 , \mb{n}_2}  + 
\sum_{\mb{n}'_1} \left(\begin{smallmatrix}
\lambda_3  \\ 0 
\end{smallmatrix}\middle|
\begin{smallmatrix}
 \lambda_2 & \lambda_1 \\  \mb{n}_2 & \mb{n}'_1
\end{smallmatrix}\right) 
R_{%(\se_i),
\mb{n}'_1}^{\lambda_1 , \mb{n}_1}  ~.
\ee
To lighten notation, let us 
drop the first column in the notation 
$(\cdot|\cdot \cdot)$, whose first column will be everywhere 
$(\begin{smallmatrix} \lambda_3 \\ 0 \end{smallmatrix})$
and where $\lambda_3$ is fully determined by the other 
weights.

\subsection{Clebsch-Gordan coefficients for generic weights $\lambda$}

For generic weights $\lambda$,  
equation \eqref{eq:generalCGrel} can be used iteratively 
to determine the CGC 
$\left(\begin{smallmatrix}
 \lambda_2 & \lambda_1 \\ \bn_2 & \bn_1
\end{smallmatrix}\right) $
in terms of 
$\left(\begin{smallmatrix}
 \lambda_2 & \lambda_1 \\ \bn & 0
\end{smallmatrix}\right) $ 
by constructing $\bn_1$ as 
$\bn_1= n_{1,1}\bs_1+n_1\bs_{12}+n_{1,2}\bs_2$. 
Below we first describe the calculation for $\bn_1\in\{\bs_1,
\bs_2,\bs_{12},\bs_3\}$, which represent the first 
two steps of the iteration, and then describe the general inductive step. 
Beginning with equation 
\eqref{eq:generalCGrel}, $\mb{n}_1=0$ and $i=1,2$  
\bea  
0 &=&
\left(\begin{smallmatrix}
 \lambda_2 & \lambda_1 \\  \ss_k & \bs_i
\end{smallmatrix}\right)  
R_{\bs_i}^{\lambda_1,0}  
+ 
\left(\begin{smallmatrix}
 \lambda_2 & \lambda_1 \\  \ss_k +\bs_i & 0
\end{smallmatrix}\right) 
\fq^{(e_i,\lambda_{1})}
R_{ \ss_k+\bs_i}^{\lambda_2, \ss_k}  + 
\left(\begin{smallmatrix}
 \lambda_2 & \lambda_1 \\ \ss_{k+1} + \bs_i & 0
\end{smallmatrix}\right) 
\fq^{(e_i,\lambda_{1})}
R_{\ss_{k+1} + \bs_i}^{\lambda_2,\ss_k} ~. \qqq\quad 
\eea
The resulting system is under-determined, so we solve for 
a one-parameter family of solutions with the constraint 
\eqref{normcond}
\bea \label{eq:relaCG1}
\left(\begin{smallmatrix}
 \lambda_2 & \lambda_1 \\ \ss_k & \bs_i
\end{smallmatrix}\right)_k  &=& -
\fq^{(e_i,\lambda_{1})}
\left(R_{\bs_i}^{\lambda_1,0}  \right)^{-1} 
R_{\ss_k+\bs_i}^{\lambda_2,\ss_k} ~,~
\left(\begin{smallmatrix}
 \lambda_2 & \lambda_1 \\  \ss_{k-1} & \bs_i
\end{smallmatrix}\right)_k  = -
\fq^{(e_i,\lambda_{1})}
\left(R_{\bs_i}^{\lambda_1,0}  \right)^{-1} 
R_{\ss_k+\bs_i}^{\lambda_2,\ss_{k-1}} ~.~~~~
\eea
The second step sets $\bn_1'=\bs_{12},\bs_3$ and we use \eqref{eq:generalCGrel} with $(\bn_1=\bs_2,i=1)$ and $(\bn_1=\bs_1,i=2)$
\bea \label{eq:relaCG4a}
0 &=&
\sum_{j=3,12}
\left(\begin{smallmatrix}
 \lambda_2 & \lambda_1 \\  \ss_l & \bs_{j}
\end{smallmatrix}\right)_k  
R_{\bs_{j}}^{\lambda_1,\bs_3-\bs_i}  
+ 
\fq^{1+(e_i,\lambda_{1})}\sum_{l'=0,1}
\left(\begin{smallmatrix}
 \lambda_2 & \lambda_1 \\  \ss_{l+l'}+\bs_i & \bs_3-\bs_i
\end{smallmatrix}\right)_k 
R_{\ss_{l+l'}+\bs_i}^{\lambda_2,\ss_l}  ~.
\nn
\eea
These equations alone are not sufficient to fully determine 
$\left(\begin{smallmatrix}
 \lambda_2 & \lambda_1 \\ \bn_2 & \bs_3
\end{smallmatrix}\right)_k $, 
$\left(\begin{smallmatrix}
 \lambda_2 & \lambda_1 \\ \bn_2 & \bs_{12}
\end{smallmatrix}\right)_k $. One first 
needs to insert the solutions \eqref{eq:relaCG1} into 
the right-most coefficients, thereby arriving at a pair of coupled equations 
\begin{align} \label{eq:relaCG5a}
&
\fq^{-1-(e_{12},\lambda_{1})} 
R_{\bs_i}^{\lambda_1,0}
\sum_{j=3,12}
\left(\begin{smallmatrix}
 \lambda_2 & \lambda_1 \\  \ss_l & \bs_{j}
\end{smallmatrix}\right)_k 
R_{\bs_{j}}^{\lambda_1,\bs_i}    
= 
\left(\begin{smallmatrix}
 \lambda_2 & \lambda_1 \\  \ss_l+\bs_3 & 0
\end{smallmatrix}\right)_k 
R_{\ss_l+\bs_3-\bs_i}^{\lambda_2,\ss_l}
R_{\ss_l+\bs_3}^{\lambda_2,\ss_l+\bs_3-\bs_i} 
\\
& \qqq\qqq\qqq\qqq    
+
\left(\begin{smallmatrix}
 \lambda_2 & \lambda_1 \\  \ss_{l+1}+\bs_3 & 0
\end{smallmatrix}\right)_k  
\left(
R_{\ss_l+\bs_3-\bs_i}^{\lambda_2,\ss_l} 
R_{\ss_{l+1}+\bs_3}^{\lambda_2,\ss_l+\bs_3-\bs_i} 
+
R_{\ss_{l+1}+\bs_3-\bs_i}^{\lambda_2,\ss_l} 
R_{\ss_{l+1}+\bs_3}^{\lambda_2,\ss_{l+1}+\bs_3-\bs_i} 
\right)   
\nn\\
& \qqq\qqq\qqq \qqq
+
\left(\begin{smallmatrix}
 \lambda_2 & \lambda_1 \\  \ss_{l+2}+\bs_3 & 0
\end{smallmatrix}\right)_k 
R_{\ss_{l+1}+\bs_3-\bs_i}^{\lambda_2,\ss_l}
R_{\ss_{l+2}+\bs_3}^{\lambda_2,\ss_{l+1}+\bs_3-\bs_i} ~
\nn\end{align}
with three solutions $\left(\begin{smallmatrix}
 \lambda_2 & \lambda_1 \\ \ss_l & \bs_i
\end{smallmatrix}\right)_k$ distinguished by $l\in\{k,k-1,k-2\}$.

\paragraph{Inductive step:} Let us consider a pair of equations 
labeled by two integers $(s_1,s_2)$ which enter the weight 
$\nu_1'=\lambda_1 - s_1 e_1 - s_2 e_2 < \lambda_1$. If we denote by 
$w'$ the sum $w'=s_1+s_2$ of these labels, which 
enter the CGC under the weight $\lambda_1$  
\be \label{eq:induction1}
\left(\begin{smallmatrix}
 \lambda_2 & \lambda_1 \\ \bn & \bs_m
\end{smallmatrix}\right)_k ~, \qqq 
\left(\begin{smallmatrix}
 \lambda_2 & \lambda_1 \\ \bn & \bs_{m+1}
\end{smallmatrix}\right)_k ~,
\ee
then these coefficients are determined by the coupled pair of linear 
equations with label $(s_1,s_2)$ in terms of coefficients with 
$w=w'-1$ and for which 
$\nu_1= \lambda_1 - s_1 e_1 - s_2 e_2 + e_i > \nu_1'$.
It is thus possible to set up an induction procedure through the pair of equations
\be \label{eq:induction2}
\sum_{m'=0,1} 
\left(\begin{smallmatrix}
 \lambda_2 & \lambda_1 \\  \bn & \ss_{m+m'} 
\end{smallmatrix}\right)_k 
R_{\ss_{m+m'}}^{\lambda_1 , \ss_m - \bs_i}
= 
-\fq^{(e_i,\nu_1)}
\sum_{\bn'\in\{\bn+\bs_i, \bn+\bs_{12}-\bs_{j\neq i}\}}
\left(\begin{smallmatrix}
 \lambda_2 & \lambda_1 \\  \bn' & \ss_m - \bs_i
\end{smallmatrix}\right)_k 
R_{\bn'}^{\lambda_2,\bn}    ~
\ee
for $i=1,2$, which corresponds to the actions of the $\se_1,\se_2$ 
generators respectively. Note that the parameter $j$ can also only 
take values $1,2$ depending on the particular value of $i$. A 
possible obstruction to inverting this system and solve for the CGC 
in \eqref{eq:induction1} could arise if the matrix of  
coefficients in equations \eqref{eq:induction2} 
\be 
\left(\begin{smallmatrix}
 R_{\ss_{m}}^{\lambda_1 , \ss_m -\bs_1} & R_{\ss_{m+1}}^{\lambda_1 , \ss_m -\bs_1} \\  R_{\ss_{m}}^{\lambda_1 , \ss_m -\bs_2} & R_{\ss_{m+1}}^{\lambda_1 , \ss_m -\bs_2}
\end{smallmatrix}\right) ~
\ee
were not invertible. This is however not the case for generic weights 
$\lambda_1,\lambda_2$, given the explicit expressions \eqref{explicit-matrix-coefficients}.

\subsection{Clebsch-Gordan coefficients for the case $\lambda_2=\omega_1$}\label{simpleClebsch}

When one of the highest 
weights is a fundamental weight, specifically 
$\lambda_2=\omega_1$, the CGC no longer come in families 
parametrized by some integer $k$ and they can be computed in a few steps. For example 
\bea 
0 =
\left(\begin{smallmatrix}
 \omega_1 & \lambda \\  \bs_1 & 0
\end{smallmatrix}\right) 
\fq^{(e_1 , \lambda)} 
R_{\bs_1}^{\omega_1 , 0}  + 
\left(\begin{smallmatrix}
 \omega_1 & \lambda \\  0 & \bs_1
\end{smallmatrix}\right) 
R_{\bs_1}^{\lambda , 0}  
\quad \Rightarrow \quad  
\left(\begin{smallmatrix} \omega_1 & \lambda\\ 0 &\bs_1 \end{smallmatrix}\right) 
=
\fq^{-1} [-(e_1,\lambda)]^{-1}  
\left(\begin{smallmatrix} \omega_1 & \lambda \\ \bs_1 &0
\end{smallmatrix}\right) ~.~~~~~
\eea 
When $\bn_1'=\bs_{12},\bs_3$, one finds 
from equation \eqref{eq:generalCGrel}
\bea  \label{eq:CGconstraint-deg-1}
0 &=&
\left(\begin{smallmatrix}
 \omega_1 & \lambda \\  \bs_i & \bs_3-\bs_i
\end{smallmatrix}\right) 
\fq^{1+(e_i,\lambda)}
R_{\bs_i}^{\omega_1,0} 
 + 
\left(\begin{smallmatrix}
 \omega_1 & \lambda \\  0 & \bs_3
\end{smallmatrix}\right)R_{\bs_3}^{\lambda,\bs_3-\bs_i}+ 
\left(\begin{smallmatrix}
 \omega_1 & \lambda \\  0 & \bs_{12}
\end{smallmatrix}\right)
R_{\bs_{12}}^{\lambda,\bs_3-\bs_i} ~,~~ i=1,2 \quad 
\eea
and
\bea  \label{eq:CGconstraint-deg-3}
0 &=&
\left(
\begin{smallmatrix}
 \omega_1 & \lambda \\  \bs_{12} & 0
\end{smallmatrix}\right) 
 \fq^{(e_{12},\lambda)}
 R_{\bs_{12}}^{\omega_1,0}  
+
\left(\begin{smallmatrix}
 \omega_1 & \lambda \\  \bs_3 & 0
\end{smallmatrix}\right) 
 \fq^{(e_{12},\lambda)}
R_{\bs_3}^{\omega_1,0}   
\nn\\
&&  + 
\left(\begin{smallmatrix}
 \omega_1 & \lambda \\  0 & \bs_3
\end{smallmatrix}\right) 
R_{\bs_3}^{\lambda,0}  + 
\left(\begin{smallmatrix}
 \omega_1 & \lambda \\  0 & \bs_{12}
\end{smallmatrix}\right)  
R_{\bs_{12}}^{\lambda,0}  
+ (1-\fq^2)
\left(\begin{smallmatrix}
 \omega_1 & \lambda \\  \bs_1 & \bs_2
\end{smallmatrix}\right)  
\fq^{(e_1,\lambda)}
R_{\bs_2}^{\lambda,0} R_{\bs_1}^{\omega_1,0}
~.  \qqq\quad~~
\eea
The solution to this system is given by 
\be 
\left(
\begin{smallmatrix}
 \omega_1 & \lambda \\  0 & \bs_{12}
\end{smallmatrix}\right) = -[-(e_2,\lambda)] 
\fq^{1+(e_{12},\lambda)} 
\fc_{\lambda}^{-1} ~,\quad 
\left(
\begin{smallmatrix}
 \omega_1 & \lambda \\  0 & \bs_3
\end{smallmatrix}\right) =
\fq^{1+(e_{12},\lambda)} 
\fc_{\lambda}^{-1} 
\ee
\be 
\left(
\begin{smallmatrix}
 \omega_1 & \lambda \\  \bs_1 & \bs_2
\end{smallmatrix}\right) = \fq^{(e_2-e_1,\lambda)} 
([-(e_2,\lambda)]-[1+(e_1,\lambda)]) 
\fc_{\lambda}^{-1} ~,
\ee
for the normalization 
$ 
\left(
\begin{smallmatrix}
 \omega_1 & \lambda \\  \delta_i & 0
\end{smallmatrix}\right) = 1 %~\nn
$ and where 
\bea 
\fc_{\lambda}= 
\fq^{-(e_{12},\lambda)}[(e_2,\lambda)]\left(
\fq^{2+2(e_1,\lambda)}(1+[-(e_{12},\lambda)])
+
(1-\fq^2)
\left([-(e_2,\lambda)]-[1+(e_1,\lambda)]\right)
\right)~.\nn
\eea

%%%%%%%%%%%%%%%%%%%%%%%%%%%%%%%%%%%%%%%%%%%%%%%%%%%%%%%%%%%%
%%%%%%%%%%%%%%%%%%%%%%%%%%%%%%%%%%%%%%%%%%%%%%%%%%%%%%%%%%%%

\section{Braid matrix derivation}
\label{sec:app:braiding2}
\setcounter{equation}{0}
%%%%%%%%%%%%%%%%%%%%%%%%%%%%%%%%%%%%%%%%%%%%%%%%%%%%%%%%%%%%
%%%%%%%%%%%%%%%%%%%%%%%%%%%%%%%%%%%%%%%%%%%%%%%%%%%%%%%%%%%%

In this section we collect a few of the more technical 
points that occur in the calculations of the braiding of screened vertex 
operators, as was depicted in Figure \ref{Braidop} and 
reproduced below in Figure \ref{Braidop2}.
\begin{figure}[h!]
\centering
\includegraphics[width=0.5\textwidth]{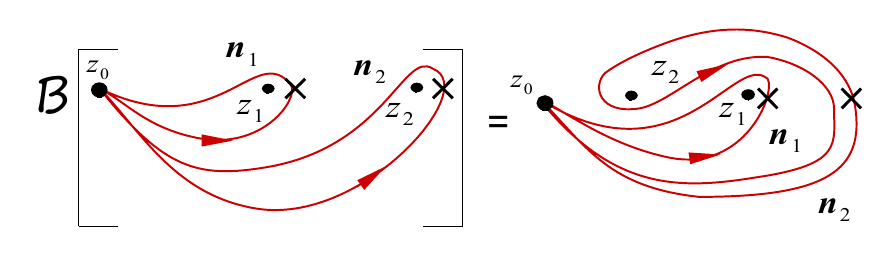}
\caption{{\it Braiding of two generic multi-contours.  
}}
\label{Braidop2}
\end{figure}
This procedure was described in section \ref{sec:braidingSVO1} as an induction process to 
deform contours around $z_2$ in Figure 
\ref{Braidop2} to pass to linear combinations of the 
types of loops illustrated in Figure \ref{Braid1B} and 
then deform the resulting auxiliary loops into a set 
of basis contours like those in Figure \ref{Contours1}.
\begin{figure}[h!]
\centering
\includegraphics[width=0.3\textwidth]{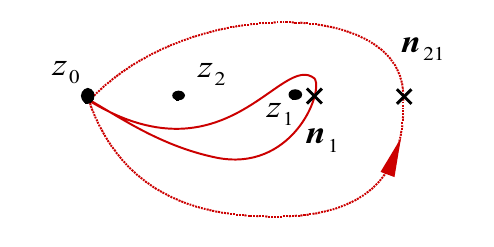}
\caption{{\it A set of auxiliary contours for the braiding calculation.}}
\label{Braid1B}
\end{figure}
\begin{figure}[h!]
\centering
\includegraphics[width=1.0\textwidth]{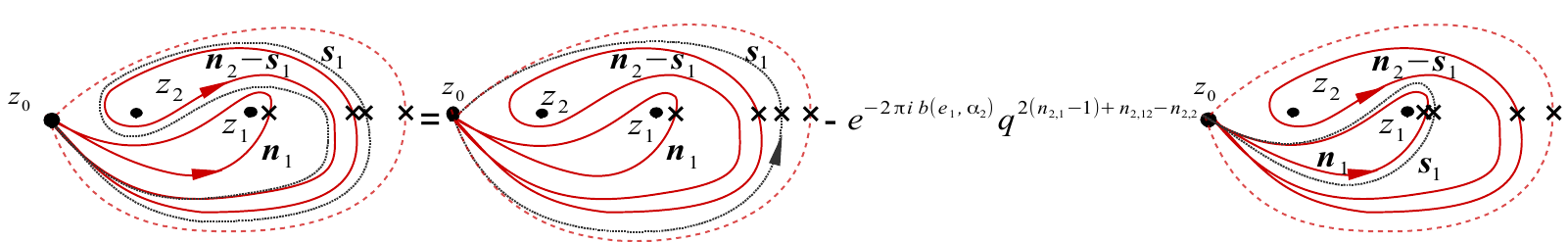}
\caption{{\it Deformation of the outermost of the contours encircling $z_2$. 
}}
\label{Braid2B}
\end{figure}

\paragraph{Part 1:} The first part of the braiding 
calculation specifically deforms the contours 
around $z_2$ from Figure \ref{Braidop2} to linear 
combinations of loops like in Figure \ref{Braid1B}. 
An intermediate step of this procedure is depicted in 
Figure \ref{Braid2B}. 
The ordered marked points $\mathsf{x}$ represent 
normalization points for the integral over the 
indicated contours, where the integrand is real.  
The gray contour being deformed in Figure 
\ref{Braid2B} is associated with a $Q_1$ 
screening charge and carries the label $\bs_1$. 
On the right hand side, this loop turns into a sum 
of two types of contours. The first is a loop 
around both points $z$ and has normalization point 
$\mathsf{x}$ at the same position as on the original 
contour. 
%Since this therefore carries no additional multiplicative factors. 
The second type however, is a loop around only the point 
$z_1$, whose normalization point is related by analytic 
continuation to the marked point $\mathsf{x}$ on the original gray contour. This means that the relative order of the marked points $\mathsf{x}$ 
has changed and as a consequence of the exchange relation 
\eqref{eq:exchangeRel1}, a braiding factor gets generated.

This factor contains a full monodromy 
$\ex^{-2 \pi \ii b (e_1,\alpha_2)}$ due to 
%moving the normalization point $\mathsf{x}$ on the original gray curve around $z_2$, which is equivalent to 
moving the insertion point of the screening current $S_1$ 
counter-clockwise around $z_2$. The factor $\fq^{2(n_{2,1}-1)+n_{2,12}-n_{2,2}}$ then arises 
because one needs to relate the respective integrands defined by using two different choices of normalization points with the help of the exchange relation 
\eqref{eq:exchangeRel1}.

\begin{figure}[H]
\centering
\includegraphics[width=1.0\textwidth]{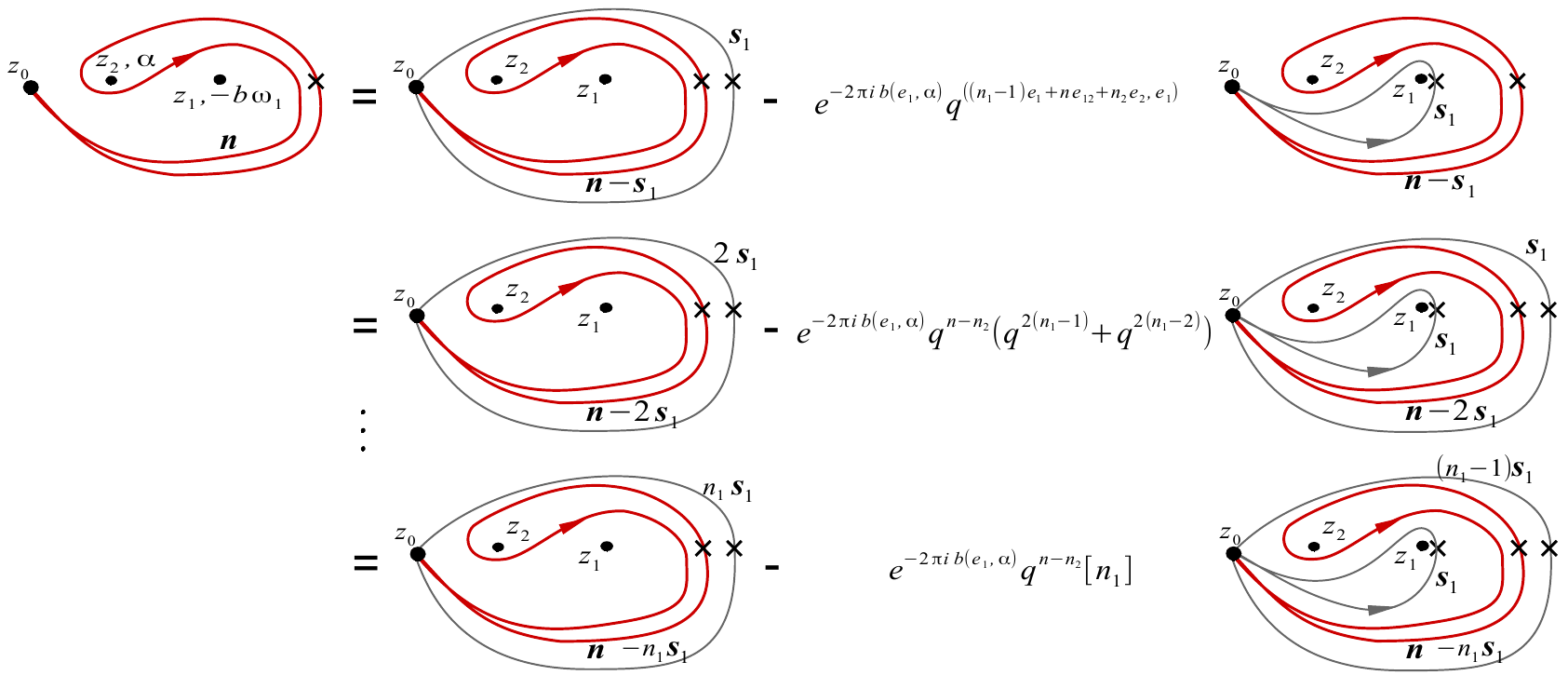}
\caption{{\it Example illustrating how quantum numbers arise in the braiding calculation.}}
\label{Example}
\end{figure}

Repeating the step described in Figure \ref{Braid2B} 
generates quantum numbers 
$[n] = \frac{1-\fq^{2n}}{1-\fq^2}$, as illustrated in 
Figure \ref{Example}, which has a generic vertex operator 
at $z_2$ and a degenerate operator at $z_1$. 
Let us explain a few more steps in this example, as it is relevant for one of the 
braiding calculations in which we are interested. 
We may denote 
$\gamma_1^{\bs_i}$ the loop around $z_1$ which is 
associated with screening charges $Q_1$ and $Q_{12}$ for 
$i=1$ and $12$, respectively,  and with no charge for $i=0$.
If we then denote $\beta_2^\bn$ the deformed loop around $z_2$ and $\beta_{21}^{\bn_{21}}$ the loop around both 
points $z$ and denote 
$I[\beta_{21}^{\bn_{21}} \beta^\bn \gamma_1^{\bs_i}]$ 
the integral over the indicated contours, then the  
intermediate results after resolving the contours associated with each type of screening charge are 
\begin{align}
I[\beta_2^{\mb{n}}\gamma_1^{\bs_0}] &= 
I[\beta_{21}^{n_1\bs_1}\beta_2^{\bn-n_1\bs_1}\gamma_1^{\bs_0}] - 
\ex^{-2\pi\ii b(e_1,\alpha)} \fq^{n-n_2} [n_1]
I[\beta_{21}^{(n_1-1)\bs_1}\beta_2^{\bn-n_1\bs_1}\gamma_1^{\bs_1}]  
\\
&= I[\beta_{21}^{\bn-n_2\bs_2}\beta_2^{n_2\bs_2}\gamma_1^{\bs_0}] 
-  
\ex^{-2\pi\ii b(e_{12},\alpha)} \fq^{-2+n_2} [n] I[\beta_{21}^{n_1\bs_1+(n-1)\bs_{12}}\beta_2^{n_2\bs_2}\gamma_1^{\bs_{12}} ] 
- 
\nn\\
&  \quad ~~
-\ex^{-2\pi\ii b(e_1,\alpha)} (\fq^{-1}-\fq) \fq^{-n_2} [n] I[\beta_{21}^{n_1\bs_1+(n-1)\bs_{12}+\bs_2}\beta_2^{n_2\bs_2}\gamma_1^{\bs_1} ] 
- 
\nn\\
& \quad ~~
-\ex^{-2\pi\ii b(e_1,\alpha)} \fq^{n-n_2} 
[n_1]  
I[\beta_{21}^{(n_1-1)\bs_1+n\bs_{12}}\beta_2^{n_2\bs_2}\gamma_1^{\bs_1}]  
\nn\\
&=
I[\beta_{21}^{\bn}\gamma_1^{\bs_0}] 
- 
\ex^{-2\pi\ii b(e_1,\alpha)} (\fq^{-1}-\fq) \fq^{-n_2} [n] I[\beta_{21}^{\bn-\bs_{12}+\bs_2}\gamma_1^{\bs_1} ] 
- 
\nn\\
& \quad ~~
-\ex^{-2\pi\ii b(e_{12},\alpha)} [n] \left( 
\fq^{-2+n_2} + (\fq^{-1}-\fq) \fq^{-1-n_2} [n_2]
\right) I[\beta_{21}^{\bn-\bs_{12}}\gamma_1^{\bs_{12}} ]  -
\nn\\
& \quad ~~
-\ex^{-2\pi\ii b(e_1,\alpha)} \fq^{n-n_2} 
[n_1] 
I[\beta_{21}^{\bn-\bs_1}\gamma_1^{\bs_1}]  - 
\ex^{-2\pi\ii b(e_{12},\alpha) \fq^{n-1-n_2}} [n_1][n_2]
I[\beta_{21}^{\bn-\bs_3}\gamma_1^{\bs_{12}}] ~.
\nn
\end{align}

\paragraph{Note:} There are  two additional technical observations to note here: {\it i)} contours associated with the composite screening 
charge $Q_{12}=Q_1Q_2-\fq Q_2Q_1$ are most safely treated when  
considered as linear combinations of contours for simple 
charges and {\it ii)} one must take care of the ordering of 
the newly generated contours, in particular when they 
are associated with the screening charges $Q_{12}$ and 
$Q_2$. Any reordering of contours implies a permutation 
of their associated normalization points, which 
generates braiding factors  
%and possibly additional terms. 
as illustrated for example by
\be \label{eq:monodromy-MP2}
\int_{\gamma} dy S_1(y) V_\alpha^{\mb{n}} (z) = 
V_\alpha^{\mb{n}+\bs_1} (z) 
\qqq\mathrm{vs.}
\qqq
\int_{\gamma} dy S_{12}(y) V_\alpha^{\mb{n}} (z) = 
\fq^{n_1} V_\alpha^{\mb{n}+\bs_{12}} (z)
~. 
\ee 
The case where a screening charge $Q_2$ is added has one additional feature. Changing the order of contours such that the $\bs_2$ contour is inside of the loops with labels 
$\bs_1$ means commuting an $S_2$ screening current past 
$S_1$, which by the definition of composite screening charges generates a current $S_{12}$ and thus 
%the  screened vertex operator $V_\alpha^{\bn+\bs_{12}-\bs_1}$.  
\be \label{eq:monodromy-MP22}
\int_{\gamma} dy S_2(y) V_\alpha^{\mb{n}} (z) 
=
\fq^{n_{12}-n_1 } V_\alpha^{\mb{n}+\bs_2} (z) - 
\fq^{-n_1}[n_1] 
V_\alpha^{\mb{n}+\bs_{12}-\bs_1} (z)~.
\ee

\paragraph{Part 2:} Resolving all of the deformed 
contours around the point $z_2$ in Figure \ref{Braidop2}, 
that are generated by the braiding procedure, creates a  combination of the types of loops in Figure \ref{Braid1B}.
The second part of the induction procedure described in 
section \ref{sec:braidingSVO1} is then the deformation of 
any such auxiliary 
loops that enclose both $z_1$ and $z_2$ in Figure 
\ref{Braid1B} into combinations of the basis 
contours in Figure \ref{Contours1}. An intermediate step is depicted in Figure 
\ref{Co-product2}. 
The same considerations as above apply here when deforming the gray contour on the left hand side. The 
resulting gray loops on the right have normalization points $\mathsf{x}$ 
that are either at the same position as on the original curve or 
moved to a different position and the integrands associated to different choices of normalization points are related by analytic continuation, which gives rise to a 
braiding factor. Furthermore, if the gray loop being deformed like in Figure \ref{Co-product2}
carries labels $\bs_{12}$ or $\bs_2$, it is necessary to 
reorder the resulting contours like we did in equations \eqref{eq:monodromy-MP2}, \eqref{eq:monodromy-MP22} so that they match the order prescribed in Figure \ref{Contours1}.

\begin{figure}[H]
\centering
\includegraphics[width=1.0\textwidth]{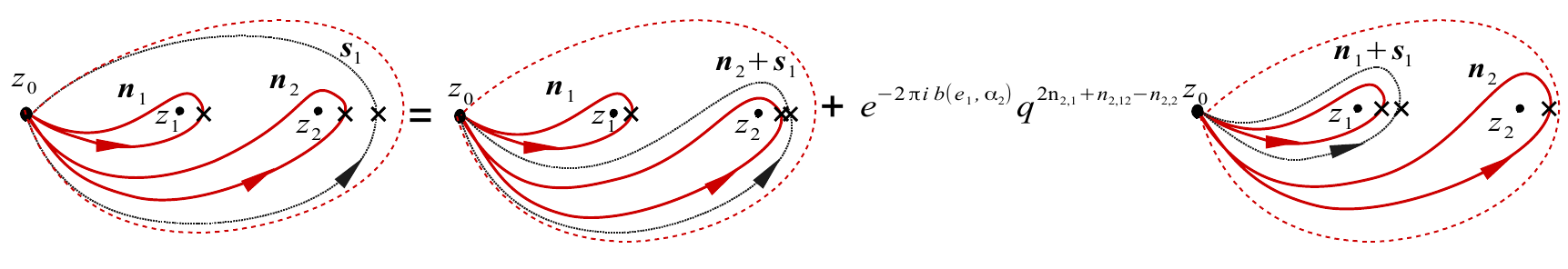}
\caption{{\it Decomposition of a simple contour associated 
with the root $e_1$ around two punctures into contours 
enclosing only one of the punctures. }}
\label{Co-product2}
\end{figure}

Let us now take the contours in Figure \ref{Braid1B} with 
a generic 
vertex operator inserted at the position $z_2$ and a 
degenerate  operator at $z_1$ and deform the loop 
enclosing both points $z$ into a combination of basis 
contours like  in Figure \ref{Contours1}.  
This case is relevant to the braiding calculation of interest here.  %We may call 
If we call 
$\gamma_1^{\bs_i}$ the loop around $z_1$ 
%which is 
%associated with screening charges $Q_1$ and $Q_{12}$ for 
%$i=1$ and respectively $12$ and with no charge for $i=0$.
%If we then name 
and $\beta_{21}^{\bn_{21}}$ the loop around both 
points $z$, 
%and $I[\beta_{21}^{\bn_{21}}\gamma_1^{\bs_i}]$ 
%the integral over the indicated contours, 
then the decomposition of integrals over $\beta_{21}$-type contours 
%in Figure \ref{Braid1B} 
leads to  
\begin{align} \label{eq:sample:A1}
& I[\beta_{21}^{\bn_{21}} \gamma_1^{\bs_0}] = 
\fq^{1-n}[n]
I[\gamma_2^{\bn_{21}-\bs_{12}}\gamma_1^{\bs_{12}}] + 
\fq^{1-n_1-n}[n_1] 
I[\gamma_2^{\bn_{21}-\bs_1}\gamma_1^{\bs_1}] + 
\fq^{-n_1-n} 
I[\gamma_2^{\bn_{21}}\gamma_1^{\bs_0}] , \\
& I[\beta_{21}^{\bn_{21}} \gamma_1^{\bs_1}] 
=\fq^{n_1-n_2} 
I[\gamma_2^{\bn_{21}}\gamma_1^{\bs_1}]
-\fq^{n-n_2} [n_2]
I[\gamma_2^{\bn_{21}-\bs_2}\gamma_1^{\bs_{12}}]
-
\fq^{-1-n_2}(1-\fq^2) [n] 
I[\gamma_2^{\bn_{21}+\bs_1-\bs_{12}}\gamma_1^{\bs_{12}}] , \nn\\
&I[\beta_{21}^{\bn_{21}} \gamma_1^{\bs_{12}}] = 
\fq^{n_2+n} I[\gamma_2^{\bn_{21}}\gamma_1^{\bs_{12}}] 
,~~ \nn
%\qqq\qqq\qqq\qqq
\end{align}
with notation 
$I[\beta_{21}^{\bn_{21}}\gamma_1^{\bs_i}]$ 
for the integral over the indicated contours and
where the loop $\gamma_2^{\bs_i}$ is a basis contour 
enclosing the point $z_2$. 
Combining the two parts of the braiding calculation, we 
reach the main result.

\paragraph{Main result:} The two types of braid 
matrices that we compute following the steps outlined 
above have a degenerate vertex operator inserted at 
either the point $z_1$ or $z_2$ and are explicitly  
\bea \label{eq:braiding-Vo1}
\sB [ W_{\alpha , -b \omega_1}^{\ss_k - \delta_\imath , \mb{d}_\imath} (z_2,z_1)] 
&=&
\sum_{\jmath=1}^3 (\sb_+)_{\imath\jmath} 
W_{-b \omega_1 , \alpha}^{\mb{d}_\jmath , \ss_k - \delta_\jmath} (z_1,z_2) 
\eea
\be \label{eq:braiding-Vo2}
\sB [ W_{-b \omega_1 , \alpha}^{\mb{d}_\imath , \ss_k - \delta_\imath } (z_2,z_1) ] = \sum_{\jmath=1}^3 (\sb_-)_{\imath\jmath} 
W_{\alpha , -b \omega_1}^{\ss_k -\mb{d}_\jmath , \mb{d}_\jmath} (z_1,z_2)~
\ee
using the notation $\ss_k=(s_1-k,k,s_2-k)$ and 
$\delta_1=(0,0,0),\delta_2=(1,0,0)$ and 
$\delta_3=(1,0,1)$. 
The matrix elements of $\sb_+$ are 
\bea \label{eq:braiding-Vo10}
\mathsf{b}_{11}^+ &=& 
\fq^{(h_1, \nu )} ~, \qqq 
\mathsf{b}_{22}^+ = 
\fq^{(h_2, \nu + e_1 )} ~, \qqq 
\mathsf{b}_{33}^+ = 
\fq^{(h_3, \nu + e_{12} )}
\\
\mathsf{b}_{12}^+ &=& 
\fq^{(h_2, \nu + e_1 )} 
(\fq-\fq^{-1})\fq^{-s_2-(e_1,\lambda)}
\left(
\fq^{1+2k} [s_1-k][1-s_1+s_2-k+(e_1,\lambda)] + 
\sV^{-1} [k]
\right)
\nn\\
\mathsf{b}_{23}^+ &=& 
\fq^{(h_3 , \nu + e_{12} )} 
(\fq^{-1}-\fq) \left(
[s_2-k][1-s_2+k+(e_2,\lambda)]\fq^{-(e_2,\lambda)}
- \sV^{-1} [k]\fq^{1-2 s_2 + (e_2,\lambda)}
\right)
\nn
\eea
and 
\begin{align}
\mathsf{b}_{13}^+ =
\fq^{(h_3 ,\nu + e_{12} )}
&(1-\fq^2)
\Big(
 [s_1-k][s_2-k][s_2-1-k-(e_2,\lambda)]
\fq^{1+4k-3s_2+(e_2-e_1,\lambda)}\\  
& +\sV^{-1}[k] %\times
\big([s_2-2-(e_{12},\lambda)]\fq^{2-2k-s_2+(e_{12},\lambda)}
-
[s_2-k]\fq^{-2-s_2-(e_{12},\lambda)} \nn\\
 & \qquad\qquad+
[s_1-k]\fq^{2k-3 s_2+(e_2-e_1,\lambda)}  
+ [s_2-k+1]\fq^{2k-2-3 s_2+(e_2-e_1,\lambda)}
\big)\nn \\
 & -\fq^{-1} \sV^{-2} [k][k-1]\fq^{-3s_2-(e_2-e_1,\lambda)}
\Big)~,
\nn\end{align}
while those of $\sb_-$ are
\bea \label{eq:braiding-Vo11}
 \mathsf{b}_{11}^- &=& \fq^{(h_1, \nu )} ~, \qqq 
 \mathsf{b}_{22}^- = 
\fq^{(h_2, \nu + e_1 )} ~, \qqq 
\mathsf{b}_{33}^- = 
\fq^{(h_3 , \nu + e_{12} )}
\\
 \mathsf{b}_{21}^- &=& 
\fq^{(h_1, \nu )} (\fq-\fq^{-1}) ~, \qqq 
\mathsf{b}_{32}^- = 
\fq^{(h_2, \nu + e_1 )}
(\fq^2-1)\fq^{-s_1+1}\left(\sV\fq^{2k} [s_1-1-k] - \fq^{2k} \right)
\nn\\
\mathsf{b}_{31}^- &=& 
\fq^{(h_1, \nu )}
(\fq-\fq^{-1}) \sV \fq^{s_1-1} ~
\qqq\qqq\qqq\qqq\qqq\qqq\qqq\qqq\qqq\qqq\qqq\qqq \nn
\eea
for $\alpha=-b\lambda$ and $\nu=\lambda-s_1 e_1 - s_2 e_2$. 
% The example calculations presented in greater detail in the paragraphs above provide the ingredients to compute the matrix elements $\mathsf{b}^+_{1\bullet}$.

%%%%%%%%%%%%%%%%%%%%%%%%%%%%%%%%%%%%%%%%%%%%%%%%%%%%%%%%%%%%

%%%%%%%%%%%%%%%%%%%%%%%%%%%%%%%%%%%%%%%%%%%%%%%%%%%%%%%%%%%%
%%%%%%%%%%%%%%%%%%%%%%%%%%%%%%%%%%%%%%%%%%%%%%%%%%%%%%%%%%%%

\section{Realisation of the generators $\se_i$  on screened vertex operators}
\label{sec:app:eaction}
\setcounter{equation}{0}
%%%%%%%%%%%%%%%%%%%%%%%%%%%%%%%%%%%%%%%%%%%%%%%%%%%%%%%%%%%%
%%%%%%%%%%%%%%%%%%%%%%%%%%%%%%%%%%%%%%%%%%%%%%%%%%%%%%%%%%%%

In this section we will indicate how to derive the identification between the 
action of $L_{-1}$ on screened vertex operators 
\eqref{def:SVoperator}, which removes one contour of 
integration, 
and the action of $\se_i$ generators on tensor products 
of representations \eqref{block-module}. 
Using notations similar to the ones introduced in section \ref{sec:W-comm-qgrp} for
the simpler case of $\cU_q(\fsl_2)$, 
we find that 
\be \label{eq:appD1}
\big[\,L_{-1}\,,\, {V}_n^\alpha (z)\,\big] - 
\partial_{z}{V}_n^\alpha (z)%+(\text{boundary terms})
= S(z_0)
(\fq-\fq^{-1}) \sk^{-1}\se {V}_{n-1}^\alpha (z),
\ee
%with boundary terms proportional to 
%$(\fq-\fq^{-1}) \sk^{-1}\se {V}_n^\alpha (z)$,  
by the following considerations. For screening number 
$n=2$, there are only two contours of integration. 
$L_{-1}$ acts by the Leibniz rule  on each screening current in ${V}_n^\alpha (z)$, producing
total derivative terms. 
%
%by differentiation on each 
%integration variable, where the corresponding contour 
%first needs to be deformed into a segment that connects 
%the base point $z_0$ and the insertion point $z$ of the 
%vertex operator $V_\alpha(z)$, a loop that encircles the 
%point $z$ creating a monodromy and another segment that 
%returns to the base point along the same path as the first segment. This produces 
The contributions from the two boundaries of the integration contour are related by %multiplicative 
monodromy factors. In this way it is not hard to arrive at  the following equation
\be 
\big[\,L_{-1}\,,\, V_2^\alpha (z)\,\big] =
\oint dy (1-\ex^{-4\pi\ii b(\alpha + b)})S(z_0)S(y) + 
(1-\ex^{-4\pi\ii b\alpha}) S(y)S(z_0)
V_\alpha (z) + \partial_z 
V_2^\alpha (z) ~. \nn
\ee
By factoring off the operator $S(z_0)$, this simplifies to 
\be 
\big[\,L_{-1}\,,\, V_2^\alpha (z) \,\big] 
=
\fq^{2(1-\lambda_\alpha)}(\fq-\fq^{-1})^{-1}
(\fq^{2\lambda_\alpha-1}-\fq^{-2\lambda_\alpha+1})(
\fq^2-\fq^{-2}) 
S(z_0) V_1^\alpha (z)
+ \partial_z V_2^\alpha (z)  ~. \nn
\ee
More generally, for arbitrary $n\in\mathbb{N}$ units of screening charge, we find by iterating these steps 
\be \label{eq:L-1action}
\big[\,L_{-1} \,,\,V^n_\alpha (z)\,\big]   = 
\partial_z V_n^\alpha (z)
+
\fq^{2(n-1-\lambda)}(\fq-\fq^{-1})^{-1}
(\fq^{2\lambda-n+1}-\fq^{-2\lambda+n-1})(\fq^n-\fq^{-n}) 
S(z_0) V_{n-1}^\alpha (z) ~.\nn
\ee
To compare now with the actions of the $\se$ and $\sk$ generators, notice that these are
\be
\se ~ e^\lambda_n = 
(\fq-\fq^{-1})^{-2}
(\fq^{2\lambda-n+1}-\fq^{-2\lambda+n-1})(\fq^n-\fq^{-n}) e^\lambda_{n-1} ~,
\quad
\sk^{-1} ~ e^\lambda_{n-1} = \fq^{2(n-1-\lambda)} e^\lambda_{n-1} ~,\nn
\ee
so we arrive indeed at equation \eqref{eq:appD1}. 

At higher rank, for $\cU_q(\fsl_3)$, one can similarly show   
for example
\be 
\big[\,L_{-1}\,,\, {V}_\bn^\alpha (z)\,\big] - 
\partial_{z}{V}_\bn^\alpha (z) \sim 
(\fq-\fq^{-1}) 
\left( \sk_1^{-1}\se_1 + \sk_2^{-1}\se_2 \right) 
{V}_\bn^\alpha (z)
\ee
The final observation to make in order to identify 
the actions of
$L_{-1}$ and $\se$ is that the action of $L_{-1}$ on 
composite screened vertex operators correctly reproduces 
the action of  the coproduct 
\begin{equation}
\big[\,L_{-1}\,,\, {V}_{\mb{n}_m , \ldots , \mb{n}_1}^{\alpha_m , \ldots , \alpha_1} (z_m , \ldots , z_1)\,\big]
-\sum_{l=1}^m\partial_{z_l}
{V}_{\mb{n}_m , \ldots , \mb{n}_1}^{\alpha_m , \ldots , \alpha_1} (z_m , \ldots , z_1)
\sim 
(\Delta(x)
{V}_{\mb{n}_m , \ldots , \mb{n}_1}^{\alpha_m , \ldots , \alpha_1} )(z_m , \ldots , z_1),\nn
\end{equation} 
 for 
\begin{equation}
x=(q-q^{-1})(\sk_1^{-1}\se_1+\sk_2^{-1}\se_2),\nn
\end{equation}
which we verified by direct computation. 

In order to see that the boundary terms occurring in the commutators with arbitrary generators of the $W$-algebra 
admit a similar representation in terms of the generators $\se_i$ 
the main point to observe is that all commutators between screening currents and $W$-generators 
can be represented as total derivatives.


\begin{thebibliography}{99}




\bibitem[AGT]{AGT}
L.~F.~Alday, D.~Gaiotto, and Y.~Tachikawa,
{\em Liouville Correlation Functions from Four-dimensional Gauge Theories},
Lett. Math. Phys. {\bf 91} (2010) 167--197.
% e-Print: arXiv:0906.3219



\bibitem[AGGTV]{AGGTV}
L. F. Alday, D. Gaiotto, S. Gukov, Y. Tachikawa, H. Verlinde,
{\em Loop and surface operators in $\mathcal{N}=2$ gauge theory and
Liouville modular geometry}, J. High Energy Phys. {\bf 1001} (2010) 113.

\bibitem[BMPTY]{BMPTY}
L. Bao, V. Mitev, E. Pomoni, M. Taki, F. Yagi,
{\it Non-Lagrangian Theories from Brane Junctions},
JHEP {\bf 1401} (2014) 175. 
%e-Print: arXiv:1310.3841


\bibitem[BFS]{BFS} R. Bezrukavnikov, M. Finkelberg, V. Schechtman,
{\it Factorisable sheaves and quantum groups},
Lecture Notes in Mathematics {\bf 1691}, Springer Verlag, Berlin, 1998.

\bibitem[BMP]{BMP} P. Bouwknegt, J. McCarthy, K. Pilch,
{\it Free field approach to two-dimensional conformal field theories}, 
Prog. Theor. Phys. Suppl. {\bf 102} (1990) 67-135. 

\bibitem[Bu]{Bu} M. Bullimore,
{\it  Defect Networks and Supersymmetric Loop Operators},
JHEP {\bf 1502} (2015) 066.
%e-Print: arXiv:1312.5001
 
\bibitem[BT1]{BT}
A.~G.~Bytsko and J.~Teschner,
{\em $R$-operator, co-product and Haar-measure for the modular double of 
$U_q(\fsl(2,\Bbb R))$}. 
Comm. Math. Phys. {\bf  240}  (2003) 171--196.

\bibitem[CF]{CF} L.O. Chekhov, V. Fock: {\it A quantum Teich\-m\"uller space},
  Theor. Math. Phys. {\bf 120} (1999) 1245--1259 
%(Preprint arXiv:math/9908165)

\bibitem[CGT]{CGT}
I. Coman, M. Gabella, J. Teschner,
{\it Line operators in theories of class $\mathcal{S}$, quantized moduli space of flat connections, 
and Toda field theory},
JHEP {\bf 1510} (2015) 143. 



\bibitem[DGOT]{DGOT} N. Drukker, J. Gomis, T. Okuda, J. Teschner,
{\em Gauge Theory Loop Operators and Liouville Theory},
J. High Energy Phys. {\bf 1002} (2010) 057.

\bibitem[F95]{Fa1} L. D.~Faddeev,
{\em Discrete Heisenberg-Weyl Group and Modular Group},
Lett. Math. Phys. {\bf 34} (1995) 249-254.

\bibitem[F99]{Fa2} L. D.~Faddeev,
{\em Modular double of a quantum group}, 
Conférence Moshé Flato 1999, Vol. I (Dijon),  149--156, 
Math. Phys. Stud., {\bf 21}, Kluwer Acad. Publ., Dordrecht, 2000,
{\tt arXiv:math/9912078}.

\bibitem[FL]{FL}
V. A. Fateev, A. V. Litvinov,
{\it Correlation functions in conformal Toda field theory I},
JHEP {\bf 0711} (2007) 002.

\bibitem[FW]{FW}
G. Felder, C. Wiecerkowski,
{\it Topological representations of the quantum group $U_q(\fsl(2))$},
Comm. Math. Phys. {\bf 138} (1991) 583-605. 


\bibitem[FG1]{FG1}
V.V. Fock, A. Goncharov,
{\it Moduli spaces of local systems and higher Teichm\"uller theory.} 
Publ. Math. Inst. Hautes \'Etudes Sci. {\bf 103} (2006) 1--211.


\bibitem[FG2]{FG2}  Fock, V. V.; Goncharov, A. B. 
{\it The quantum dilogarithm and representations of 
quantum cluster varieties.} 
Invent. Math. {\bf 175} (2009) 223--286.

\bibitem[FI]{FI} I. Frenkel, I. Ip,
{\it Positive representation of split real quantum groups and future perspectives},
International Mathematics Research Notices, {\bf 2014} (8) (2014) 2126-2164 



\bibitem[Ga]{Ga} D. Gaiotto,  {\it N=2 dualities},
JHEP {\bf 1208} (2012) 034. 

\bibitem[GMN]{GMN} D. Gaiotto, G. Moore and A. Neitzke,
{\it Wall-crossing, Hitchin systems, and the WKB approximation}.  
Adv. Math.  {\bf 234} (2013) 239--403.

\bibitem[Go]{Go} W. Goldman,
{\it Convex real projective structures on compact surfaces}, 
J. Differential Geom.  {\bf 31} (1990) 126-159.

\bibitem[GS]{GS} C. Gomez, G. Sierra,	
{\it The Quantum Symmetry of Rational Conformal Field Theories},
 Nucl.Phys. {\bf B352} (1991) 791-828 .
 
  \bibitem[GLF1]{GLF1} J. Gomis, B. Le Floch,
{\it 't Hooft Operators in Gauge Theory from Toda CFT},
 JHEP {\bf 1111} (2011) 114.
 % arXiv:1008.4139
 
 \bibitem[GLF2]{GLF2} J. Gomis, B. Le Floch,
{\it  M2-brane surface operators and gauge theory dualities in Toda},
 JHEP {\bf 1604} (2016) 183.
 % arXiv:1407.1852
 

 
 \bibitem[GOP]{GOP} J. Gomis, T. Okuda and V. Pestun, 
{\it Exact results for 't Hooft loops in gauge theories on $S^4$},
JHEP {\bf 1205} (2012) 141.
 
\bibitem[Hi]{Hi} N. Hitchin,
{\it Lie group and Teichm\"uller space},
Topology {\bf 31} (1992) 449--473.


\bibitem[HK]{HK} L. Hollands, O. Kidwai,
{\it Higher length-twist coordinates, generalized Heun's opers, and twisted superpotentials},
Preprint arXiv:1710.04438.

\bibitem[HN]{HN} L. Hollands, A. Neitzke,
{\it Spectral Networks and Fenchel--Nielsen Coordinates}, 
Lett. Math. Phys. {\bf 106} (2016) 811--877.
%e-Print: arXiv:1312.2979

 \bibitem[ILT]{ILT}
 N. Iorgov, O. Lisovyy, J. Teschner,
 {\it Isomonodromic tau-functions from Liouville conformal blocks},
 Comm. Math. Phys. {\bf 336} (2015) 671-694.
 

\bibitem[IMP]{IMP} M. Isachenkov, V. Mitev, E. Pomoni,
{\it  Toda 3-Point Functions From Topological Strings II},
 JHEP {\bf 1608} (2016) 066. 
%e-Print: arXiv:1412.3395

 \bibitem[IOT]{IOT}
Y. Ito, T. Okuda and M. Taki,
{\it Line operators on $S^1\times R^3$ and quantization of 
the Hitchin moduli space},
JHEP {\bf 1204} (2012) 010.


\bibitem[Ji]{Ji} M. Jimbo, 
{\it Monodromy problem and the boundary condition for some Painlevé equations}. 
Publ. Res. Inst. Math. Sci. {\bf 18} (1982) 1137?1161 (1982)

\bibitem[Ka1]{Ka1} R.M. Kashaev: 
{\it Quantization of Teichm\"uller spaces and 
  the quantum dilogarithm,} 
Lett. Math. Phys. {\bf 43} {(1998)} {105-115}.
%Preprint q-alg/9705021


\bibitem[Ka2]{K} R. M. Kashaev, {\it The quantum dilogarithm and Dehn 
  twists in quantum Teich\-m\"uller theory}.  Integrable structures of 
  exactly solvable two-dimensional models of quantum field theory 
  (Kiev, 2000),  211--221, NATO Sci. Ser. II Math. Phys. Chem., 35,
  Kluwer Acad. Publ., Dordrecht, 2001.
  
  \bibitem[Ki]{Ki} H.C. Kim,
  {\it The symplectic global coordinates on the moduli space of real projective structures},
  J. Differential Geom. {\bf 53} (1999) 359-401.
  
  \bibitem[La1]{L1}
S. Lawton,
{\it Generators, Relations and Symmetries in Pairs of 3x3 Unimodular Matrices},
J. Algebra 313 (2007), no. 2, 782--801,
Preprint math/0601132.

\bibitem[La2]{L2}
S. Lawton,
{\it Poisson Geometry of $SL(3,C)$-Character Varieties Relative to a Surface with Boundary},
Trans. Amer. Math. Soc. {\bf 361} (2009) 2397--2429,
Preprint math/0703251.

 
  \bibitem[LF]{LF} B. Le Floch,
  {\it S-duality wall of SQCD from Toda braiding},
 Preprint  arXiv:1512.09128.

  
\bibitem[MP]{MP} V. Mitev, E. Pomoni,
{\it  Toda 3-Point Functions From Topological Strings}
JHEP {\bf 1506} (2015) 049.
%  e-Print: arXiv:1409.6313


\bibitem[NRS]{NRS} N. Nekrasov, A. Rosly, S. Shatashvili,
{\em Darboux coordinates, Yang-Yang functional, and gauge theory}, 
Nucl. Phys. Proc. Suppl.~\textbf{216} (2011) 69--93.

\bibitem[NS]{NS}  N. Nekrasov, S. Shatashvili,
{\it Quantization of Integrable Systems and Four Dimensional Gauge Theories}, 
Proceedings of the 16th International Congress on Mathematical Physics, Prague, August 2009, P. Exner, Editor, pp.265-289, World Scientific 2010.
% e-Print: arXiv:0908.4052 

\bibitem[NT]{NT} I. Nidaiev, T. Teschner,
{\it On the relation between the modular double of $U_q(\fsl(2,R))$ and the quantum Teichmueller theory},
Preprint arXiv:1302.3454.

%\bibitem[NW]{NW} N. Nekrasov, E. Witten,
%{\it The Omega Deformation, Branes, Integrability, and Liouville Theory},
%JHEP {\bf 1009} (2010) 092.

\bibitem[Pe]{Pe} R.C. Penner, 
{\it The decorated Teichm\"uller space of punctured surfaces,} Comm. Math. Phys. {\bf 113} (1987) 299--339.


\bibitem[PT1]{PT1}
B.~Ponsot and J.~Teschner,
{\em Liouville bootstrap via harmonic analysis on a noncompact quantum group}, {\tt arXiv:hep-th/9911110}.

\bibitem[PT2]{PT2}
B.~Ponsot and J.~Teschner,
{\em Clebsch-Gordan and Racah-Wigner coefficients for a continuous series of representations of $U_q(\fsl(2,\mathbb{R}))$}, Commun.\ Math.\ Phys.\  {\bf 224} (2001) 613--655.


\bibitem[RRR]{RRR} C. Ramirez, H. Ruegg, M. Ruiz-Altaba,
{\it The Contour picture of quantum groups: Conformal field theories.} 
Nucl.Phys. {\bf B364} (1991) 195-233.


\bibitem[SS1]{SS1} G. Schrader, A. Shapiro, 
{A cluster realization of $U_q(\mathfrak{sl}_n)$ from quantum character varieties},
Preprint  arXiv:1607.00271.

\bibitem[SS2]{SS2} G. Schrader, A. Shapiro, 
{\it Continuous tensor categories from quantum groups I: algebraic aspects},
Preprint  arXiv:1708.08107.

\bibitem[T01]{T01}
J.~Teschner, {\em Liouville theory revisited}, 
Class.\ Quant.\ Grav.\  {\bf 18} (2001) R153--R222.
% arXiv:hep-th/0104158 


\bibitem[T03]{T03} J. Teschner,
{\it On the relation between quantum Liouville theory and the quantized Teichm\"uller spaces},
Int. J. Mod. Phys. {\bf A19S2} (2004) 459--477.
%arXiv:hep-th/0303149 

\bibitem[T10]{T10} J. Teschner,
{\it Quantization of the Hitchin moduli spaces, Liouville theory,
and the geometric Langlands correspondence I}.
Adv. Theor. Math. Phys.  {\bf 15}  (2011) 471--564.
%arXiv:1005.2846






\bibitem[T17]{TLect}  J. Teschner
{\it A guide to two-dimensional conformal field theory},
to appear 
in the proceedings of the Les Houches Summer School, Session 106: 
{\it Integrability: from Statistical Systems to Gauge Theory}, Les Houches, France, June 6--July 1, 2016,
Preprint arXiv:1708.00680.

\bibitem[TV]{TV13} J. Teschner, G. S. Vartanov,
{\it Supersymmetric gauge theories, quantization of 
moduli spaces of flat connections, and conformal field theory}.
Adv. Theor. Math. Phys. {\bf 19} (2015) 1--135. 

\bibitem[TW]{TW} Y. Tachikawa, Watanabe,	
{\it On skein relations in class S theories},
 JHEP {\bf 1506} (2015) 186.
%  e-Print: arXiv:1504.00121

\bibitem[ZZ]{ZZ}
A.B. and Al.B. Zamolodchikov, 
{\it Structure constants and conformal bootstrap in Liouville field theory,} 
Nucl. Phys. {\bf B477} (1996), 577--605.


\end{thebibliography}
\end{document}